\RequirePackage{fix-cm}
\documentclass[smallextended,natbib]{svjour3}       % onecolumn (second format)
\smartqed  % flush right qed marks, e.g., at end of proof
\usepackage{graphicx}
\usepackage[usenames,dvipsnames]{xcolor}
\usepackage{aas_macros}
\usepackage{url}
\usepackage{bm}
\usepackage[colorlinks=true, linktocpage, linkcolor={blue!50!black}, citecolor={blue!50!black}, urlcolor={blue!50!black}]{hyperref}
\usepackage{amssymb, amsmath, txfonts}
\usepackage{adjustbox}

\newcommand{\rhoc}{\mathrel{\rho_{\rm crit}}}
\newcommand{\Msol}{\mathrel{M_\odot}}

\newcommand{\sigT}{\mbox{$\sigma_{\mbox{\tiny T}}$}}
\newcommand{\Tcmb}{\mbox{$T_{\mbox{\tiny CMB}}$}}
\newcommand{\Kcmb}{\mbox{$K_{\mbox{\tiny CMB}}$}}
\newcommand{\kB}{\mbox{$k_{\mbox{\tiny B}}$}}

\newcommand{\taue}{\mbox{$\tau_{\mbox{\tiny e}}$}}
\newcommand{\yksz}{\mbox{$y_{\mbox{\tiny kSZ}}$}}

\newcommand{\Lamee}{\mbox{$\Lambda_{\rm ee}$}}

\newcommand{\LCDM}{\mbox{$\Lambda$CDM}}
\newcommand{\arcsec}{\mbox{{$^\prime$}{$^\prime$}}}
\newcommand{\arcmin}{\mbox{$^\prime$}}
\newcommand{\mue}{\mbox{$\mu_{\mbox{\tiny e}}$}}
\newcommand{\proton}{\mbox{$m_{\mbox{\tiny p}}$}}

\newcommand{\rs}{\mbox{$r_{\mbox{\scriptsize s}}$}}
\newcommand{\expf}[1]{{{\rm e}^{#1}}}
\newcommand{\Te}{{T_{\rm e}}}
\newcommand{\Tmw}{{T_{\mbox{\tiny mw}}}}
\newcommand{\Tx}{{T_{\mbox{\tiny X}}}}
\newcommand{\keV}{{{\rm keV}}}
\newcommand{\Ne}{{n_{\rm e}}}
\newcommand{\me}{{m_{\rm e}}}
\newcommand{\Pe}{{P_{\rm e}}}
\newcommand{\id}{{{\rm d}}}
\newcommand{\vek}[1]{{\bm #1}}

\begin{document}

\setcounter{tocdepth}{2}
\setcounter{secnumdepth}{4}

\title{Astrophysics with the Spatially and Spectrally Resolved Sunyaev-Zeldovich Effects}
%placed here. General acknowledgements should be placed at the end of the article.}
%}
\subtitle{A Millimetre/Submillimetre Probe of the Warm and Hot Universe}

%\titlerunning{Astrophysics with the Spatially and Spectrally Resolved Sunyaev-Zeldovich Effect}        % if too long for running head

\author{
Tony Mroczkowski \and
Daisuke Nagai  \and
Kaustuv Basu \and 
Jens Chluba \and
Jack Sayers \and
R\'emi Adam \and 
Eugene Churazov \and 
Abigail Crites \and
Luca Di Mascolo \and 
Dominique Eckert \and
Juan Macias-Perez \and
Fr\'ed\'eric Mayet \and
Laurence Perotto \and
Etienne Pointecouteau \and
Charles Romero \and
Florian Ruppin \and 
Evan Scannapieco \and
John ZuHone
}


\institute{
Tony Mroczkowski \at
European Southern Observatory (ESO), Karl-Schwarzschild-Str. 2, D-85748 Garching, Germany\\
\email{tony.mroczkowski@eso.org}
\and 
Daisuke Nagai \at
Department of Physics, Yale University, PO Box 208101, New Haven, CT, USA \\
Yale Center for Astronomy and Astrophysics, PO Box 208101, New Haven, CT, USA 
\and
Kaustuv Basu \at 
Argelander-Institut f\"{u}r Astronomie, Universit\"{a}t Bonn, Auf dem H\"{u}gel 71, D-53121 Bonn, Germany
\and 
Jens Chluba \at
Jodrell Bank Centre for Astrophysics, School of Physics and Astronomy,
The University of Manchester, Manchester, M13 9PL, U.K.
\and
Jack Sayers  \at
Division of Physics, Math, and Astronomy, California Institute of Technology, Pasadena, CA 91125, USA
\and
R\'emi Adam \at
Laboratoire Leprince-Ringuet (LLR), Avenue Chasles, 91120 Palaiseau, France
%Centro de Estudios de F\'isica del Cosmos de Arag\'on (CEFCA), Plaza San Juan, 1, planta 2, E-44001, Teruel, Spain
\and 
Eugene Churazov \at
Max-Planck-Institut f\"{u}r Astrophysik, Karl-Schwarzschild-Str. 1, D-85748 Garching, Germany\\
Space Research Institute, Profsoyuznaya 84/32, Moscow 117997, Russia
\and
Abigail Crites \at 
Department of Physics, California Institute of Technology, 1216 California Blvd., Pasadena, CA
\and
Luca Di Mascolo \at
Max-Planck-Institut f\"{u}r Astrophysik, Karl-Schwarzschild-Str. 1, D-85748 Garching, Germany
\and 
Dominique Eckert \at
Department of Astronomy, University of Geneva, Ch. d'Ecogia 16, CH-1290 Versoix, Switzerland
%Max-Planck-Institut f\"{u}r extraterrestrische Physik, Giessenbachstrasse 1, 85748 Garching, Germany
\and
Juan Mac{\'{\i}}as-P\'erez \at
Univ. Grenoble Alpes, CNRS, Grenoble INP, LPSC-IN2P3, 53, av. des Martyrs, 38000 Grenoble, France
\and
Frederic Mayet  \at
Univ. Grenoble Alpes, CNRS, Grenoble INP, LPSC-IN2P3, 53, av. des Martyrs, 38000 Grenoble, France
\and
Laurence Perotto  \at
Univ. Grenoble Alpes, CNRS, Grenoble INP, LPSC-IN2P3, 53, av. des Martyrs, 38000 Grenoble, France
\and
Etienne Pointecouteau  \at
IRAP, Universit\'e de Toulouse, CNRS, CNES, UPS, Toulouse, France
\and
Charles Romero  \at
University of Pennsylvania, 209 South 33rd Street, Philadelphia, PA 19104-6396, USA
\and
Florian Ruppin  \at
Univ. Grenoble Alpes, CNRS, Grenoble INP, LPSC-IN2P3, 53, av. des Martyrs, 38000 Grenoble, France
Kavli Institute for Astrophysics and Space Research, Massachusetts Institute of Technology, Cambridge, MA 02139, USA
\and 
Evan Scannapieco  \at
School of Earth and Space Exploration, Arizona State University, P.O. Box 876004, Tempe, AZ 85287, USA
\and
John ZuHone  \at
Smithsonian Astrophysical Observatory, 60 Garden St. MS-03, Cambridge, MA 02138, USA
}

\date{Received: date / Accepted: date }

\maketitle

\begin{abstract}
In recent years, observations of the Sunyaev-Zeldovich (SZ) effect have had significant cosmological implications and have begun to serve as a powerful and independent probe of the warm and hot gas that pervades the Universe. 
As a few pioneering studies have already shown, SZ observations both complement X-ray observations -- the  traditional tool for studying the intra-cluster medium -- and bring unique capabilities for probing astrophysical processes at high redshifts and out to the low-density regions in the outskirts of galaxy clusters.
Advances in SZ observations have largely been driven by developments in centimetre-, millimetre-, and submillimetre-wave instrumentation on ground-based facilities, with notable exceptions including results from the {\it Planck} satellite.  Here we review the utility of the thermal, kinematic, relativistic, non-thermal, and polarised SZ effects for studies of galaxy clusters and other large scale structures, incorporating the many advances over the past two decades that have impacted SZ theory, simulations, and observations. 
We also discuss observational results, techniques, and challenges, and aim to give an overview and perspective on emerging opportunities, with the goal of highlighting some of the exciting new directions in this field.

\end{abstract}

\tableofcontents
%\bigskip

\section{Introduction}
\label{sec:intro}

In the hierarchical picture of structure formation, hot, X-ray emitting plasmas with temperatures exceeding $\simeq 10^5$~K trace the largest overdensities to form in large scale structure: galaxy clusters, galaxy groups, galaxies, and intergalactic filaments.
While X-ray observations are the main tool for probing the emission from these structures, particularly at temperatures $\gtrsim 10^7$~K (or $ \kB \Te \gtrsim 1$~keV), it has long been known that the free electrons in the ionised plasma scatter cosmic microwave background (CMB) photons to, on average, higher energies, producing several unique spectral signatures referred to as the Sunyaev-Zeldovich (SZ) effects, the beginnings of which were first calculated nearly 50 years ago \citep{Zeldovich1969,Sunyaev1970}. 
The thermal SZ effect, which is the dominant effect due to up-scattering of CMB photons by the hot intra-cluster medium (ICM), was proposed in \cite{SZ1972}, shortly after the discoveries of X-ray emission from clusters (\citealt{Byram1966,Bradt1967}; see e.g.\ \citealt{Sarazin1986} and other chapters of these proceedings for reviews), and of the CMB itself \citep{Penzias1965}, as a test of whether the CMB was truly cosmic in origin.  

Over the past half century, observational studies relying on the SZ effects have matured from early detections and imaging of known, massive clusters to using SZ measurements as a powerful tool for detection in wide-field surveys and detailed astrophysical studies.  In parallel, many more facets of the SZ effect have been discovered, and our understanding has been refined, while both simulations and deep X-ray observations have transformed our view of the intra-cluster, intra-group, circumgalactic, and warm-hot filamentary structures on cosmic scales.  
Yet we have only begun to scratch the surface.  We take this opportunity to review the advances of the past two decades, since the highly cited and seminal reviews of \cite{Birkinshaw1999} and \cite{Carlstrom2002}, and discuss opportunities for the near future and beyond.

\subsection{A brief history of SZ observations}
\label{sec:SZobs_in_a_nutshell}

The first attempts to measure the thermal SZ focused on the decrement, and were performed using single-dish telescopes operating at radio wavelengths near $15$~GHz  \citep{Pariiskii1973, Gull1976, Lake1977, Birkinshaw1978}.
By the mid-1980's, measurements with the Owens Valley Radio Observatory (OVRO) 40-meter started to produce more reliable detections in several tens of hours of integration time \citep{Birkinshaw1984}.
Interferometric observations were performed later with the Very Large Array \citep[VLA;][]{Moffet1989} and the Ryle telescope \citep{Jones1993, Grainge1993}. These two interferometers also provided the first SZ imaging and purported detections based solely on an SZ signal, including the so-called ``dark clusters'' (i.e., clusters with no X-ray counterparts), which upon further investigation appeared to be spurious \citep{Richards1997, Jones1997}. 
Subsequent higher frequency (30--90~GHz) interferometric measurements with the OVRO 10-meter array and the Berkeley-Illinois-Maryland Array \citep[BIMA;][]{Carlstrom1996} and its successors the Sunyaev-Zeldovich Array and the Combined Array for mm-wave Astronomy \cite[SZA \& CARMA, respectively;][]{Muchovej2007}), the Arcminute MicroKelvin Interferometer \citep[AMI;][]{Zwart2008} and the Array for Microwave Background Anisotropy \citep[AMiBA;][]{Lin2009} began interferometric imaging of large statistical samples of known clusters. 
Around the same time, photometric instruments using bolometers such as the Sunyaev-Zeldovich Infrared Experiment \citep[SuZIE;][]{Wilbanks1994} and Diabolo \citep{Desert1998}, mounted at the focus of large single dish telescopes, allowed mm-wave measurements at $\sim$150 GHz, where the SZ decrement is strongest.
The next generation of photometric imaging arrays, including Bolocam \citep{Glenn1998} and the Atacama Pathfinder Experiment Sunyaev-Zeldovich instrument \citep[APEX-SZ;][]{Schwan2003}, were able to build large statistical samples of cluster images using single dish measurements. 

The first measurement of the SZ increment toward a cluster---A2163---was performed by the PRONAOS balloon-borne experiment at 350~GHz.  When combined with Diabolo and SuZIE measurements, it delivered the first spectro-photometric characterisation of the SZ spectrum \citep{Lamarre1998}. 
Soon after, the first successful imaging of the SZ increment was demonstrated using SCUBA on the JCMT \citep{Komatsu1999}.
In the following decade, higher-quality SZ increment imaging was made possible using the LABOCA bolometer camera, once again for the cluster A2163 \citep{Nord09}, while further SCUBA observations enabled the first large statistical study in this part of the spectrum \citep{Zemcov2007}. 
Shortly thereafter, using the {\it Herschel}-SPIRE photometer, the first detection of the SZ effect at $\gtrsim600$~GHz was published by \cite{Zemcov2010}. Building upon these spectro-photometric measurements, the first high-resolution SZ spectral measurement was performed towards the cluster RX~J1347.5-1145 using the Z-Spec instrument \citep{Zemcov2012}. Subsequent multi-band photometric measurements towards a high-velocity sub-cluster in MACS~J0717.5+3745 showed the first high-significance deviation from the classical tSZ effect spectrum \citep{Sayers2013, Adam2017ksz}, confirming the low-significance, $\lesssim 2\sigma$ indication of a non-zero kSZ signal reported in \cite{Mroczkowski2012}.

The first high-spatial resolution SZ images were produced by the Diabolo camera on the IRAM 30-meter telescope and the NOBA camera on the NRO 45-meter telescope. Targeting RX~J1347.5-1145---which quickly superseded A2163 as the SZ target of choice---Diabolo delivered a $5\arcmin \times 5\arcmin$ map of with $\sim$20$\arcsec$ resolution \citep{Pointecouteau1999, Pointecouteau2001}, and NOBA delivered a $2\arcmin \times 2\arcmin$ maps with $\sim$13$\arcsec$ resolution \citep{Komatsu1999,Komatsu2001}. Both maps exhibit an excess SZ signal to the southeast of the X-ray peak and central active galactic nucleus (AGN).  This peak was not seen in the X-ray by ROSAT PSPC \citep{Schindler1995}, but was soon confirmed by {\it Chandra} \citep{Allen2001}. At the same time, OVRO/BIMA obtained $\approx$20$\arcsec$ resolution in some of their interferometric images, including hints of an offset in the SZ centroid RX~J1347.5-1145 to the southeast, but these features were never analysed in detail, and the cluster was instead used for determinations of the angular diameter distance \citep[e.g.][]{Reese2002,Bonamente2006}. The next significant improvement in angular resolution came with the MUSTANG camera on the 100-meter Green Bank Telescope \citep[GBT;][]{Dicker2008}, which imaged shock-heated gas and pressure substructures in a handful of clusters at $\sim 9\arcsec$ resolution \citep[e.g.][]{Mason2010,Korngut2011}. As discussed in more detail in \S~\ref{sec:ICMstructures}, recent interferometric imaging with the Atacama Large Millimeter/Submillimeter Array (ALMA) has been used to probe one of the shock features in the `El Gordo' cluster at a resolution of $3.5\arcsec$ through observations of the SZ effect \citep{Basu2016}, while ALMA and Atacama Compact Array (ACA, \S~\ref{sec:alma_B3}) further improved constraints on RX~J1347.5-1145 \citep{Kitayama2016,Ueda2018}.

Nearly a decade ago, the South Pole Telescope (SPT) and the Atacama Cosmology Telescope (ACT) obtained sufficiently high mapping speeds to detect previously unknown clusters in wide-field surveys based on their thermal SZ effect signals \citep{Staniszewski2009, Menanteau2010}. Subsequent to these ground-based surveys, the {\it Planck} satellite surveyed the full sky in 9 photometric bands spanning the range 30--850~GHz, delivering a final catalogue of roughly 2000 SZ-selected clusters \citep{Planck2016_XXVII}. The {\it Planck} survey data have also been used to measure the SZ effect spectrum with the broadest frequency coverage to date ({\it e.g.}, \citealt{Planck2011_VII, Hurier2016rSZ, Erler2018}).

It must be noted that the backdrop to the developments of the past two decades has been the transformational X-ray observations of the ICM with the {\it Chandra} and {\it XMM-Newton} X-ray observatories.
Understanding the physical processes inside the ICM relates directly to our ability to find galaxy clusters in X-ray or SZ surveys, and to model the ICM accurately to infer cluster masses for cosmological applications. SZ observations -- especially deep, targeted observations -- are also helpful in probing a wide range of physical phenomena and the imprints of merger events, such as shocks and cold fronts, energy dissipation through turbulence, and random bulk motions within the ICM. 
It is thus important to continue along this trajectory and develop the next generation cm/mm/submm tools for both surveys and targeted observations, particularly to deliver higher spatial and spectral resolution.

\subsection{Theoretical and computational developments}

As mentioned previously, the multiple manifestations of the SZ effect did not come as one discovery, but as a long series of discoveries and refinements.  Much of this progress was made in parallel with observational and instrumental developments, as well as advances in simulations and our understanding of ICM physics.  We discuss the many aspects of the SZ effect in \S~\ref{sec:sz_theory}, aiming to provide a comprehensive but concise guide for the reader, while the references within indicate the long history of developments.

\subsection{Organisation of this review}

This review is presented as follows.
In \S~\ref{sec:sz_theory}, we discuss the fundamental theory behind the multiple manifestations of, and nuances to, the SZ effects. In \S~\ref{sec:ICMthermodynamics}, we discuss how the SZ effects can be used to study the thermodynamics of ICM. 
In \S~\ref{sec:ICMstructures}, we present recent advances in our understanding of ICM structure, such as AGN feedback, turbulence, and discontinuities. In \S~\ref{sec:obscons}, we discuss the practical considerations of performing observations of the SZ effects. In \S~\ref{sec:instrumentation}, we highlight several of what we consider to be the main subarcminute resolution observatories today, and discuss a few near-term instruments under development and some long-term instruments and projects under study. In \S~\ref{sec:conclusion} we conclude with an outlook on what the future may bring.

%%%%%%%%%%%%%%%%%%%%%%%%%%%%%%%%%%%%%%%%%%%%%%%%%%%%%%%%%%%%%%%%%%
\section{Overview of the SZ effects}
\label{sec:sz_theory}
%------
The Sunyaev-Zeldovich (SZ) effects \citep{SZ1972,Sunyaev1980} are caused by the scattering of CMB photons with the free electrons residing in the potential wells of clusters of galaxies and, more broadly speaking, the diffuse plasma at large scales.\footnote{For now we shall ignore the presence of primordial CMB temperature anisotropies, which will become relevant when considering polarisation effects (\S~\ref{sec:pSZ}).  We will also neglect multiple-scattering effects, only briefly discussing them in \S~\ref{sec:msSZ}.} This leads to several CMB signals in the direction of clusters that can be used to learn about ICM physics and cosmology.

The physics behind the SZ signals is quite simple. Electrons at rest with respect to the isotropic CMB produce no net effect, as the number of photons scattered in and out of the line of sight is the same. However, moving electrons can transfer some of their kinetic energy to the CMB photon field through the Doppler effect. This can be appreciated by studying the Compton scattering relation for the ratio of the scattered to initial photon frequency \citep[e.g.,][]{Jauch1976}
%------------------------------------------------------------------------------------
\begin{equation}
\frac{\nu'}{\nu}=\frac{1-\beta \mu}{1-\beta \mu'+\frac{h\nu}{\gamma\me c^2}(1-\mu_{\rm sc})}
\approx \frac{1-\beta \mu}{1-\beta \mu'}
%\approx 1-\beta (\mu-\mu')+(\beta\mu')^2+\mathcal{O}(\beta\mu')^3
.
\label{eq:nup_nu_relation}
\end{equation}
%------------------------------------------------------------------------------------
Here $\beta=\varv/c$ is the speed of the scattering electron with Lorentz factor $\gamma=1/\sqrt{1-\beta^2}$ in units of the speed of light, $c$; $\me$ is the electron mass; $h$ is the Planck constant; $\mu$ and $\mu'$ are respectively the direction cosines of the incoming and scattered photon with respect to the incoming electron; and $\mu_{\rm sc}$ is the corresponding direction cosine between the incoming and scattered photons. 

Since the typical energy of the CMB photons is very small when compared with that of the electrons, $h\nu\ll \gamma \me c^2$, one can neglect the corresponding photon recoil correction, which is usually dominant in the classical Compton effect. Hence, only Doppler and aberration terms are relevant (no Lorentz factors appear in the right hand side of Eq.~\ref{eq:nup_nu_relation}), and Klein-Nishina corrections, $\mathcal{O}(h\nu/\me c^2)$, can be omitted\footnote{In the rest-frame of the moving electron, the scattering event can actually be calculated using the Thomson limit for the differential cross section provided that $h\nu\ll \me c^2$ in this frame.}. The maximal photon energy after the scattering event is, $\nu'_{\rm max}=\nu\,(1+\beta)/(1-\beta)>\nu$, for a photon being back-scattered in a head-on collision with the electron. Similarly, the minimal scattered photon energy is $\nu'_{\rm min}=\nu\,(1-\beta)/(1+\beta)<\nu$ for the initial photon travelling in the same direction as the incoming electron and then being back-scattered. However, these scattering events occupy a small phase-space volume, and for any given $\mu$, up-scattering of the CMB photon occurs when $\mu'>\mu$, or equivalently, when the scattered photon deflects {\it towards} the direction of the incoming electron. Furthermore, the angular distributions of the incoming electrons and photons play a crucial role for the net energy exchange, as we will see below.

As these simple arguments already illustrate, scattering by moving electrons leads to a change of the CMB intensity in the direction of galaxy clusters, with the spectral shape of the signal depending on the {\it velocity distribution} of the electrons. Thermal electrons, described by an isotropic (relativistic) Maxwell-Boltzmann distribution, give rise to the so-called thermal SZ (tSZ) effect (\S~\ref{sec:tSZ_detail}), while the cluster's bulk motion ($\delta$-function in velocity space) causes the kinematic SZ (kSZ) effect (\S~\ref{sec:kSZ_detail}). Depending on the characteristic speed of the electrons, relativistic corrections can become important, yielding the relativistic SZ (rSZ) effect (\S~\ref{sec:rSZ_detail}). Finally, non-thermal velocity distributions (e.g., in the cocoons of radio galaxies, or turbulence and magnetic fields) can create the non-thermal SZ (ntSZ) effect (\S~\ref{sec:ntSZ_detail}).

Although the physical origin of all these SZ signals is due to simple electron scattering, each of them has slightly different spectral and spatial dependence across the cluster. With future multi-frequency observations, covering both the low- ($\nu \lesssim 150\,{\rm GHz}$) and high-frequency ($\nu \gtrsim 220\,{\rm GHz}$) parts of the CMB blackbody, we are thus in principle able to distinguish them. This will provide an exciting opportunity for extracting valuable information about the structure of the cluster's atmosphere and its gas physics. We now explain each of the SZ signals in turn, highlighting how to compute them and detailing which physical parameters they can inform.

\begin{figure}
\begin{center}
  \includegraphics[trim=5cm 0cm 5cm 0cm, clip=true, width=0.75\textwidth]{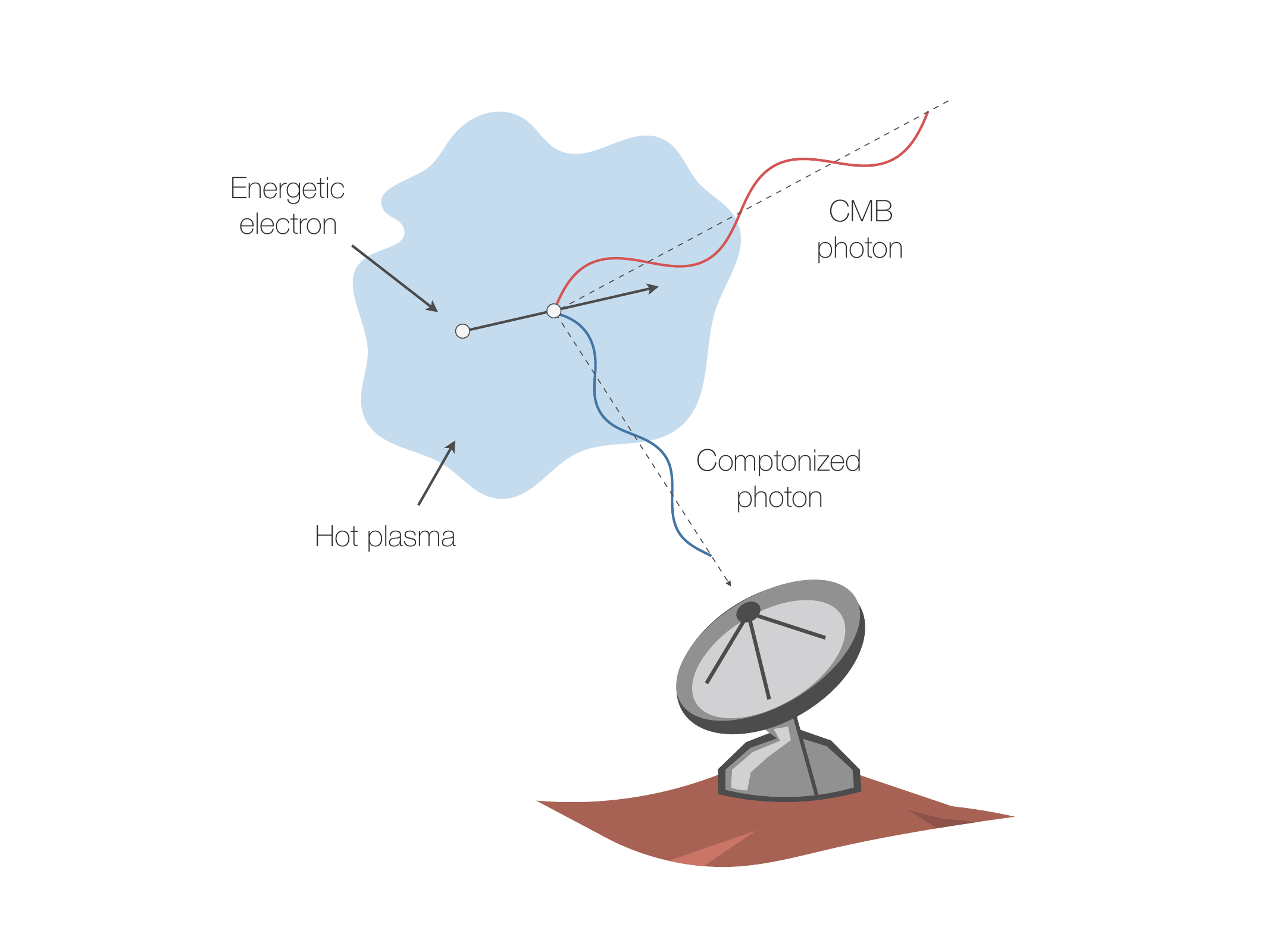}
\end{center}
\caption{Updated illustration based on the classic L.\ van Speybroeck SZ diagram adapted by J.\ E.\ Carlstrom. A CMB photon (red) enters the hot ICM (light blue) from an arbitrary angle, and on average is up-scattered to higher energy (blue) by an electron (black). The largest energy is imparted on the photon when it is scattered into the direction of the incoming electron, and it is minimal when deflected into the direction opposite to the incoming electron. However, on average scattering constellations with $\simeq 90^\circ$ angles between the particles are most relevant for the tSZ. The total momentum in the interaction is conserved, so the electron is essentially undeflected by the interaction.}
\label{fig:sz}
\end{figure}

\subsection{The thermal SZ effect}
\label{sec:tSZ_detail}
%------------------------------------------------------------------------------------
As CMB photons pass through regions of hot thermal gas (see schematic representation in Figure \ref{fig:sz}), inverse Compton scattering moves them from the low-frequency region of the blackbody spectrum towards higher energies. In single-scattering events with electrons at speed $\beta$ drawn from an isotropic velocity distribution there is no net effect, as the gains and losses average out to leading order, leaving a second order term. The average energy gained by a CMB photon in each scattering is determined by $\Delta \nu/\nu\simeq (4/3)\, \beta^2\simeq 4 k\Te/\me c^2$ \citep[e.g.,][]{Rybicki1979, Sazonov2000}. In the last step, we used $\beta_{\rm th}^2/3 \approx k\Te/\me c^2$ for a thermal (non-relativistic) velocity distribution. Similarly, a narrow photon line broadens by $\Delta \nu/\nu\simeq \sqrt{(2/3)\, \beta^2} \simeq \sqrt{2 k\Te/\me c^2}$ in each scattering event.
In the non-relativistic limit, both effects can be incorporated using the Kompaneets equation \citep{Kompa56}, which when applied to the case of SZ clusters\footnote{Stimulated scattering and recoil terms can be omitted.} reduces to a simple diffusion equation in frequency-space. This approach was originally used by \citet{Zeldovich1969} to compute the distinctive tSZ distortion signal, often referred to as the Compton $y$-type distortion. 
The corresponding distortion is given by\footnote{An alternative derivation uses that the superposition of blackbodies with slightly different temperatures, as indeed caused by the scattering process, is no longer a blackbody \citep{Zeldovich1972}. In this case, the $y$-parameter is related to the temperature dispersion, $y=\frac{1}{2}\,\left<\frac{\Delta T}{T}\right>^2$, induced by Doppler-shifts.}: 
%------------------------------------------------------------------------------------
\begin{equation}
\Delta I_\nu \approx I_0\,y \frac{x^4 \expf{x}}{(\expf{x}-1)^2}\,\left( x \,\frac{\expf{x}+1}{\expf{x}-1} -4 \right)\equiv I_0\,y\,g(x)
\label{eq:dI_tsz}
\end{equation}
%------------------------------------------------------------------------------------ 
in terms of the CMB intensity. Here $x= h\nu/\kB \Tcmb \approx \nu /56.8$ GHz, with \Tcmb\ denoting the temperature of the CMB,
%------------------------------------------------------------------------------------ 
\begin{equation}
\label{eq:I0_cmb}
I_0 = \frac{2 (\kB \Tcmb)^3}{(h c)^2} = 270.33 \, \left[\frac{\Tcmb}{2.7255\,{\rm K}}\right]^3~\rm MJy/sr,
\end{equation}
%------------------------------------------------------------------------------------ 
and the classical tSZ spectral function $g(x)$ is defined implicitly in Eq.~\eqref{eq:dI_tsz}.
Assuming $\Delta I_\nu/I_\nu\ll 1$, one can use the derivative with respect to temperature of the Planck function to alternatively express the signal in terms of the effective CMB temperature, yielding:
%------------------------------------------------------------------------------------
\begin{equation}
\frac{\Delta\Tcmb}{\Tcmb} \approx y \left( x \,\frac{\expf{x}+1}{\expf{x}-1} -4 \right)=y \, f(x).
\label{eq:dT_tsz}
\end{equation}
%------------------------------------------------------------------------------------ 
The function $f(x)$, defined implicitly above, is the classical tSZ spectrum in terms of $\Delta\Tcmb$.
The change in the effective CMB temperature is proportional to the Compton-$y$ parameter, which depends on the Thomson scattering optical depth, $\taue$, and temperature of the hot electron gas, $\Te$, as
%------------------------------------------------------------------------------------
\begin{equation}
y \equiv \int  \frac{\kB \Te}{\me c^2}\,\id \taue
=
\int  \frac{\kB \Te}{\me c^2}\,\Ne \sigT \id l
=
\frac{\sigT}{\me c^2} \int \Pe \,\id l.
\label{eq:Comptony}
\end{equation}
%------------------------------------------------------------------------------------
Here $\sigT$ is the Thomson cross section, $\Pe = \Ne \kB \Te$ is the pressure due to the electrons and $\Ne$ is the number density of the electrons. 
The integral is performed over the proper distance along the line of sight. Thus the magnitude of the tSZ signal is a direct measure of the integrated line of sight pressure.  

\begin{figure*}
\begin{center}
  \includegraphics[width=1.0\textwidth]{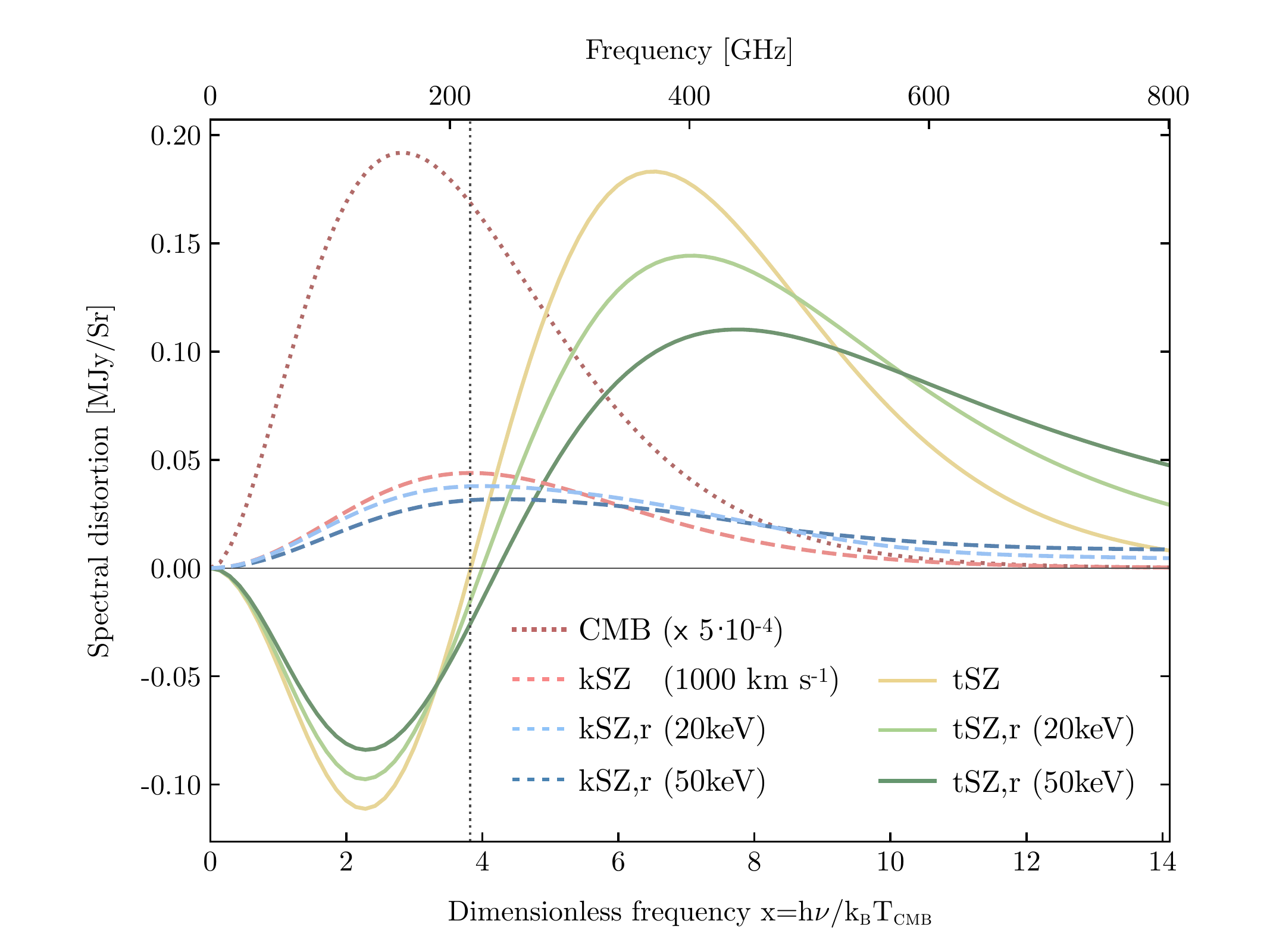}
\end{center}
\caption{Thermal (solid) and kinematic (dashed) SZ spectra, including relativistic corrections for various temperatures. We assumed an optical depth $\taue=10^{-2}$ and an overall Compton parameter $y=10^{-4}$. The dotted, dark red curve illustrates the shape of the unscattered CMB spectrum, which for comparison was scaled by a factor of $5\times 10^{-4}$.}
\label{fig:szspectrum}
\end{figure*}

Typical clusters contain electrons with $k\Te\simeq 5-10\,\keV$, or $k\Te/\me c^2 \simeq 0.01-0.02$. The central optical depth can reach $\taue=\int \Ne \sigT \id l \simeq 10^{-2}$, such that for massive clusters one can expect $y\simeq 10^{-4}$ (see e.g.\ the cluster outskirts review in these proceedings). The spectral shape of the $y$-distortion in terms of CMB intensity for $y=10^{-4}$ is illustrated in Figure~\ref{fig:szspectrum}. The signal manifests itself as a deficit in the number of photons at frequencies below $\nu_{\rm null} \approx  217$ GHz ($\approx 1.4$~mm) and an increase above $\nu_{\rm null}$ (n.b.\ photon number is conserved by scattering). In the Rayleigh-Jeans limit ($x \ll 1$) the change in the effective temperature $\Delta T$ reduces to $\Delta T/T\approx -2 y$, while at high frequencies ($x\gg 1$) one has $\Delta T/T\approx y (x-4)$.

An important property of the tSZ is its near redshift independence. The spectral shape remains unchanged and the tSZ does not suffer from redshift-dimming\footnote{The CMB intensity increases towards higher redshifts and thus the tSZ signal starts off at a higher level at $z>0$, remaining constant relative to the CMB intensity.  Alternatively, this can be understood qualitatively as the scattering events producing a fractional change in the intensity of the CMB that would be constant for an observer at any given epoch.}. This makes the tSZ a unique probe of the large-scale structure in the Universe \citep[see e.g.][for a number of excellent reviews]{Sunyaev1980b, Rephaeli1995ARAA, Birkinshaw1999, Carlstrom2002, Kitayama2014}. It can furthermore in principle be used to measure the expansion rate of the Universe through the combination with X-ray data, exploiting the differing density dependencies in their surface brightness integrals (see Equations \ref{eq:xray_sb} and \ref{eq:SB}) to infer the angular diameter distance to a cluster \citep{SilkWhite1978, Cavaliere1979, Birkinshaw1979, Hughes1998, Battistelli2003}.

\begin{figure*}
\begin{center}
  \includegraphics[trim=5cm 0cm 5cm 0cm, clip=true, width=0.75\textwidth]{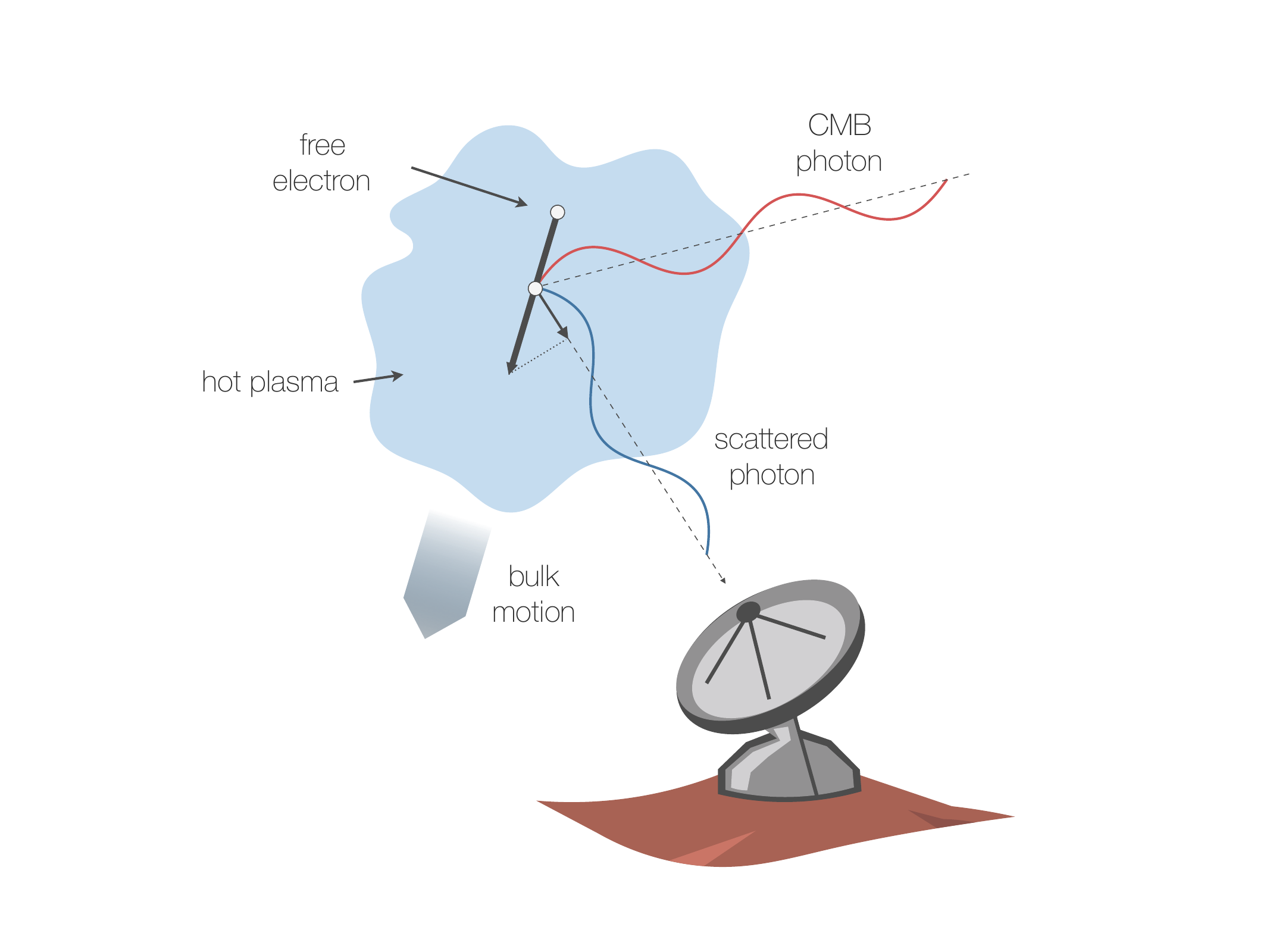}
\end{center}
  \caption{Diagram for the kinematic SZ. A CMB photon (red) enters the hot ICM (light blue) from an arbitrary angle, and, for this geometry, is up-scattered to higher energy (blue) by an electron (white dot) in a moving ICM. To first order order in $\beta=\varv/c$ only the line of sight projection of the cluster's bulk motion matters for the corresponding intensity change.}
\label{fig:ksz}
\end{figure*}

\subsection{The kinematic SZ effect}
\label{sec:kSZ_detail}

%------------------------------------------------------------------------------------
The kinematic SZ (kSZ) effect is due to scattering of CMB photons interacting with free electrons undergoing bulk motion relative to the CMB rest-frame \citep[][see the schematic diagram in Figure~\ref{fig:ksz}]{Sunyaev1980}. In contrast to the tSZ, the velocity distribution of the electrons is anisotropic in the case (i.e., mono-directional), such that, upon averaging over all scattering angles of the photons drawn from the isotropic CMB, a linear order Doppler term $\propto \beta$ remains. Similar physics play a crucial role in the formation of the CMB temperature and polarisation anisotropies \citep[e.g.,][]{Sunyaev1970, Peebles1970, Hu1995CMBanalytic}. 

The kSZ effect produces a shift in CMB temperature in the direction $\vek{n}$ of a moving cluster, which can be written as
%------------------------------------------------------------------------------------
\begin{equation}
\frac{\Delta \Tcmb}{\Tcmb} \approx - \int {\sigT} \Ne \,\vek{n}\cdot\vek{\beta}_{\rm p} \,\id l
= - \int \vek{n}\cdot\vek{\beta}_{\rm p} \,\id \taue \equiv - \yksz
\label{eq:dT_ksz}
\end{equation}
%------------------------------------------------------------------------------------
in terms of the effective shift in CMB temperature, or
%------------------------------------------------------------------------------------
\begin{equation}
\Delta I_\nu  \approx - I_0 \frac{x^4 \expf{x}}{(\expf{x}-1)^2}\,\yksz
\label{eq:dI_ksz}
\end{equation}
%------------------------------------------------------------------------------------
in terms of the CMB intensity shift (where $I_0$ is defined in Eq.\ \ref{eq:I0_cmb}). This signal is indistinguishable from that of the hot and cold spots in the primary CMB, unless scale-dependent information or correlations with other astronomical data are exploited. 
To leading order, only the line of sight component, $\beta_{\rm p,\parallel}=\vek{n}\cdot\vek{\beta}_{\rm p}=\mu_{\rm p}\,\beta_{\rm p}$, of the cluster's peculiar motion is relevant. The temperature shift is furthermore negative for a line of sight velocity away from the observer, and positive when the cluster approaches the observer (consistent with the convention that the $\vek{z}$ vector increases with redshift).
The parameter $\yksz$ (see Eq.~\ref{eq:dT_ksz}), is often defined in the literature as a kSZ analogue to the Compton-$y$ parameter \citep[e.g.][]{ruan2013}.

The typical peculiar motions of clusters in the standard cosmological model are expected to be $\beta_{\rm p}\lesssim {\rm few}\times10^{-3}$ (speed $\simeq {\rm few}\times 100\,{\rm km~s^{-1}}$), such that $\yksz \lesssim {\rm few}\times10^{-5}$ at the map peak. This is about one order of magnitude smaller than the typical tSZ $y$-parameter (see Figure~\ref{fig:szspectrum}). 
By measuring this signal, one can in principle map out the large-scale motions of baryons in the Universe, constraining the growth of structure and testing isotropy/homogeneity of the Universe \citep{Zhang2011, Hand2012}.

\subsection{Relativistic corrections to the SZ effect}
\label{sec:rSZ_detail}
%------------------------------------------------------------------------------------
The tSZ and kSZ signals discussed above were obtained assuming non-relativistic speeds for the electrons. For typical bulk motions $\beta_{\rm p}\lesssim {\rm few}\times10^{-3}$ this is rather well justified. However, electrons in a thermal gas at temperature $k\Te\simeq 5\,\keV$ have typical speeds $\beta \simeq \sqrt{3k\Te/\me c^2}\simeq 0.1-0.2$. In this case, the non-relativistic approximation (i.e., only including terms up to second order in $\beta$) for the tSZ derivation no longer suffices and (special-)relativistic corrections become relevant \citep[e.g.,][]{Wright1979, Fabbri1981, Rephaeli1995}, leading to changes of the tSZ and kSZ spectral shapes. Thus, relativistic corrections to the SZ signals (`rSZ' for short) can in principle provide additional information about the detailed temperature and velocity structure of the ICM, as we discuss below.

Relativistic corrections to both the tSZ and kSZ have been studied in detail using various methods to evaluate the Compton collision term. {\it Direct numerical integration} with various levels of analytical reductions \citep[e.g.,][]{Wright1979, Fabbri1981, Rephaeli1995, Pointecouteau1998, Molnar1999, Ensslin2000, Dolgov2001, Nozawa2009} are generally time-consuming but least prone to errors or loss of precision. For a specified temperature range, the computation can be accelerated using {\it fits} to the numerical results \citep{Nozawa2000fitting, Itoh2004fittingII} or {\it pre-computed basis functions} \citep{Chluba2012SZpack}. Insight into the physics of the problem can be gained through {\it Taylor-series approximations} to various orders in $\Theta_{\rm e}=k\Te/\me c^2$ and $\beta_{\rm p}=\varv_{\rm p}/c$ \citep{Stebbins1997, Challinor1998, Sazonov1998, Itoh98, Nozawa1998SZ, Shimon2004, Nozawa2006}. However, at $k\Te\gtrsim 5\,\keV$ the expansions start to converge quite slowly, since the width of the scattering kernel quickly rises with $\Te$. In particular in the Wien tail of the CMB spectrum this becomes problematic \citep[e.g.,][]{Stebbins1997, Itoh98, Chluba2012SZpack}, as derivatives of an exponential are not well approximated by a sum of exponentials. Another kinematic correction is due to the {\it motion of the observer}, which can be added by performing a Lorentz transformation of the SZ signal from the CMB rest frame to the observer's frame \citep{Chluba2005b, Nozawa2005}. This leads to a dipolar modulation of the cluster number counts across the sky \citep{Chluba2005b} and can also be interpreted as a distortion of the CMB dipole spectrum \citep{Balashev2015}.

After this broad-brush overview, let us discuss the physics of relativistic temperature corrections in more detail. One of the important effects is that the average energy shift and broadening per scattering both increase more quickly with temperature than in the non-relativistic limit. Including terms up to fourth order in $\Theta_{\rm e}$, one finds \citep{Sazonov2000, Chluba2012SZpack}
%------------------------------------------------------------------------------------
\begin{align}
\left<\frac{\Delta \nu}{\nu}\right>
&\approx 
4 \Theta_{\rm e}+10 \Theta^2_{\rm e}
+\frac{15}{2}\Theta^3_{\rm e}-\frac{15}{2}\Theta^4_{\rm e}+\mathcal{O}(\Theta^5_{\rm e}),
\nonumber\\
\left<\left(\frac{\Delta \nu}{\nu}\right)^2\right>
&\approx 2 \Theta_{\rm e}+47 \Theta^2_{\rm e}
+\frac{1023}{4}\Theta^3_{\rm e}+\frac{2505}{4}\Theta^4_{\rm e}
+\mathcal{O}(\Theta^5_{\rm e}),
\label{eq:moments_of_scattering_kernel}
\end{align}
%------------------------------------------------------------------------------------
for the first two moments of the scattering kernel, which illustrates the effect. At higher temperature, it is thus no longer possible to assume $|\Delta \nu/\nu|\ll 1$ in the scattering event, one of the key assumptions in the derivation of the Kompaneets equation. This implies that both higher order derivatives of the blackbody function\footnote{Derivatives of the blackbody occupation number, $n_{\rm bb}=1/(\expf{x}-1)$, can be given in closed form using Eulerian numbers, $\left<\!\begin{array}{c} k \\ m \end{array}\!\right>$, yielding $x^k\partial^k_x n_{\rm bb}=(-x)^k \expf{-x}/(1-\expf{-x})^{k+1}\,\sum_{m=0}^{k-1} \,\left<\!\begin{array}{c} k \\ m \end{array}\!\right> \,\expf{-mx}$ \citep{Chluba2012SZpack}.} and higher order moments of the scattering kernel become important.  

Overall relativistic temperature corrections cause a broadening of the tSZ signal, with a systematic shift towards higher frequencies, as is illustrated in Figure~\ref{fig:szspectrum}. Through this additional dependence it is in principle possible to directly measure the temperature of the cluster \citep{Wright1979, Fabbri1981, Rephaeli1995, Pointecouteau1998}. 
As discussed further in \S \ref{sec:rsz_temperature}, rSZ temperature determinations have been attempted for individual clusters \citep[e.g.,][]{Hansen2002, Prokhorov2012rSZ, Chluba2013} and in stacking analyses \citep{Hurier2016, Erler2018}, albeit with large errors in both. In the future this could also become possible at the tSZ power spectrum level \citep{Remazeilles2019} and for cluster number counts \citep{Fan2003}.
The relativistic tSZ in principle can also be used to measure the CMB temperature-redshift relation, which in non-standard cosmologies could depart from the standard $\Tcmb\propto(1+z)$ scaling \citep[e.g.,][]{Fabbri1978, Rephaeli1980, Battistelli2002, Luzzi2009}. Even if it is fairly difficult to create a change in the CMB temperature at late times \citep{Chluba2014} without violating CMB spectral distortion constraints from {COBE/FIRAS} \citep{Mather1994, Fixsen1996}, this is an interesting application of the rSZ.

Turning to the physics of relativistic corrections to the kSZ, just as those for the tSZ, additional higher order terms become relevant when evaluating the kernel moments. The temperature corrections to the kSZ (at leading order $\propto \beta_{\rm p}\,\Theta_{\rm e}$) are illustrated in Figure~\ref{fig:szspectrum}. As with the tSZ effect, these again lead to a broadening and systematic shift of the kSZ signal towards higher frequencies.
Mixed kinematic and temperature corrections have been considered for terms up to second order in $\beta_{\rm p}$ \citep{Sazonov1998, Nozawa1998SZ, Shimon2004, Nozawa2006, Chluba2012SZpack}. The bulk motion of the cluster breaks the isotropy of the CMB blackbody field at second order inducing a correction to the monopole $\propto \beta^2_{\rm p}$ and quadrupole $\propto \beta_{\rm p}^2 (3\mu_{\rm p}^2-1)/2$ inside the cluster's rest frame. 
Through relativistic kSZ one can therefore in principle measure {\it two projections} of the cluster's velocity. The remaining azimuthal degeneracy can in principle be broken by considering pSZ (\S~\ref{sec:pSZ}). For this the primordial quadrupole terms have to be carefully subtracted \citep{Chluba2014mSZII}. Instead of Taylor series approximations of the Compton collision term, by considering the tSZ effect in the cluster frame for an anisotropic CMB photon field, one can account for the kinematic corrections through the Lorentz transformation, an approach that also ensures the correct interpretation of the optical depth as cluster-frame optical depth \citep{Chluba2012SZpack}. Higher order corrections in $\beta_{\rm p}$ can be added by using a  multipole-dependent Kompaneets equation or anisotropic scattering kernels \citep{Chluba2012, Chluba2014mSZII}, however, these modifications are expected to be negligible.

All the aforementioned relativistic corrections (kinematic due to the cluster's and observer's motions as well as those due to temperature) can be efficiently computed using {\tt SZpack}\footnote{\url{www.chluba.de/SZpack}} \citep{Chluba2012SZpack, Chluba2013} with complementary numerical and analytical methods implemented to ensure a large amount of flexibility. {\tt SZpack} furthermore allows including the effect of spatial variations of the temperature and velocity fields (both along the line of sight and within the beam), which give rise to frequency-dependent morphological changes of the SZ signals \citep{Chluba2013}. In the future, the associated {\it moments} of the velocity and temperature field could become direct observables, and by combining with X-ray data we may be able to extract detailed information about the ICM structure.

\subsection{The non-thermal SZ effect}
\label{sec:ntSZ_detail}
%------------------------------------------------------------------------------------
In the discussion of the tSZ effect we only considered {\it thermal} distributions of electron momenta. However, the momentum distribution can be more complex and highly relativistic, e.g., having long power-law tails at high energies \citep[e.g.,][]{Ensslin2000}. In this case, we refer to the associated distortions as non-thermal SZ (ntSZ) effect.
Assuming a general\footnote{With the condition $\gamma h \nu \ll \me c^2$ such that Klein-Nishina corrections are still negligible.} isotropic momentum distribution $f(p)$, where $p$ is the dimensionless electron momentum (i.e., $p=p_{\rm phys}/\me c$ and $\gamma=\sqrt{1+p^2}$) and the distribution $f(p)$ has a normalisation $\int_0^\infty \! f(p) p^2\id p = 1$, one can write the scattered CMB signal as\footnote{Here we used a property of the scattering kernel that implies $P(s, p)\equiv P(-s, p)/\expf{-3s}$.}
%------------------------------------------------------------------------------------
\begin{equation}
\Delta I_\nu \approx I_0\,x^3\,\taue \int_0^\infty f(p) p^2\id p \int_{-s_{\rm m}(p)}^{s_{\rm m}(p)}
\,P(s, p) \,\left[n_{\rm bb}(x\,\expf{s})-n_{\rm bb}(x)\right] \id s.
\label{eq:dI_general_beta}
\end{equation}
%------------------------------------------------------------------------------------ 
Here, we introduced the blackbody occupation number, $n_{\rm bb}=1/(\expf{x}-1)$, and maximal logarithmic energy shift, $s_{\rm m}(p)=\ln\left[(1+\beta)/(1-\beta)\right]$ with $\beta(p)=p/\sqrt{1+p^2}$. The scattering kernel, $P(s, p)$, is given by \citep[e.g.,][]{Rephaeli1995, Fargion1997, Fargion1998, Sazonov2000, Ensslin2000, Shimon2002, Colafrancesco2003}
%------------------------------------------------------------------------------------
\begin{align}
P(s, p)&=\frac{3}{8}\Bigg\{
\frac{\expf{s}(1+\expf{s})}{p^5}\left[\frac{3+2p^2}{2p}\left(|s|-s_{\rm m}\right)+\frac{3+3p^2+p^4}{\sqrt{1+p^2}}\right]
\nonumber\\
&\qquad\qquad-\frac{|1-\expf{s}|}{4p^6}\left[1+(10+8p^2+4p^4) \expf{s} + \expf{2s}\right]
\Bigg\}
\label{eq:scatt_kernel}
\end{align}
%------------------------------------------------------------------------------------ 
and is normalised as $\int_{-s_{\rm m}}^{s_{\rm m}} P(s, p)\,\id s=1$. One furthermore finds the first moment, $\left<\Delta \nu/\nu\right>=\int_{-s_{\rm m}}^{s_{\rm m}} \left(\expf{s}-1\right)\,P(s, p)\,\id s=4p^2/3$. We emphasise that Eq.~\eqref{eq:dI_general_beta} is only strictly valid when anisotropies in the radiation and velocity fields can be neglected. This assumption can in principle be violated by the presence of magnetic fields \citep{Koch2003, Gopal2010JCAP}, pressure anisotropies \citep{Khabibullin18}, anisotropies in the scattering medium \citep{Chluba2014mSZI, Chluba2014mSZII}, and kSZ effects, as mentioned above.

\begin{figure}[t]
\begin{center}
  \includegraphics[width=1.0\textwidth]{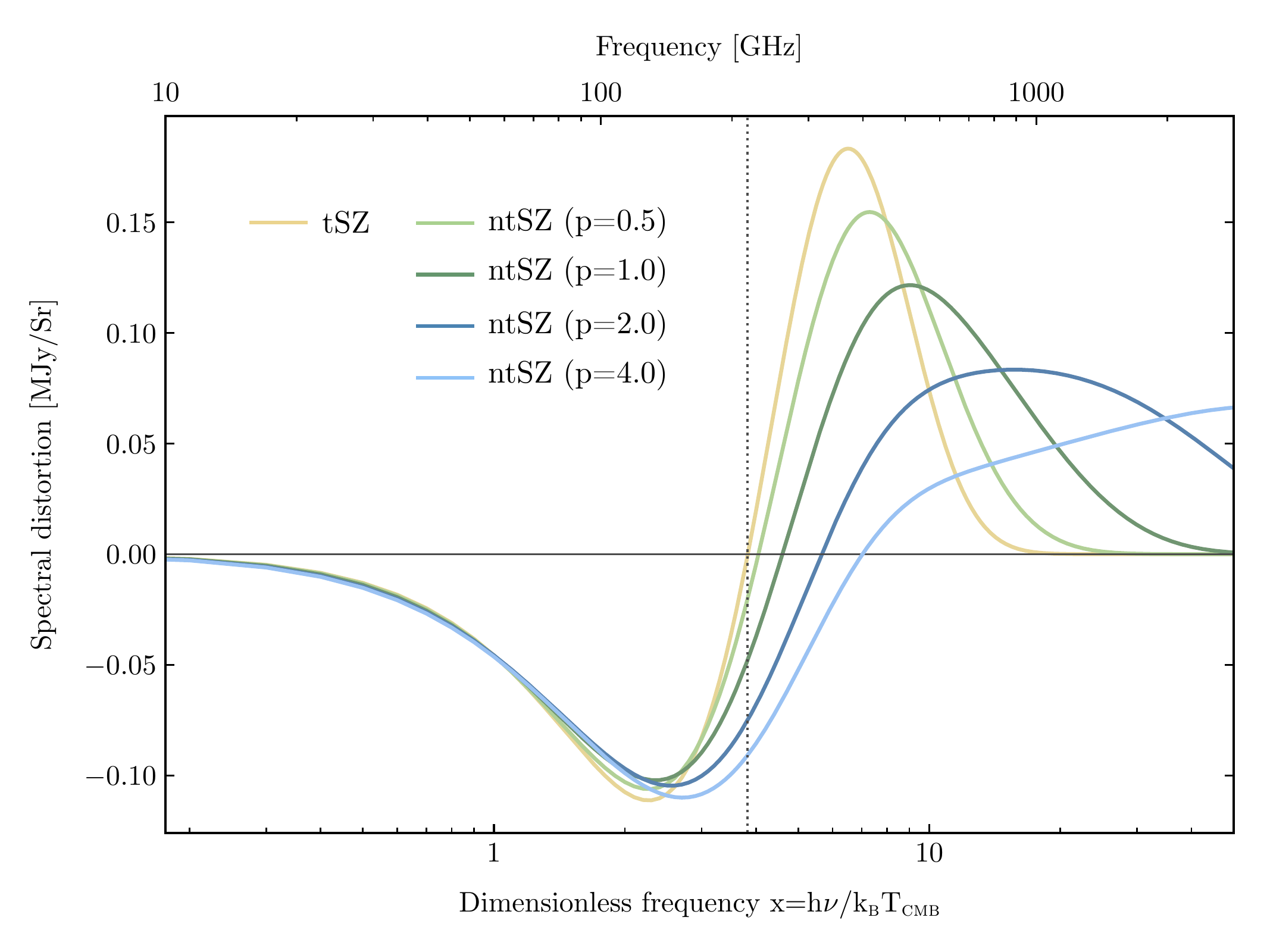}
\caption{Comparison of the non-relativistic tSZ spectrum and the non-thermal SZ spectrum for mono-energetic electrons with varying momentum, where $p$ is the dimensionless electron momentum. The $y$-parameter for the tSZ case was set to $y=10^{-4}$. In Eq.~\eqref{eq:dI_general_beta}, we set $\taue=y/[\beta^2/3]$ for the non-thermal cases to mimic a fixed overall $y$-parameter. The ntSZ contribution is expected to be a small fraction ($\lesssim 1\%-10\%$) of the tSZ signal.}
\label{fig:ntszspectrum}
\end{center}
\end{figure}

Equation~\eqref{eq:dI_general_beta} allows us to compute the scattered signal for a given $f(p)$. If we assume the electron energies follow a relativistic Maxwell-Boltzmann distribution, $f_{\rm rMB}(p, \Theta)=\expf{-\sqrt{1+p^2}/\Theta}/[\Theta K_2(1/\Theta)]$, we can reproduce the tSZ effect with relativistic temperature corrections. Here $K_2(x)$ is the modified Bessel function of the second kind, ensuring the correct normalisation of $f(p)$. For power-law distributions, analytic expressions can be given in terms of incomplete $\beta$-functions \citep{Ensslin2000, Colafrancesco2003}. Similarly, for mono-energetic electrons, $f_\delta(p)=\delta(p-p_0)/p^2$, the outer integral becomes trivial and only one numerical integral over the blackbody distribution has to be carried out. In Figure \ref{fig:ntszspectrum}, we show the ntSZ spectrum for this case. CMB photons are strongly up-scattered towards higher frequencies once the momentum exceeds $p\simeq 1$ \citep{Ensslin2000, Colafrancesco2003, Malu2017}. This effect could allow us to search for the presence of aged radio plasmas and relativistic outflows from AGN using CMB measurements. The total contribution to the Comptonisation of CMB photons caused by relativistic plasmas could reach a level equivalent to $y\simeq {\rm few}\times 10^{-6}$ \citep{Ensslin2000, Shimon2002}. However, the detailed shape of the ntSZ spectrum can become complex since it depends strongly on the energy distribution of the electrons doing the scattering. A direct comparison with the tSZ is thus not as straightforward. The ntSZ could also allow us to shed light on the nature of dark matter and annihilating particles \citep{Colafrancesco2004, Colafrancesco2006}.

\subsection{The polarised SZ effects}
\label{sec:pSZ}
%------------------------------------------------------------------------------------
It is well-known that Thomson scattering of CMB photons by free electrons inside clusters leads to a small polarisation effect \citep[e.g.,][]{Sunyaev1980, Sunyaev1981, Gibilisco1997, Kamionkowski1997b, Audit1999, Sazonov1999, Challinor2000, Itoh2000Pol, Shimon2009}. The physics behind this signal is very similar to the way the primordial CMB $E$-mode polarisation patterns are created \citep{Bond1984Pol, Seljak1997, Kamionkowski1997}, the most important ingredient being the presence of a {\it quadrupole anisotropy} in the local radiation field. However, for hot electrons inside clusters the Comptonisation of CMB photon can lead to a distinct frequency-dependence of the polarised SZ (pSZ). For this, the origin of the local quadrupole anisotropy plays a crucial role, as we discuss now.

The largest effect is due to the local primordial CMB quadrupole, leading to a polarisation amplitude $\simeq 0.1 \,\taue Q$ in units of the CMB temperature \citep{Kamionkowski1997b, Sazonov1999}. Here $Q$ is the CMB quadrupole moment at the location of the cluster. The signal thus can reach a level of $\simeq 10^{-8}$ of the primary CMB temperature for rich clusters ($\taue\simeq 0.01$ and $Q\simeq 10^{-5}$), and could allow us to measure the CMB quadrupole at different locations in the Universe, thereby in principle circumventing the cosmic variance limit \citep{Kamionkowski1997b, Portsmouth2004, Bunn2006, Yasini2016, Meyers2018}. However, the frequency-dependence of this signal is identical to that of the primordial CMB polarisation anisotropies, such that knowledge about the cluster location and its redshift are required.

The second largest pSZ signal is due to second scattering corrections of the thermal ($\simeq \taue^2 \Theta_{\rm e}$) and kinematic ($\simeq \taue^2 \beta_{\rm p}$) SZ signals \citep[e.g.,][]{Sazonov1999}. In both cases, the spectral-dependence follows that of the re-scattered tSZ and kSZ respectively. The signal related to the second scattering of the tSZ signal can be of similar order of magnitude as the one caused by the primordial CMB quadrupole, but since a scattering-induced quadrupole anisotropy is required (see \S~\ref{sec:msSZ}) it vanishes along the centre of the cluster unless asphericities or anisotropies in the medium are present \citep[e.g.,][]{Sazonov1999, Puy2000, Shimon2009}. For the second scattering pSZ induced by the kSZ only the tangential component of the clusters motion is relevant.

At second order in $\beta_{\rm p}$, a $y$-type CMB quadrupole is also induced by the cluster's motion. This is the original pSZ effect that was highlighted by \citet{Sunyaev1980}. 
Again only the tangential component of the cluster's velocity, $\beta_{\rm p,\perp}$, is relevant, causing a quadrupole pattern that leads to a polarisation signal $\simeq 0.1 \,\taue \,\beta^2_{\rm p,\perp}$.
This can in principle be used to measure the tangential velocity of the cluster's motion in the plane of the sky. 
Similarly, internal gas motions \citep[e.g.,][]{Chluba2001, Cooray2002, Chluba2002, Diego2003, Lavaux2004, Shimon2006, Maturi2007} can lead to complex polarisation patterns.

Yet another physical mechanism that could lead to polarisation of the SZ signal is associated with anisotropic distribution function of electrons \citep{Khabibullin18}. The ICM is an astrophysical example of weakly collisional plasma where the Larmor frequencies of charged particles greatly exceed their collision frequencies. In such conditions the magnetic moments of particles are conserved between collisions and the evolving magnetic fields or heat fluxes can generate pressure anisotropies of particles. Therefore, the characteristic thermal velocities of electrons can differ along and perpendicular to the direction of the magnetic field, inducing a polarisation pattern in the CMB. The signal scales linearly with the optical depth of the region containing large-scale correlated anisotropy (e.g., along ubiquitous cold fronts in clusters), and with the degree of anisotropy itself. It has the same spectral dependence as the polarisation induced by cluster motion with respect to the CMB frame (kinematic SZ effect polarisation), but can be distinguished by its spatial pattern. The magnitude of the effect is on par with majority of other SZ polarisation signals considered here \citep[for a useful summary, see Table 1 of][]{Khabibullin18}. An increase of the effective electron collisionality due to plasma instabilities will reduce the effect. Such polarisation, therefore, may be an independent probe of the electron collisionality in the ICM, which is one of the key properties of a high-$\beta$ weakly-collisional plasma\footnote{The parameter $\beta$ in this context refers to the ratio of thermal to magnetic pressure, $p_{\mbox{\tiny mag}}=B^2/(2 \mu_0)$.} from the point of view of both astrophysics and plasma theory.

Even in the future, the aforementioned pSZ signals will be challenging to extract due to their intrinsic faintness, the limited polarisation purity of existing instrumentation technologies, and the large number of polarised astrophysical signals that could contaminate them \citep[e.g.][]{Sunyaev1982}. Beam depolarisation effects will furthermore render much of the polarisation signal unobservable for unresolved clusters. Nevertheless, the various pSZ signals may provide another avenue forward for detailed studies of ICM structure, and maybe become accessible through stacking analyses.

\subsection{Multiple scattering effects}
\label{sec:msSZ}
%------------------------------------------------------------------------------------
Another subdominant correction to the SZ signals is caused by multiple scattering events inside rich clusters \citep{Sunyaev1980, Sazonov1999, Molnar1999, Dolgov2001, Itoh2001, Colafrancesco2003, Shimon2004}. This correction is usually derived in the isotropic scattering approximation (ISA), which assumes that the radiation field locally remains isotropic. In this limit, to leading order, the contribution is suppressed by a factor of $\simeq 10\, y$ relative to the single-scattering tSZ signal, rendering it a $\simeq 0.1\%$ correction \citep{Itoh2001}. In the ISA no correction $\propto \taue$ arises.

However, when the scattering-induced anisotropy in the radiation field is included, the scenario changes slightly. In this case, corrections may become detectable, which even in a constant density sphere are caused by the variations of the photon's path in different directions \citep{Chluba2014mSZI}. Hence, the spectral dependence of the multiple-scattering signal is modified, and a new contribution of order $\taue/20$ relative to the tSZ effect arises \citep{Chluba2014mSZI, Chluba2014mSZII}. The net signal depends explicitly on the considered line of sight and structure of the medium. Future measurements of the spatial and frequency dependence of the multiple-scattering SZ signal could thus help in the reconstruction of ICM density and temperature profiles. However, such measurements will remain very challenging for the foreseeable future.

%%%%%%%%%%%%%%%%%%%%%%%%%%%%%%%%%%%%%%%%%%%%%%%%%%%%%%%%%%%%%%%%%%

\section{ICM thermodynamics through the SZ effects} 
\label{sec:ICMthermodynamics}

\subsection{The ``universal'' pressure profile}
\label{sec:profiles}

One of the key insights from modern hydrodynamical cosmological simulations is that the hot X-ray emitting plasma in galaxy clusters exhibits a remarkable degree of self-similarity \citep[e.g.,][]{Nagai2007,Battaglia2010,Lau2015}, where the ICM pressure profile is well characterised by a generalised Navarro, Frenk, \& White (NFW) profile \citep{Navarro1996}. 
The use of the generalised NFW (gNFW) profile to describe pressure, the integral of which yields the tSZ signal (Eq.~\ref{eq:Comptony}), was first proposed by \citet{Nagai2007}:
\begin{equation}
{P}(x) = \frac{P_{0} } { (c_{500}x)^{\gamma}\left[1+(c_{500}x)^\alpha\right]^{(\beta-\gamma)/\alpha} } \;.
\label{eq:pgnfw}
\end{equation}
Here, the parameters $(\gamma,\beta)$ are the central slope ($r \ll \rs$) and outer slope ($r \gg \rs$), respectively.  The parameter $\alpha$ modulates how smoothly the slope changes from $\gamma$ to $\beta$ around $\rs$, where $x = r/\rs$, $\rs = r_{500}/c_{500}$, and $r_{500}$ is defined as the radius within which the average overdensity is $500\times$ greater than the critical density of the Universe at that redshift, $\rhoc(z)$.\footnote{The reference radius $r_{500}$ is a convention adopted simply as a reflection of what contemporary instrumentation circa 2007 could probe, rather than being motivated by cluster physics.}
All three of the slopes are highly correlated with the value of the scale radius $\rs$. This analytic profile has been widely used to measure pressure profiles using X-ray and SZ data. Measurements of ICM pressure profiles provide important information about the thermodynamic structure of the ICM, including the effects of AGN feedback, bulk and turbulent motions, substructures, and asphericity of clusters. 
Beyond $r_{500}$, an increasing level of non-thermal pressure support at the level of 10 -30\% is expected, depending on the dynamical state of clusters \citep[e.g.,][]{Lau2009,Battaglia2012,Nelson2014}.

Motivated by \cite{Nagai2007}, the gNFW parametrisation (given in Eq.~\ref{eq:pgnfw}) was first applied to SZ observations in \cite{Mroczkowski2009} and has since displaced the previously used $\beta$-model \citep{Cavaliere1976,Cavaliere1978} in nearly all SZ-related pressure profile studies. 
A number of SZ and X-ray studies have since sought to measure or refine the azimuthally-averaged, ``universal'' pressure profile of the ICM, often attempting to constrain the slope parameters of the gNFW profile \citep{Arnaud2010,Plagge2010,Bonamente2012,Planck2013_V,Sayers2013b,Eckert2013,Adam2015,Sayers2016b,Ghirardini2017,Romero2017,Bourdin2017,Ruppin2018}.
The gNFW parameters are highly covariant, and thus the interest is often more on the overall profile shape, or the combinations of the parameters, rather than the individual parameter values themselves. 

The majority of the above studies find agreement with the ``universal'' pressure profile presented in \citet{Arnaud2010} (A10), especially within $r_{500}$. Where disagreement has been found within $r_{500}$, systematic uncertainties and sample variance are likely to mitigate the purported tension. Beyond $r_{500}$, studies have relied on, or heavily supplemented X-ray data with, tSZ data  \citep[e.g.]{Planck2013_V,Sayers2016,Ghirardini2018}. At $r > r_{500}$, \citet{Ghirardini2018} find higher pressure relative to A10, while \citet{Sayers2016} find lower pressure. As the samples are disparate (notably in redshift range), one might take this as an early indication of evolution in the pressure profile, consistent with the hydrodynamical simulation results reported by \citet{Battaglia2012a}.

\begin{figure*}
\begin{center}
\includegraphics[height=0.36\textwidth]{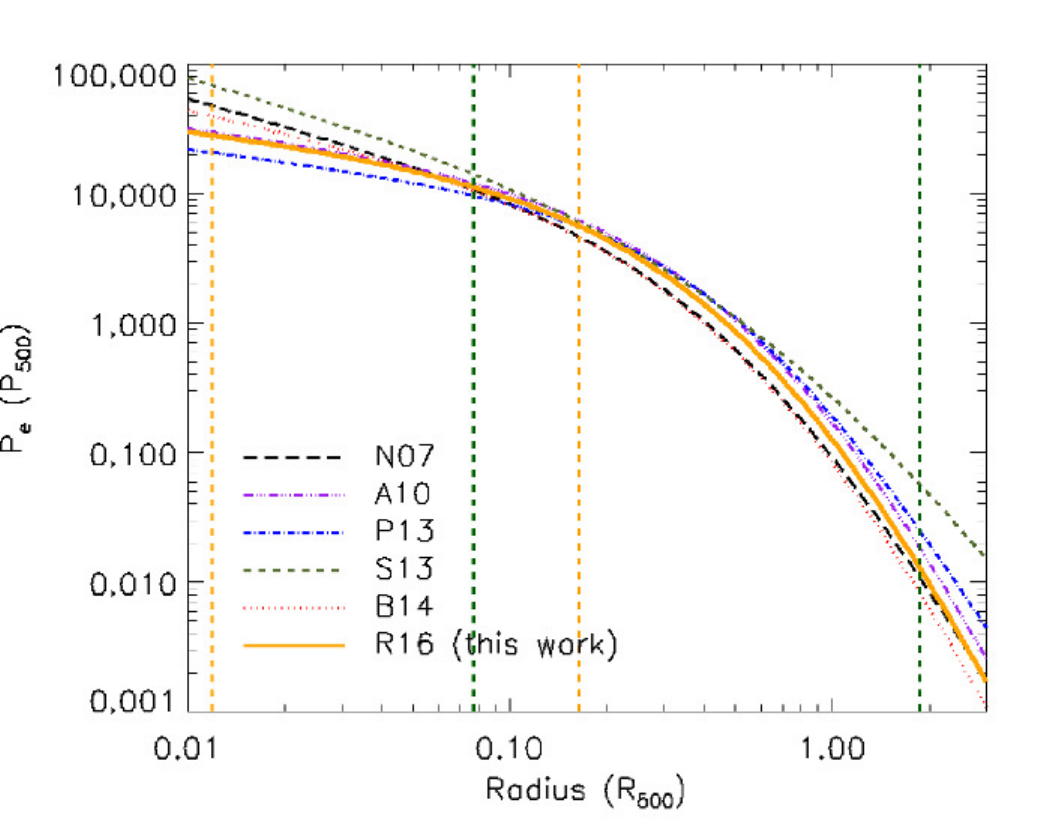}
\includegraphics[height=0.36\textwidth]{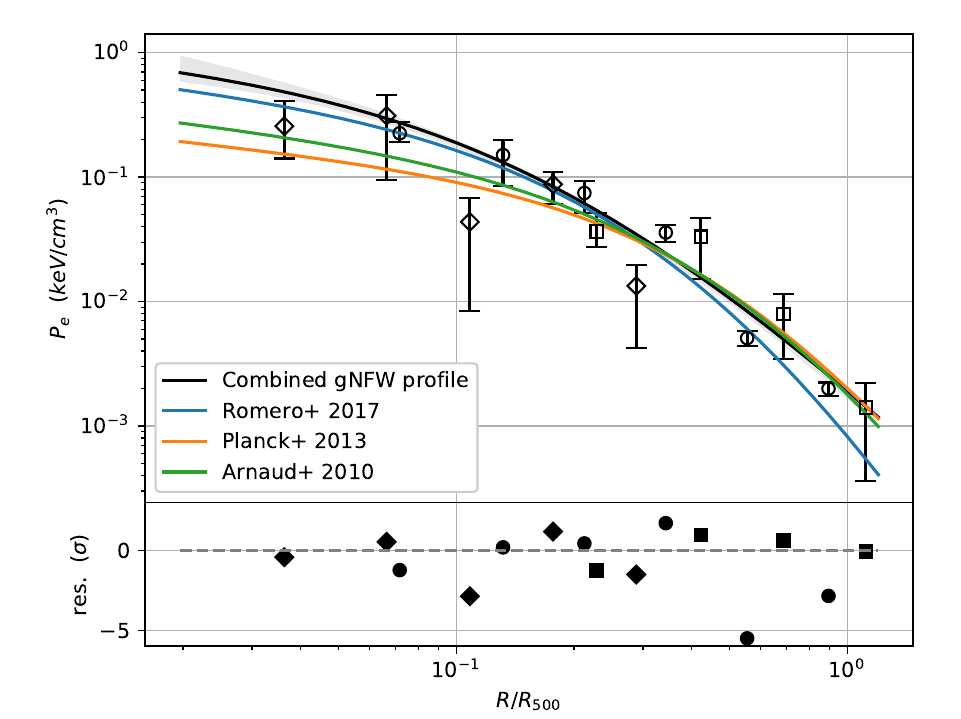}
\end{center}
\caption{  
\textit{Left:} A comparison of the joint fits of a generalised NFW profile to MUSTANG-1 and Bolocam data \citep{Romero2017} (R16), to several other fits to real and simulated data.   The profile labelled N07 is the original, theoretically-motivated gNFW profile in \citealt{Nagai2007}, using the updated parameters reported in \citealt{Mroczkowski2009}.  The fit to local X-ray selected clusters observed with {\it XMM-Newton} is from \citealt{Arnaud2010} (A10), the fit to Bolocam data is from \citealt{Sayers2013b} (S13), the fit to the {\it Planck} selected clusters is from \citealt{Planck2013_V} (P13), and the fit to the ACCEPT-2 \citep{baldi2014} pressure profiles of the sample within R16 is denoted as B14. 
\textit{Right:} Bins with error bars show non-parametric fits of one cluster using four SZ datasets: MUSTANG-1, NIKA, Bolocam, and \textit{Planck}. The black curve is \citealt{Romero2018} fit to bins from all instruments, compared to the pressure profiles found by \citealt{Romero2017} (blue), \citealt{Planck2013_V} (orange), and \citealt{Arnaud2010} (green). Figures from \citealt{Romero2017,Romero2018}, respectively.  
}
\label{fig:pressureprofs}
\end{figure*}

The gNFW profiles are often fit to binned non-parametric pressure profiles \citep[e.g.,][]{Plagge2010,Sayers2013b}, where \citet{Plagge2010} stacked their sample to recover the non-parametric pressure profiles. With higher resolution and more sensitive SZ observations, non-parametric pressure profiles have been fit to the SZ data of individual clusters \citep{Basu10,Sayers2013b,Romero2018,Ruppin2018}. An example of recent pressure profile constraints for the jointly-fit, multi-scale data in \cite{Romero2017,Romero2018} is shown in Figure \ref{fig:pressureprofs}. 

%%%%%%%%%%%%%%%
%%%%%%%%%%%%%%%

The methods of constraining pressure profiles from SZ data are non-trivial and affect the potential biases and systematic uncertainties associated with these constraints. As discussed in \S~\ref{sec:photometric}, ground-based observations suffer from atmospheric effects that typically limit the spatial scales recoverable from an observation to the instrumental field of view (FoV).
There are several ways to mitigate these filtering effects, the main tools being: parametric and non-parametric forward modelling, deconvolution, and the Abel transform \citep[see][for reference]{SilkWhite1978}.
If the noise and filtering behaviours are well characterised, one can attempt image deconvolution \citep{Basu10,Sayers2013b}. From the deconvolved images, pressure profiles may be deprojected via methods used in X-ray studies, which include an `onion-skin' method (multiple shells either jointly-fit or fit through a `peeling' method) or Abel transform \citep[e.g. as applied in][]{Basu10}. However, 
the deconvolution suffers at large angular scales, enough that \citet{Sayers2013b} opted to forward model pressure profiles calculated as power-law interpolations between several radii. Forward modelling requires, in its coarsest form, a grid search \citep[e.g.][]{romero2015}, or more often a Monte Carlo Markov Chain approach \citep[e.g.,][]{Bonamente2004,Bonamente2006,Olamaie2013,Sayers2013b,Ruppin2018}. 

Another implication of the FoV restricting the spatial scales recovered is that many SZ instruments do not access a wide dynamic range of angular scales. As high-resolution SZ instruments were developed (\S ~\ref{sec:SZobs_in_a_nutshell}), investigators soon wanted to combine SZ datasets to maximise the range of angular scales over which pressure profiles could be constrained. These studies \citep[e.g.][]{Sayers2016,Romero2017} serve both to check for agreement of pressure profile shapes between SZ and X-ray data as well as overall agreement among the SZ datasets.

Given the complexity of extracting pressure profiles from SZ data, especially ground-based data, consistency checks from joint fits or identical analyses are critical. In comparing their results to previous results, \citet{Sayers2016} note that differences in pressure profiles can derive from (1) sample selection, (2) instrumental biases, or (3) biases introduced in data processing.  For single clusters, the SZ data currently appear consistent \citep[e.g.,][]{Romero2018,Ruppin2018}, but deeper observations of an overlapping sample between mm-wave instruments may yield differences in individual clusters. Additional comparisons across samples of clusters will serve this end as well. 

Several current SZ instruments (\S~\ref{sec:instrumentation}) now recover sufficiently large angular scales so as to provide moderate resolution pressure profile constraints beyond $r_{500}$ and thus provide a clearer picture as to whether the pressure profile evolves with redshift. These instruments also benefit from the complementary X-ray data.
When combined with X-ray measurements of the ICM density, the pressure profile can also be used to infer other thermodynamic quantities such as temperature and entropy (see \S~\ref{sec:txsz} for further discussion).  However, the temperature derivation is degenerate with cluster geometry, and second-order effects like pressure clumping and helium sedimentation could bias the resulting constraints \citep{Ettori2006,Peng2009,Bulbul2011}. 

\subsection{SZ scaling relations}
\label{sec:SZscaling}

Since the thermal energy content of the cluster is determined primarily by the gravitational potential well of dark matter, the aperture (or volumetrically) integrated tSZ signal, 
\begin{equation}
Y_{\mbox{\tiny SZ}} = \int y \, d\Omega \propto d_A^2 \int \Pe \, dV ,
\label{eq:Ysz}
\end{equation}
is proportional to the thermal energy content of the ICM.  Per beam, the tSZ provides calorimetry of the ICM. Hydrodynamical cosmological simulations suggest that $Y_{\rm SZ}$ serves as one of the most robust total mass proxies for galaxy clusters, with a scatter of about 10\% \citep[e.g.,][]{motl2005,Nagai2006,Battaglia2012,Kay2012,krause2012,Yu2015}. This has motivated construction of a low-scatter core-excised X-ray mass proxy, $Y_{X} \equiv M_{gas} T_{X}$ \citep{Kravtsov2006}.

\begin{figure}
\begin{center}
\includegraphics[width=\textwidth]{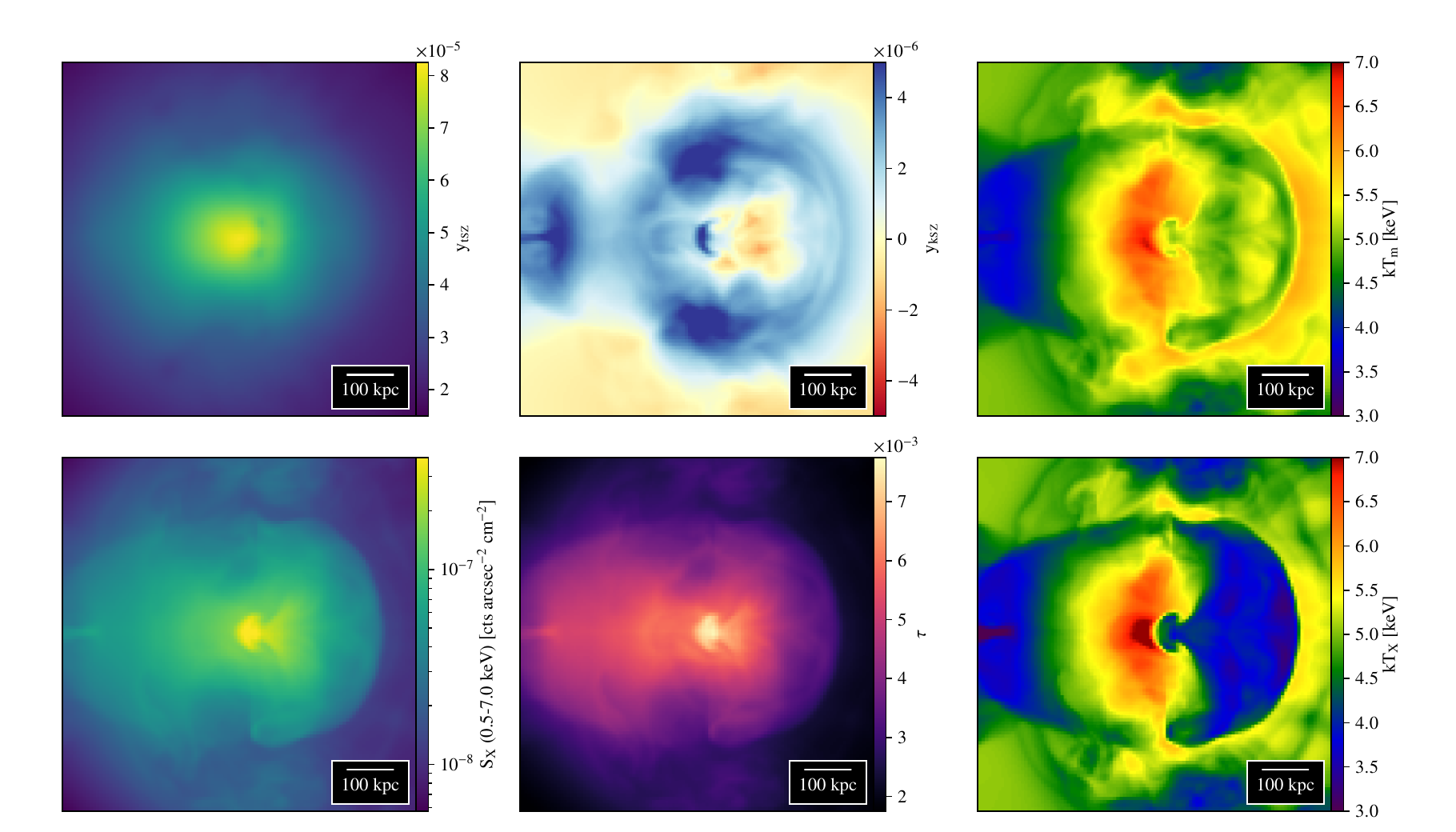}
\end{center}
\caption{The SZ effect in a cluster core which is undergoing a merger, from a simulation oriented so that the largest gas motions are predominantly within the line of sight. Certain X-ray quantities are also shown for comparison.
Shown quantities are Compton $y$ (top-left), $\yksz$ (top-middle), projected mass-weighted temperature, $\kB\Tmw$ (top-right), X-ray surface brightness, $S_{\mbox{\tiny X}}$ (bottom-left), electron optical depth $\taue$ (bottom-middle), and projected X-ray temperature, $\kB\Tx$ (bottom-right).  
Note that while cold fronts are clearly seen in projected temperature, X-ray emission, and optical depth, they are essentially invisible in the tSZ signal due to the continuity of thermal pressure across the front. However, the bulk velocity of the gas across the fronts produces a kSZ signal.}
\label{fig:sz_vs_xray}
\end{figure}

Despite the robustness of $Y_{\rm SZ}$ as a mass proxy, there can be large deviations from self-similarity, particularly in extreme cases.
Mergers between clusters of galaxies are the most energetic events in the present day Universe, and therefore not surprisingly have an effect on SZ observables (see Figure \ref{fig:sz_vs_xray}). The first and second core passages of a major merger event induce transient boosts in the tSZ signal via compression and shock heating of the cluster plasma, which are more pronounced for the maximum Comptonisation parameter $y_{\rm max}$ than for its integrated value $Y_{\rm SZ}(<R)$ \citep[][]{motl2005,poole2007,wik2008}. SZ observable-mass scaling relations involving the latter are therefore far less affected by mergers than the former or X-ray proxies such as $T_X$ and $L_X$. 

The scatter in the integrated SZ observable-mass relations originates from the non-thermal pressure provided by bulk and turbulent gas motions generated by mergers and mass accretion \citep{Yu2015} and asphericity and substructures \citep{Battaglia2012} in the ICM, while the normalisation of the $Y_{\rm SZ}-M$ relation is sensitive to the input cluster astrophysics, such as radiative cooling, star formation, and energy injection from stars and AGN feedback \citep[e.g.,][]{Nagai2006,Battaglia2012,Kay2012}.

Recent results for cosmological determinations using SZ-selected clusters and, specifically, their SZ signal as a mass proxy have found good agreement with cosmology inferred by other means, improving constraints on the dark energy equation of state and number of neutrino species, with the dominant systematic being the overall scaling of the SZ signal with cluster mass \citep[e.g.][]{Bocquet2015,Planck2016XXIV,deHaan2016,Hilton2018}.  Detailed resolved and stacked studies of the average pressure profiles and how $Y_{\rm SZ}$ scales with mass therefore provide crucial tests of the simulations.

Merging activity can induce offsets between the X-ray and tSZ peaks, since the former strongly tracks the maximum gas density ($S_X \propto \Ne^2$) and the latter is determined by the maximum integrated line-of-sight pressure \citep[][]{molnar12,zhang14}, while the peaks in the X-ray and SZ surface brightnesses should be coincident for relaxed clusters. Comparison of observed offsets to simulations can be used to estimate merger parameters such as the mass ratio and relative velocity.

Cluster-cluster and cluster-group mergers often exhibit velocities of several thousand km~s$^{-1}$ during core passage, and can therefore also produce a strong kSZ signal in some circumstances (e.g. if close to the line of sight). If the velocities are large, the kSZ signal may even dominate over the tSZ signal during core passages \citep[][]{ruan2013}. Multi-frequency observations of these systems are essential in order to distinguish between the two effects. Note, however, that due to the linear dependence of the kSZ effect on velocity, projection effects can complicate the interpretation if multiple velocity components are contributing to the line-of-sight kSZ signal.

%%%%%%%%%%%%%%%
%%%%%%%%%%%%%%%

\subsection{The complementarity of X-ray and SZ measurements}
\label{sec:XSZ}

X-ray and tSZ measurements are independent and highly complementary probes of the thermodynamics and kinematics of the ICM. Joint X-ray/tSZ studies, for example, allow one to reconstruct the thermodynamic state of the ICM by exploiting the different line of sight density dependences in the surface brightness integrals for each. The X-ray surface brightness $S_{\mbox{\tiny X}}$, commonly expressed in $\rm cts ~ arcmin^{-2} ~ s^{-1}$, is:
\begin{equation}
\label{eq:xray_sb}
S_{\mbox{\tiny X}} = \frac{1}{4\pi (1+z)^3} \! \int \!\! \Ne^2(\ell) \Lamee(\Te(\ell),Z) \, \id \ell,
\end{equation}
where $\Ne(\ell)$ and $\Te(\ell)$ are respectively the electron density and temperature along sight line $\ell$, 
$\Lamee(\Te,Z)$ (in $\rm cts ~ cm^{5} ~ s^{-1}$) is the X-ray emissivity measured by the 
instrument within the energy band used for the observation, $z$ is the cluster's redshift, and $Z$ is the metallicity. In units of  $\rm erg ~ cm^{5} ~ s^{-1}$, there is an additional factor of $(1+z)^{-1}$, consistent with the standard $(1+z)^{-4}$ dependence of cosmological dimming.
Thus, while the Comptonisation parameter ($y \propto \int \Ne \Te d\ell$; see Equation \ref{eq:Comptony}) is linear in density, X-ray surface brightness varies as the square ($S_{\mbox{\tiny X}} \propto \int \Ne^2 d\ell$).

While X-ray emission suffers from cosmological dimming, it is useful to note that scaling relations in \LCDM\ cosmology predict that, for a given mass defined with respect to overdensity, clusters at higher redshift are hotter, denser and therefore more X-ray luminous than their local counterparts. As a result, the observable X-ray flux (at fixed mass) may not decrease strongly at $z\gtrsim 1$, resembling a way that is similar to the SZ signal \citep[e.g.][]{Churazov2015}.

\subsubsection{ICM temperature from joint X-ray/tSZ studies}
\label{sec:txsz}

SZ and X-ray observations of clusters can in principle be used in combination to derive ``mass-weighted'' temperatures by using the estimate of the gas pressure from the SZ and that of the gas density from the X-ray \citep[e.g.,][]{Adam2017}. This technique offers the possibility to estimate the temperature, entropy and mass profiles determined through the assumption of hydrostatic equilibrium (HSE) to high redshift ($z\gtrsim 1$), where X-ray spectroscopic measurements are very challenging due to very low photon counts, and hence characterise a redshift evolution of the ICM profiles throughout the epoch of cluster formation.

However, some caution should be exercised on interpreting the effective weights of temperatures derived from joint tSZ/X-ray analyses. First, the tSZ signal is proportional to the gas mass multiplied by the mass-weighted temperature, $\Tmw$, while the temperature of the hot ICM inferred by fitting the X-ray spectrum with a thermal emission model is a spectroscopic temperature $T_{\rm spec}$ \citep{Mazzotta2004,Vikhlinin2006}. 
Hydrodynamical simulations predict that there are discrepancies between $\Tmw$ and $T_{\rm spec}$ \citep[][see also Figure~\ref{fig:sz_vs_xray} for the difference between the $\Tmw$ and $T_{\rm spec}$ maps of one of the idealised cluster merger simulations]{Mathiesen2001, Nagai2007, Piffaretti2008, Rasia2014}.

In addition, it is well known that the density profile results can be biased in the presence of significant gas clumping, or projection from a triaxial ICM distribution when spherical models are assumed \citep[see e.g.][]{Mathiesen1999, DeFilippis2005, Simionescu2011, Nagai2011, Bonamente2012, Limousin2013, Vazza2013, Battaglia2015a, Eckert2015, Umetsu2015, Rossetti2016}. If these density biases are not accounted for, then the joint tSZ/X-ray temperature estimate will not yield the sought-after $\Tmw$ value. While it is possible in principle to use the combination of tSZ/X-ray data, including spectroscopic temperature estimates from high X-ray photon counts, to measure cluster triaxiality out to high redshifts \citep{Sereno2012}, once again the degeneracy between the triaxiality parameters and gas clumping/substructures, as well as the challenges of X-ray spectroscopy at high-$z$, might limit this method's applicability.

Observationally, this technique was pioneered in the first decade of this millennium using, for example, the combination of Nobeyama Telescope/SCUBA and {\it Chandra} data \citep{Kitayama04}, in joint SZA + {\it Chandra} observations \citep{Mroczkowski2009}, and in APEX-SZ + {\it XMM-Newton} observations \citep{Nord09,Basu10}, although the statistical uncertainties were still very large. The combination of wide FoV and high sensitivity to the SZ signal brought about by {\it Planck} allowed a leap forward in the use of the joint tSZ/X-ray technique, as demonstrated in the analyses by \citet{Eckert13b,Eckert2013}, who exploited ROSAT and {\it Planck} data to probe out to $\approx 1.5 \times r_{500}$ in a sample of nearby massive clusters.  More recently, this effort has been extended using either the archival {\it Planck} or Bolocam datasets.
These analyses include 1) a joint analysis of Bolocam and {\it Chandra} data for a large sample of 45 clusters probing the thermodynamics out to $r_{500}$ \citep{Shitanishi2018}, 2) a detailed analysis using a sub-sample of 6 clusters including data from Bolocam, {\it Chandra}, the {\it Hubble} Space Telescope (HST), and Hyper Suprime-Cam (HSC) lensing data \citep{Siegel2018} to probe non-thermal pressure support out to $r_{500}$, and 3) a large effort by the XMM Cluster Outskirts Project (XCOP) to study the thermodynamics, non-thermal pressure support, and outskirts ($>r_{500}$) of the 13 of the most significant {\it Planck} detections \citep[see][]{Eckert2017b,Eckert2019,Ettori2019,Ghirardini2019}.

\begin{figure*}
\begin{center}
\hbox{\includegraphics[width=0.55\textwidth]{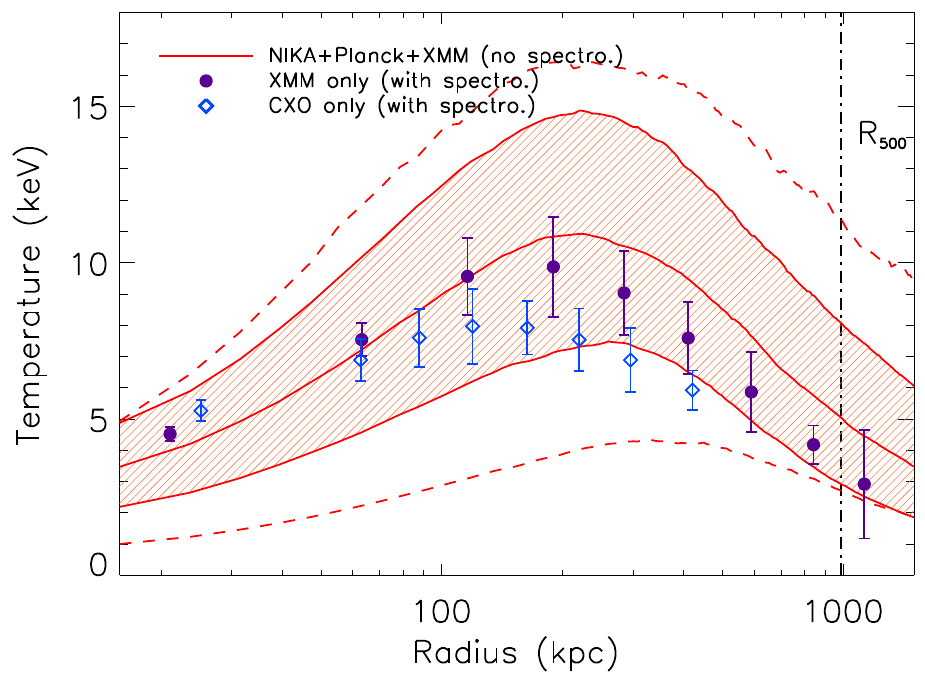}
\includegraphics[width=0.45\textwidth]{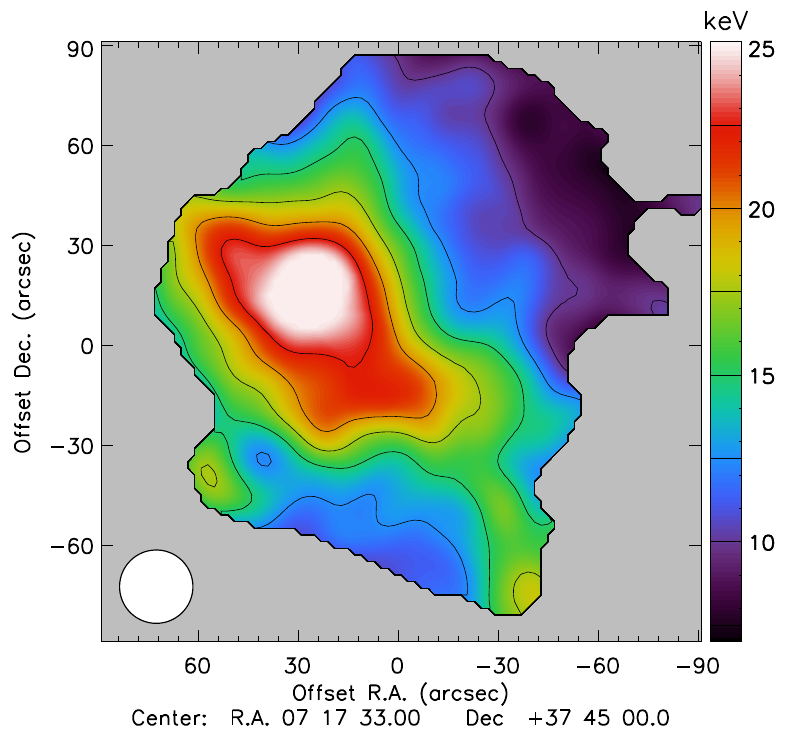}}
\end{center}
\caption{Temperature measurement of the ICM using tSZ (NIKA) + X-ray ({\it XMM-Newton}) imaging and comparison to X-ray spectroscopic measurements.
\textit{Left:} SZ+X-ray deprojected temperature profile towards MACS~J1423.8+2404 (red dashed region) and comparison to {\it XMM-Newton} (purple, filled dots) and {\it Chandra} (blue, open diamonds) spectroscopic measurements. 
\textit{Right:} SZ+X-ray temperature map of the hot gas toward the galaxy cluster MACS~J0717.5+3745 obtained from combining {\it XMM-Newton} density and NIKA pressure imaging. Figures from \citet{Adam2016,Adam2017}.}
\label{fig:temperature_SZX}
\end{figure*}

Considerable effort is now on-going to extend these tools to high resolution (subarcminute) SZ samples.  For instance, a legacy project using NIKA2 observations of a large sample ($\sim50$) clusters at a resolution of $\sim15\arcsec$ shows potential for NIKA2/{\it XMM-Newton} analyses. This potential is presented in \citet{Ruppin2018}, following the developments made with the pathfinder camera NIKA \citep{Adam2015,Adam2016,Ruppin17a}, and the first two-dimensional temperature map reconstruction with this method was recently published by \citet{Adam2017} in MACS~J0717.5+3745 ($z=0.55$). See Figure \ref{fig:temperature_SZX} for illustrations. The authors also directly compared the temperatures recovered through tSZ/X-ray imaging with the spectroscopic X-ray temperatures measured with {\it Chandra} and {\it XMM-Newton}. The SZ-derived temperature measurement is about 10\% larger than the spectroscopic one from {\it XMM-Newton}, which is within the calibration uncertainties of both instruments, but may also include systematic effects driven by assumptions about the gas line-of-sight geometry and clumpiness.

%%%%%%%%%%%%%%%
%%%%%%%%%%%%%%%

\subsubsection{ICM temperatures from the relativistic SZ effect}
\label{sec:rsz_temperature}

A more direct, spectroscopic SZ estimate of the ICM temperature is possible using multi-frequency tSZ measurements to separate the rSZ and classical tSZ contributions (\S~\ref{sec:rSZ_detail}). While theoretical progress has been made in computing the relativistic terms accurately (e.g., using {\tt SZpack}, \citealt{Chluba2012SZpack}), actual measurements of the cluster temperatures through the relativistic corrections have been more challenging. Only recently, using all-sky data from the {\it Planck} satellite, which covers the tSZ spectrum almost entirely, has it been possible to constrain the rSZ spectral distortion in a large stacked sample of clusters \citep{Hurier2016rSZ}. With future ground- and space-based instruments it is expected that measurement of the rSZ effect will turn into a robust technique for inferring cluster temperatures, and hence their masses, for both astrophysical and cosmological analyses \citep{Erler2018}. 

The primary requirement for rSZ effect measurements is multi-frequency coverage of the tSZ spectrum spanning both the decrement and increment. Early attempts were made by combining data from several different experiments, e.g., by \citet{Hansen2002rSZ} and \citet{Nord09}, but limited sensitivity did not break the degeneracy between the rSZ and velocity-induced kSZ contributions. Using data from the Z-spec grating spectrometer and Bolocam, \citet{Zemcov2012} could constrain the temperature of the hot cluster RX~J1347.5-1145 to high accuracy, but only by neglecting the kSZ contribution.
Similarly, a combination of X-ray and SZ measurements were used to quantify higher order rSZ contributions for the Bullet cluster \citep{Prokhorov2012rSZ, Chluba2013}.

%%%%%%%%%%%%%%%
\begin{figure*}
\begin{center}
\includegraphics[width=0.5\textwidth]{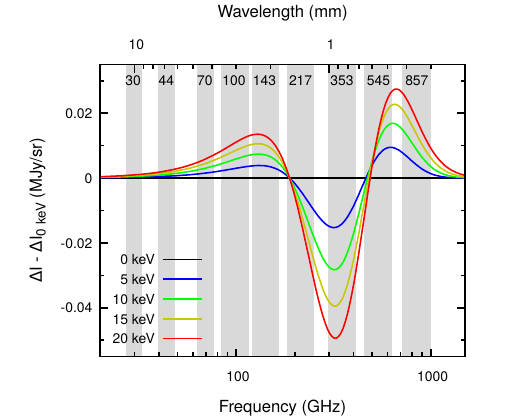}%
\includegraphics[width=0.5\textwidth]{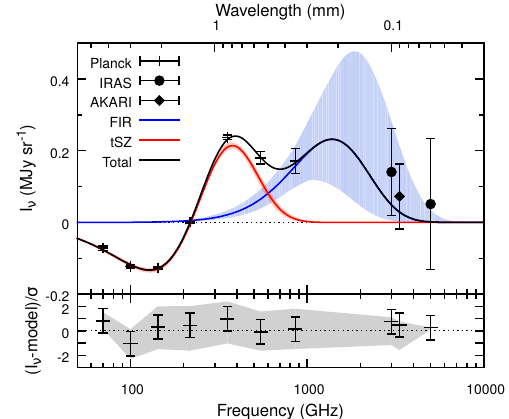}
\end{center}
\caption{Spectral distortions due to the relativistic SZ (rSZ) effect and its current state-of-measurement with {\it Planck} data.   
\textit{Left:} The difference of the tSZ spectrum computed for different temperatures and a fixed Comptonisation value $y=10^{-4}$, to the tSZ spectrum in the non-relativistic limit (equivalent to $k\Te\rightarrow 0$). This plot shows at which frequencies the relativistic spectral distortion effects are most prominent. Overplotted in grey bands are the nine channels of the {\it Planck} satellite. 
\textit{Right:} Result of modelling the tSZ spectrum and measuring the mean cluster temperature in a sample of 772 {\it Planck}-selected clusters via stacking. The data are from the {\it Planck}, AKARI, and IRAS satellites, after matched-filtering and stacking individual maps. The red and blue lines are the best-fitting tSZ and far infra-red spectrum and the shaded regions indicate 68\% confidence regions. A $2.2\sigma$ detection of the mean $y$-weighted cluster temperature ($4.4^{+2.1}_{-2.0}$~keV) is made via stacking, but future experiments (see \S \ref{sec:futureCMB}) will improve this accuracy by more than an order-of-magnitude. 
Both figures from \citet{Erler2018}.  
}
\label{fig:planck_rsz}
\end{figure*}
%%%%%%%%%%%%%%%

The situation improved considerably with data from the {\it Planck} satellite, which provided well-calibrated all-sky measurements for galaxy clusters in the frequency range $30-860$ GHz. It is also possible to ignore the kSZ contribution by averaging (stacking) the data from several hundred clusters, as the kSZ signal can be both positive or negative due to the random nature of peculiar motions. This approach was explored by \citet{Hurier2016rSZ} and \citet{Erler2018} to constrain the average $y$-weighted temperature from cluster samples. 

In Figure \ref{fig:planck_rsz} the rSZ modelling results from \citet{Erler2018} are shown. The left panel is an illustration of the rSZ spectral distortions for a fixed value of Compton-$y$ ($10^{-4}$ in this case) and the relative positions of the nine frequency bands of {\it Planck}. As can be seen, the entire tSZ spectrum is covered by {\it Planck}, although for better sensitivity and higher resolution only the $70-860$ GHz data were used in \citet{Erler2018}. In the right panel of  Figure~\ref{fig:planck_rsz}, the result of stacking images of 772 {\it Planck} clusters are shown. The impact of the kSZ effect is averaged out and the thermal spectrum is constrained to sufficient accuracy to make a measurement of the mean sample temperature to $4.4^{+2.1}_{-2.0}$ keV. This temperature is found to be slightly lower than the mass-weighted average of X-ray spectroscopic temperatures, ($T_\mathrm{X} = 6.91 \pm 0.07$ keV), although the tension was only at the level of $1.3\sigma$. This difference can potentially indicate a low level of gas clumping in galaxy clusters which causes the density-squared weighted X-ray temperature to stay above the rSZ-derived $y$-weighted temperature. Figure~\ref{fig:planck_rsz}
also demonstrates the importance of modelling the cluster-centric far infra-red (FIR) emission simultaneously to obtain an accurate estimate of the rSZ signal. 

Future high resolution, multi-frequency data can open the possibility of direct rSZ measurements of cluster temperature profiles or post-shock electron temperatures. This rSZ-derived mean projected temperature is very close to Compton $y$-weighted \citep{Hansen2004,Kay2012,Morandi2013,Erler2018}.\footnote{Ideally one would measure mass-weighted temperature, $\Tmw \equiv \frac{\int \Te(\ell) \, \Ne(\ell) \, \id\ell}{\int \Ne(\ell) \, \id\ell} \propto y/\taue$, rather than one weighted by Compton-$y$} Nonetheless, as with X-ray spectroscopic temperatures, rSZ-derived temperatures are not free of systematics. In the case of rSZ, these can be due to the effects of mixing from multiple temperature components along the line of sight, where first and higher moments of the temperature and density distribution can present a bias \citep{Chluba2013}. 

\subsubsection{Mass Profiles}
\label{sec:mass}

After deprojection of the X-ray and tSZ surface brightness profiles, the derived quantities can be combined through the ideal gas equation to recover the gas temperature $k\Te=P_{\mbox{\tiny SZ}}/n_{\mbox{\tiny X}}$ and the entropy $K=P_{\mbox{\tiny SZ}} n_{\mbox{\tiny X}}^{-5/3}$, where $P_{\mbox{\tiny SZ}}$ is the electron pressure inferred from the tSZ data and $n_{\mbox{\tiny X}}$ is the electron number density inferred from the X-ray data.
It is then possible to reconstruct the gravitating mass profiles of the ICM under the assumption of hydrostatic equilibrium by using the equation \citep[e.g.,][]{Zaroubi98,Puchwein06,Ameglio2009}:
\begin{equation}
\frac{\id P_{\mbox{\tiny SZ}}}{\id r}=-\rho_{\mbox{\tiny X}}\frac{GM(<r)}{r^2},
\end{equation}
where $\rho_{\mbox{\tiny X}} = \mue \proton n_{\mbox{\tiny X}}$ is the gas density inferred from the X-ray data, and $\mue \approx 1.18$ is the average weight per electron and $\proton$ is the proton rest mass.
The hydrostatic mass is known to be slightly biased (see review on cluster mass estimation in these proceedings).

Over the years, high-quality radial (1-D) constraints on ICM properties have been obtained \citep[e.g.,][]{LaRoque2006,Mroczkowski2009}, the most recent examples of which combined either {\it XMM-Newton} or {\it Chandra} imaging (or both) and {\it Planck} or Bolocam SZ data for various cluster samples \citep[e.g.,][]{Tchernin16,Ghirardini2018,Ghirardini2019,Shitanishi2018}. Given the linear dependence of the SZ signal on the gas density, measurements obtained through this technique extend farther in radius than typical spectroscopic X-ray measurements and allow one to probe the relatively unexplored outskirts of local clusters out to their virial radius \citep{Ghirardini18b}. The wide radial range accessible in this study also brings high-quality measurements of hydrostatic mass profiles (typical statistical uncertainty $\sim3\%$ at $r_{500}$) to probe the internal structure of dark matter halos \citep{Ettori2019}. In the majority of cases, the underlying mass distribution was found to be well described by an NFW profile. 

%%%%%%%%%%%%%%%%%%%%%%%%%%%%%%%%%%%%%%%%%%%%%%%%%%%%%%%%%%%%%%%%%%

\subsection{tSZ power spectrum constraints on ICM thermodynamics}
\label{sec:tsz_powerspec}

Measurements of the tSZ angular power spectrum from wide-field surveys, as opposed to high-resolution measurements of the pressure fluctuations within selected, high-mass clusters (discussed in \S \ref{sec:sz_powerspec}), present a complementary approach to constrain several important astrophysical parameters. Due to the redshift independence of the tSZ effect, this contribution to the CMB power spectrum comes from cluster (and galaxy group) ICM at all redshifts as well as from the unvirialised gas in cosmic filaments. 

The original application for the tSZ power spectrum measurement was intended for cosmology, due to its strong dependence on several cosmological parameters such as $\Omega_M$, the matter density, or $\sigma_{\mbox{\tiny 8}}$, the amplitude of matter power spectral fluctuations on 8~Mpc~h$^{-1}$ scales \citep[e.g.,][]{White1993,Komatsu2002}. 
Originally predicted to be dominated by massive clusters at intermediate redshift \citep{ColeKaiser1988}, the main contribution to the tSZ power at the angular scales relevant for ground-based CMB instruments (i.e., $\ell\sim 2000-3000$) in fact originates from low-mass groups and clusters, and it was realised that the uncertain astrophysics (e.g., non-thermal pressure support and energy feedback) of these low-mass systems at intermediate and high redshifts will make the goal of using the tSZ power spectrum for cosmology more difficult to attain \citep{Shaw2010SZ,Trac2011,Battaglia2012}.
If, on the other hand, the cosmological parameters are well known from other probes and can be restricted a priori, then the tSZ power spectrum measurement can be used to constrain ICM astrophysics. 
It was also recently shown that relativistic temperature corrections affect the inference of cosmological parameters from the SZ power spectrum analysis. These are less important at high-$\ell$ (i.e., $\ell\gtrsim 2000$) but are relevant on the scales probed by {\it Planck} \citep{Remazeilles2019}.

One of the main challenges to using the tSZ power spectrum to infer cluster astrophysics lies in the fact that, for arcminute resolution instruments, the tSZ power is buried under cosmic infrared background (CIB) fluctuations, especially at the smallest angular scales accessible to such low-resolution probes ($\ell \gtrsim 1000$, see e.g., \citealt{Reichardt2012SPT}).
Using the bandpower measurements from the SPT-SZ survey and a model for the CIB contribution, \citet{Ramos-Ceja2015} showed that the current-generation tSZ power spectrum measurements are already useful in constraining the outer slope of the ICM pressure profile, or departures from self-similar evolution, for fixed cosmology. However, the drawback is that several astrophysical effects can conspire to produce similar changes in the tSZ power such that it will be difficult to break the associated parameter degeneracies (e.g., between the amount of non-thermal pressure support and its redshift evolution).
%, see \citealt{Ramos-Ceja2015}). 
It is also expected that a better knowledge of ICM astrophysics and the tSZ-CIB cross-correlation will help to alleviate some of the current tensions between the {\it Planck} and SPT-SZ measurements (e.g., \citealt{McCarthy2014}, \citealt{Dolag2016}).
Future high-precision, high-resolution measurements of the tSZ and CIB power down to smaller angular scales are therefore required. This will likely be one of the (many) science drivers for future ground-based CMB measurements with large-aperture telescopes, such as those described in \S \ref{sec:atlast} and \S \ref{sec:csst}. 

%%%%%%%%%%%%%%%%%%%%%%%%%%%%%%%%%%%%%%%%%%%%%%%%%%%%%%%%%%%%%%%%%%
%%%%%%%%%%%%%%%%%%%%%%%%%%%%%%%%%%%%%%%%%%%%%%%%%%%%%%%%%%%%%%%%%%%%%%%%%%%%%

\section{SZ view of ICM (sub)structure}
\label{sec:ICMstructures}

\subsection{Shock fronts}
\label{sec:shocks}

Shock fronts, both from mergers and cosmic accretion, should be readily observable in the tSZ at high spatial resolution due to their pressure jumps \citep[][]{Markevitch2007,molnar2009,ruan2013}. 
For an ideal gas with adiabatic index $\gamma$, the relation between the pressure in the pre- and post-shock regions can be derived from the standard Rankine-Hugoniot jump condition, and is given by (e.g., \citealt{Landau1959})
\begin{equation}
\dfrac{P_\mathrm{post}}{P_\mathrm{pre}} = \dfrac{2\gamma {\cal M}^2 - (\gamma-1)}{(\gamma+1)},
\label{eq:P-shock}
\end{equation}
where ${\cal M}$ is the sonic Mach number of the shock (${\cal M} \equiv \varv/c_s$, where $c_s$ is the sound speed), and the instantaneous electron pressure in the post-shock (downstream) and pre-shock (upstream) regions as $P_\mathrm{post}$ and $P_\mathrm{pre}$, respectively. The value of adiabatic index is typically assumed to be that of an ideal monoatomic or fully ionised gas, $\gamma=5/3$, but this can vary if there is a significant energy density contribution from non-thermal particles or magnetic fields, or if there is strong magnetic field parallel to the shock front (\citealt{Helfer1953}, \citealt{Sarazin2016}, \citealt{Guo-Sironi2018}). 
Note that equation \eqref{eq:P-shock} implicitly assumes a single component fluid with equal electron and ion temperature ($\Te = T_{\mbox{\tiny ion}}$), while the tSZ signal can only be used to infer the electron pressure. This immediately shows that if the post-shock thermalisation is not instantaneous, but instead for example follows Coulomb interactions, then ${\cal M}$ derived from tSZ measurements will give a lower estimate than from the density compression value, and a similar bias will also show up from the X-ray spectroscopic temperature ratio.
Typically, one would need to analyse both X-ray and tSZ data with the same geometrical considerations to get a complete picture of the processes involved in thermalisation following a shock.

It can be seen from Equation \ref{eq:P-shock} that the pressure ratio scales roughly as ${\cal M}^2$, unlike the density compression ratio ${n_\mathrm{post}}{n_\mathrm{pre}}$, which saturates at a value of 4 at ${\cal M} \gtrsim 5$. In addition, two advantages of the tSZ effect position it as very attractive for studying shocks in the ICM. First, the redshift independence of the SZ effect means shock-like features can be studied in clusters from arbitrarily high redshifts, unlike X-ray temperature measurements which has to contend with a rapidly diminishing number of photons. Second, the linear dependence on gas density for the tSZ signal amplitude means that it can be better suited to study merger or accretion shocks in the low density cluster outskirts.
A third and practical advantage of tSZ observation compared to X-ray spectroscopy is that for a relatively strong shock in the hot ICM, the post-shock temperature will reach a high value (in the range $20-40$~keV), well beyond the spectral window of most X-ray satellites. This can make modelling of important physical processes like the thermal equilibration timescale very difficult. This is not a concern for tSZ data, particularly when it is modelled jointly with the X-ray density compression. These considerations also bring SZ-based shock measurements within observational reach for the cluster outskirts region, e.g.\ for shocks associated with radio relics \citep[for a review of diffuse radio emission, see][]{Feretti2012} or the tentative detection of an accretion shock near the cluster splashback radius \citep{Hurier2019}.

%%%%%%%%%%%%%%%
\begin{figure*}
\begin{center}
\includegraphics[width=1.00\textwidth]{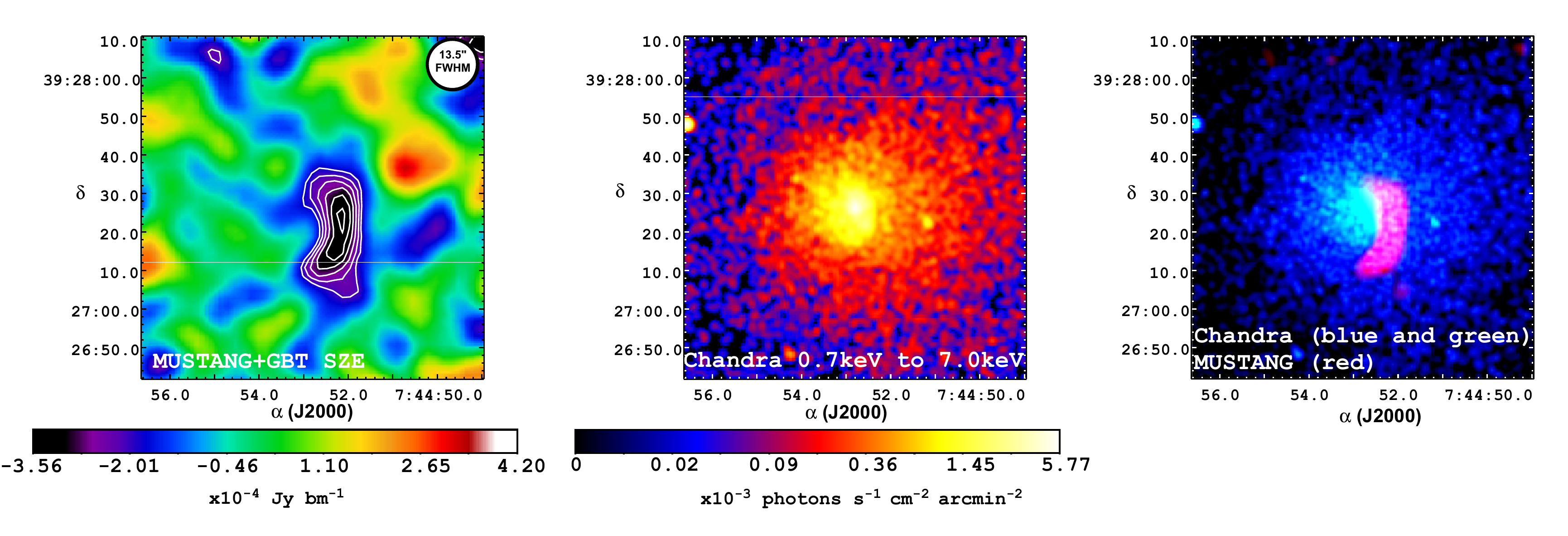}\\
\includegraphics[width=0.495\textwidth]{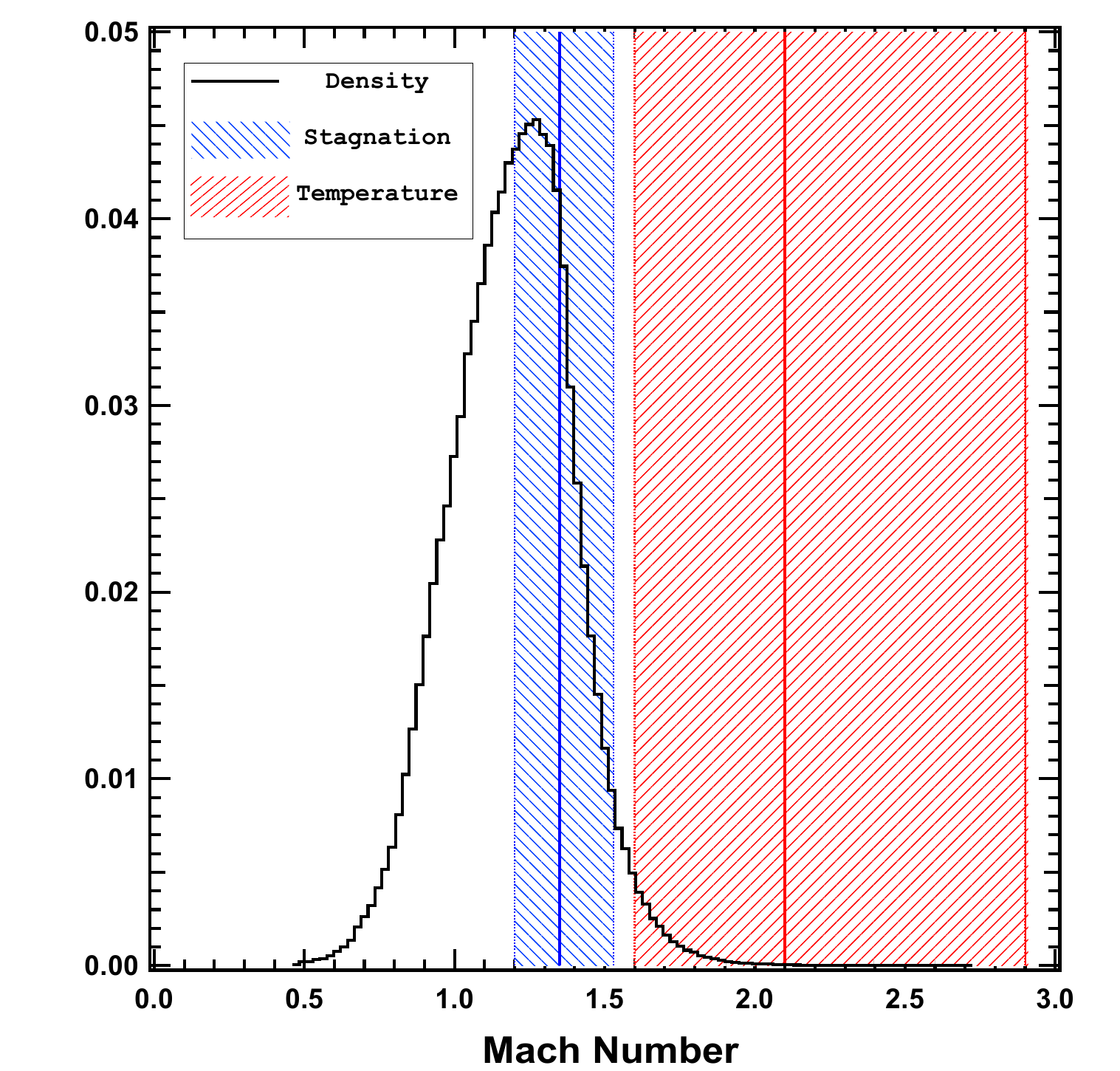}
\includegraphics[width=0.45\textwidth]{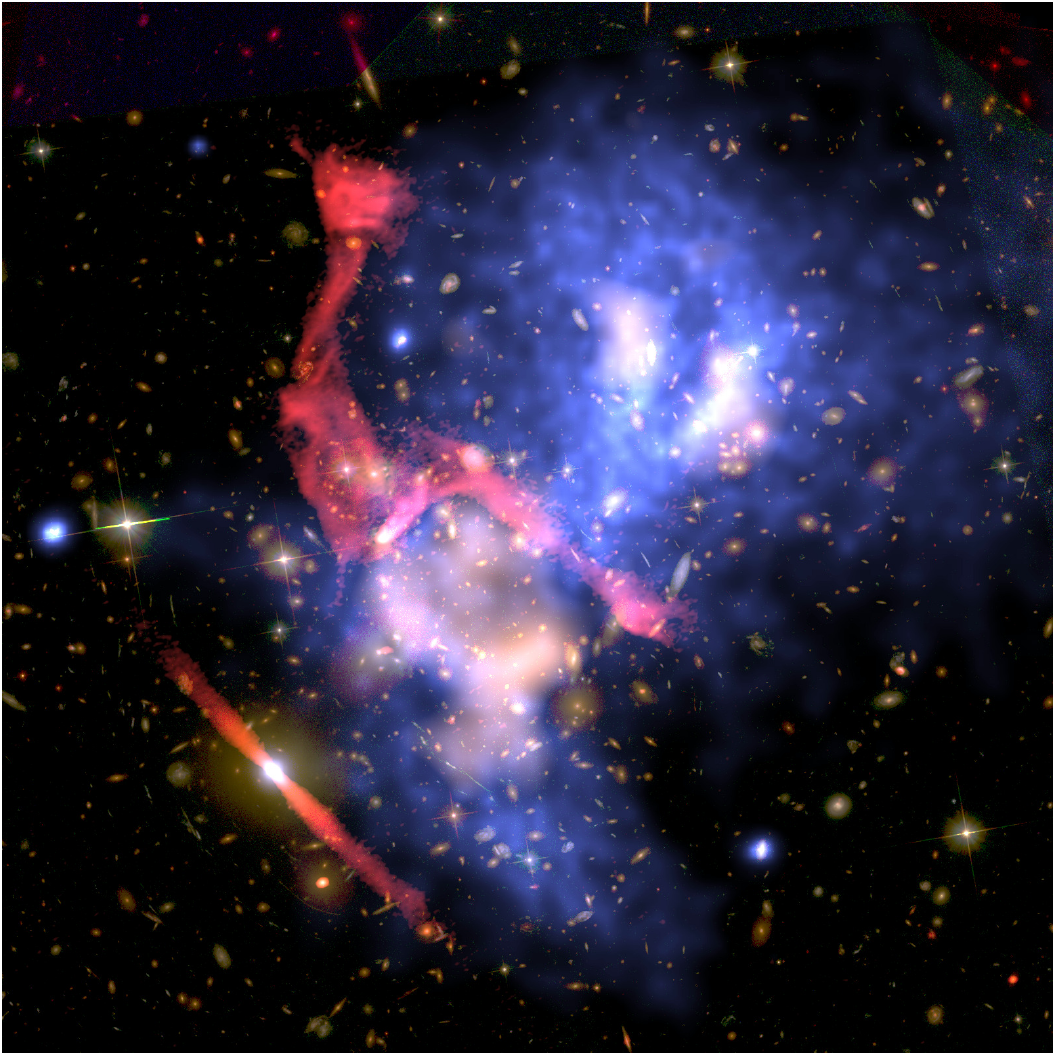}
\end{center}
\caption{A comparison of MUSTANG measurements of the shock-heated gas in the galaxy clusters MACSJ0744.8+3927 and MACSJ0717.5+3745 with multi-wavelength data from the {\it Chandra} X-ray Observatory, the Very Large Array (VLA), and the {\it Hubble} Space Telescope (HST).
{\it Top-left:} MUSTANG tSZ measurement at 90 GHz with effective resolution of 13.5$\arcsec$ (due to smoothing). 
{\it Top-middle:} {\it Chandra} X-ray image of MACSJ0744.8+3927.
{\it Top-right:} {\it Chandra} X-ray measurement in blue with the tSZ signal overplotted in magenta. 
{\it Lower Left:} Fitting for the shock Mach number in MACSJ0744 with three different methods.  The fitting was performed on the {\it Chandra} X-ray data using regions selected by the MUSTANG imaging. See \cite{Korngut2011} for details.
{\it Lower-right:} Multi-wavelength MUSTANG+{\it Chandra}+VLA+HST images of MACSJ0717.5+3745. The background image is HST imaging as part of the Frontier Fields project. The red overlay on both panels is S-band (2-4 GHz) radio data taken with the VLA, showing the radio relic associated with re-accelerated AGN tails.  The blue overlay is {\it Chandra} X-ray imaging.  The peach-coloured overlay is the MUSTANG SZ decrement at $> 3\sigma$, where the dominant feature appears to be a pressure overdensity shaping the southern half of the radio relic. All figures apart from the lower right panel are from \cite{Korngut2011}.  
Lower-right panel is courtesy Reinout van Weeren, using MUSTANG data presented in \citet{Mroczkowski2012} and {\it Chandra} + VLA data presented in \citet{vanWeeren2017}.  
}
\label{fig:macs0744-shock}
\end{figure*}
%%%%%%%%%%%%%%%

%%%%%%%%%%%%%%%
\begin{figure*}
\begin{center}
\includegraphics[height=0.38\textheight]{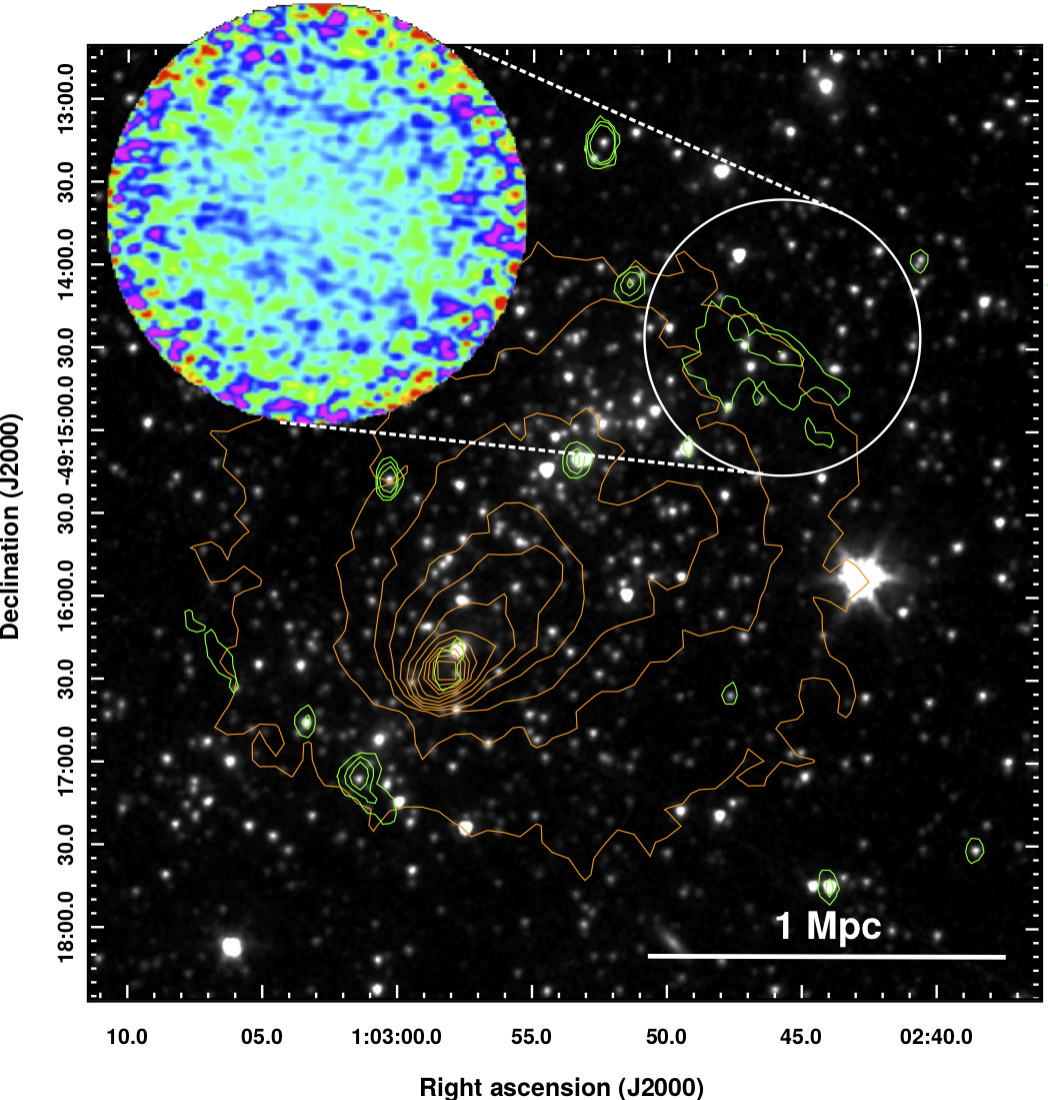}%
\includegraphics[height=0.38\textheight]{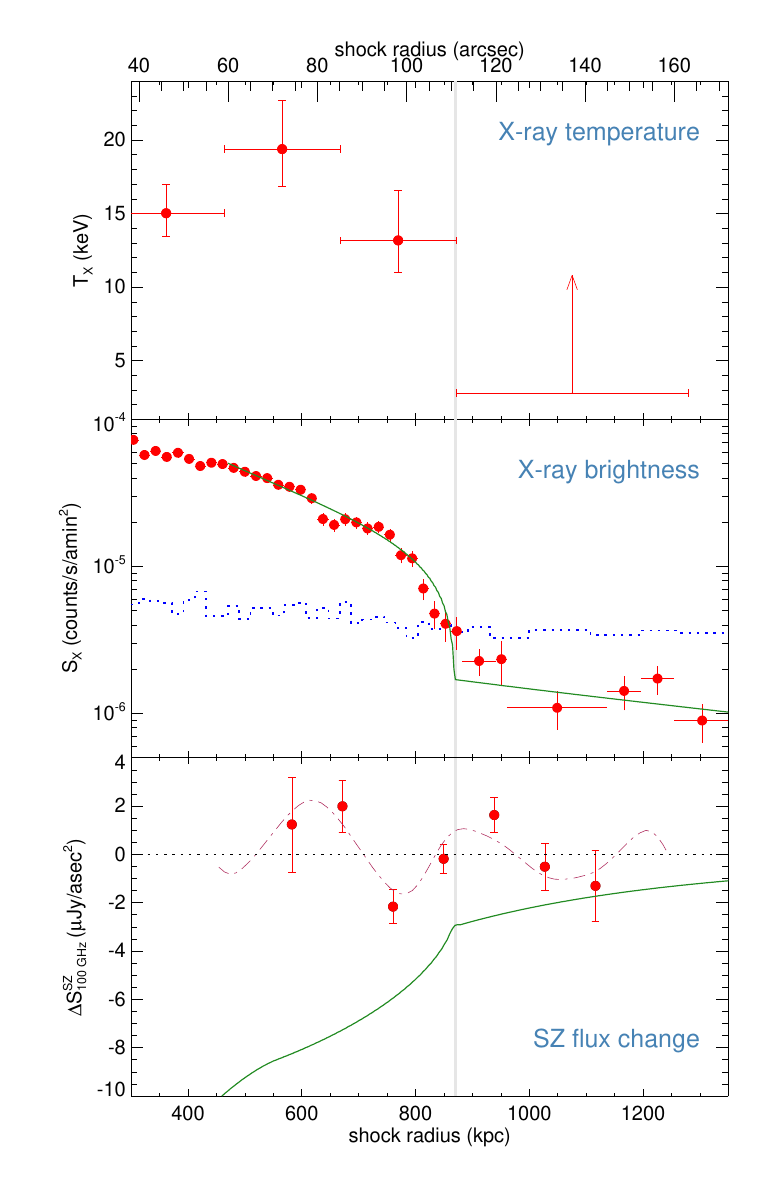}
\end{center}
\caption{An ALMA tSZ effect measurement of a shock front coincident with a bright radio relic, in the massive El Gordo galaxy cluster at $z=0.87$. 
\textit{Left:} Multi-wavelength view of the shock, with background image from a {\it Spitzer} IRAC mosaic at $3.6 ~\mu$m, and orange and green contours marking the {\it Chandra} X-ray $0.5-2$ keV X-ray emission (see also \citealt{Botteon2016}) and the 2.1~GHz radio emission \citep{Lindner2014}, respectively. The observed SZ intensity distribution from an ALMA deconvolved image is shown in the zoom-out inset.
\textit{Right:} The shock front as seen from both X-ray and tSZ measurements. The upper panel shows the temperature, where the very faint pre-shock region in this high-$z$ cluster do not provide enough photons to constrain the X-ray spectroscopic temperature. In the middle panel the surface brightness jump in X-ray clearly reveals a shock (or a cold front). Blue-dotted line is the mean X-ray background. In the bottom panel the ALMA SZ data indicates a modulated flux (red dot-dashed line) which is the result of interferometric imaging of a step function-like shock signal (shown with the green solid line for the best-fit model). Figures adapted from \citet{Basu2016}.  
}
\label{fig:alma_elgordo}
\end{figure*}
%%%%%%%%%%%%%%%

Realising these potentials requires good sensitivity and sufficiently high angular resolution, ideally well below the $\sim$1\arcmin\ beam which is the typical resolution for ground-based survey telescopes like SPT and ACT. An early example of tSZ imaging of putative shock-heated substructure in the ICM was provided by \cite{Kitayama04} using NOBA data. The first sub-arcminute detection of a shock feature was provided by \cite{Korngut2011}, which used the pathfinder MUSTANG-1 bolometer array on the 100-meter Green Bank Telescope (GBT) to image MACS J0744.8$+$3927.  The SZ-identified feature was then used to guide the region selection for X-ray analysis using {\it Chandra}, which provided constraints on the shock Mach number. Illustrations from this measurement are shown in Figure~\ref{fig:macs0744-shock}. It should be noted that the limited FoV of the MUSTANG-1 camera meant that only the shocked region could be detected, while the rest of the ICM contribution is filtered by the instrumental response. There is a same problem with interferometric observations, such as those performed with ALMA, as discussed below. This makes it impractical to image a large number of cluster targets to search for SZ shocks. Typically, X-ray observation are used to overcome the spatial filtering limitation or choose targets with known shock features. Low-resolution {\it Planck} all-sky survey data have also been used to study shocks, in the nearby Coma cluster \citep{Planck2013Coma}, in combination with {\it XMM-Newton}. An alternative approach has been to target radio relics in merging clusters, which are elongated synchrotron emitting regions thought to be associated with shock fronts. The first tSZ shock measurement on a radio relic was provided by \cite{Erler2015}, also for the Coma cluster.

Perhaps the most promising instrument for detailed tSZ studies of known shock fronts in clusters is the ALMA interferometric array. Despite its very limited FoV ($< 1\arcmin$ at 90 GHz; see \S \ref{sec:interferometric}), ALMA provides unprecedented sensitivity and resolution ($\lesssim 3\arcsec$ at 90 GHz) for accurate imaging of shock fronts. With future low-frequency additions to the ALMA suite of receivers (ALMA Bands 1 \& 2 in particular; see \S~\ref{sec:alma_B3}) the problem of limited FoV will be partially mitigated. This field is still in its early years, with the first ALMA observation of a merger shock being published by \citet{Basu2016} for the notorious galaxy cluster `El Gordo' at $z=0.87$ (Figure \ref{fig:alma_elgordo}). This high-$z$ target demonstrates several of the advantages mentioned earlier, namely a strong complementarity with X-ray data (here X-ray photon count is too low to make a measurement of the pre-shock temperature) and a measurement of shocks in the cluster outskirts at high-$z$. The inset image in the left panel of Figure \ref{fig:alma_elgordo} shows an image-equivalent (``CLEANed image'') of the ALMA interferometric data, where a wave-like intensity pattern indicates the presence of an underlying tSZ shock that roughly looks like a step function. The accepted approach for constraining a shock (or any other) model in such cases in not via the reconstructed images, which have biased noise properties (not uncorrelated), but rather by using a direct Bayesian analysis of the visibility data (or $uv$-data). This can also benefit from the combination with X-ray data to constrain the large scales inaccessible to ALMA.  The salient results of the shock analysis from \citet{Basu2016} are reproduced in the right-hand panels of Fig.\ref{fig:alma_elgordo}. The remarkable consistency of the shock location from tSZ and X-ray measurements, and the quality of the ALMA observation that took only a fraction of the time spent with {\it Chandra} for the same object, signify a coming-of-age of tSZ measurements for such astrophysical applications. 

Additional ALMA observations of shocks in other galaxy clusters have been made, including the well-known bow shock of the Bullet cluster, and the number of  results will grow in the coming years. Using multi-frequency ALMA observations, or by combining data from several instruments, an SZ-only decomposition of cluster shock fronts could soon become a realistic possibility. For example, the kSZ effect will reveal directly the velocity discontinuity at the shock front, although projection effects can limit its observability, since most known shocks are seen edge-on. 
One interesting application for future multi-frequency SZ measurements will be to determine post-shock temperatures via the rSZ effect, since the high temperatures involved will guarantee a large departure from the classical tSZ signal, potentially producing a detectable rSZ component (as well as a ntSZ component if a non-negligible population of relativistic electrons is present).  This rSZ component would provide a temperature estimate that is completely independent of X-ray data (tSZ modelling of temperature requires the X-ray density information). Future large-aperture single-dish telescopes like the AtLAST project (\S \ref{sec:atlast}) promise revolutionary new developments in this field.

\subsection{Cold fronts}
\label{sec:cold_fronts}

The tSZ view of cold fronts, which are contact discontinuities frequently seen in X-ray observations, is quite different from that of shocks (\S \ref{sec:shocks}). Cold fronts manifest as sharp jumps in X-ray surface brightness and temperature, but in opposite directions that roughly maintain thermal pressure balance. Simulations indicate that such contact discontinuities form through the subsonic bulk motions of gas. Due to the fact that cold fronts as seen in simulations are pressure-continuous, they should be invisible in the tSZ if the pressure is predominately thermal. There are two exceptions to this: a) if the ram pressure of the surrounding medium is large enough or b) if there is a strong magnetic draping layer either directly above or below the front surface. Regarding the latter possibility, these layers may form due to shear amplification of the weak ambient field during cold front formation and evolution. Such layers should be observable as a thin deficit in the tSZ signal in the vicinity of the front as seen in X-rays, but a high angular-resolution instrument will be required to detect them. In general, the thermal pressure {\it gradient} across cold fronts is not continuous in simulations, and therefore should be detectable in the tSZ.

The kSZ effect could also reveal the tangential (line of sight) gas motion across a cold front located predominantly in the plane of the sky, even if the thermal pressure were continuous across the front.
At the smallest scales the motions of electrons and ions in the weakly magnetised ICM are inherently anisotropic due to the fact that the Larmor radii of the electrons and ions are many orders of magnitude smaller than their respective mean free paths. This can manifest itself as a pressure anisotropy due to evolving magnetic fields via adiabatic invariance and other processes. \citet{Khabibullin18} showed that such a pressure anisotropy could produce polarisation in the tSZ signal due to the different thermal velocities of the electrons in the directions perpendicular and parallel to the local field line geometry, which they estimated to be produced at a level of $\sim$10~nK. Such polarisation might be prominent near cold front surfaces where the magnetic field is increasing rapidly due to shear amplification which would in turn drive a local pressure anisotropy. 

%%%%%%%%%%%%%%%
\begin{figure*}
\begin{center}
\includegraphics[width=0.49\textwidth]{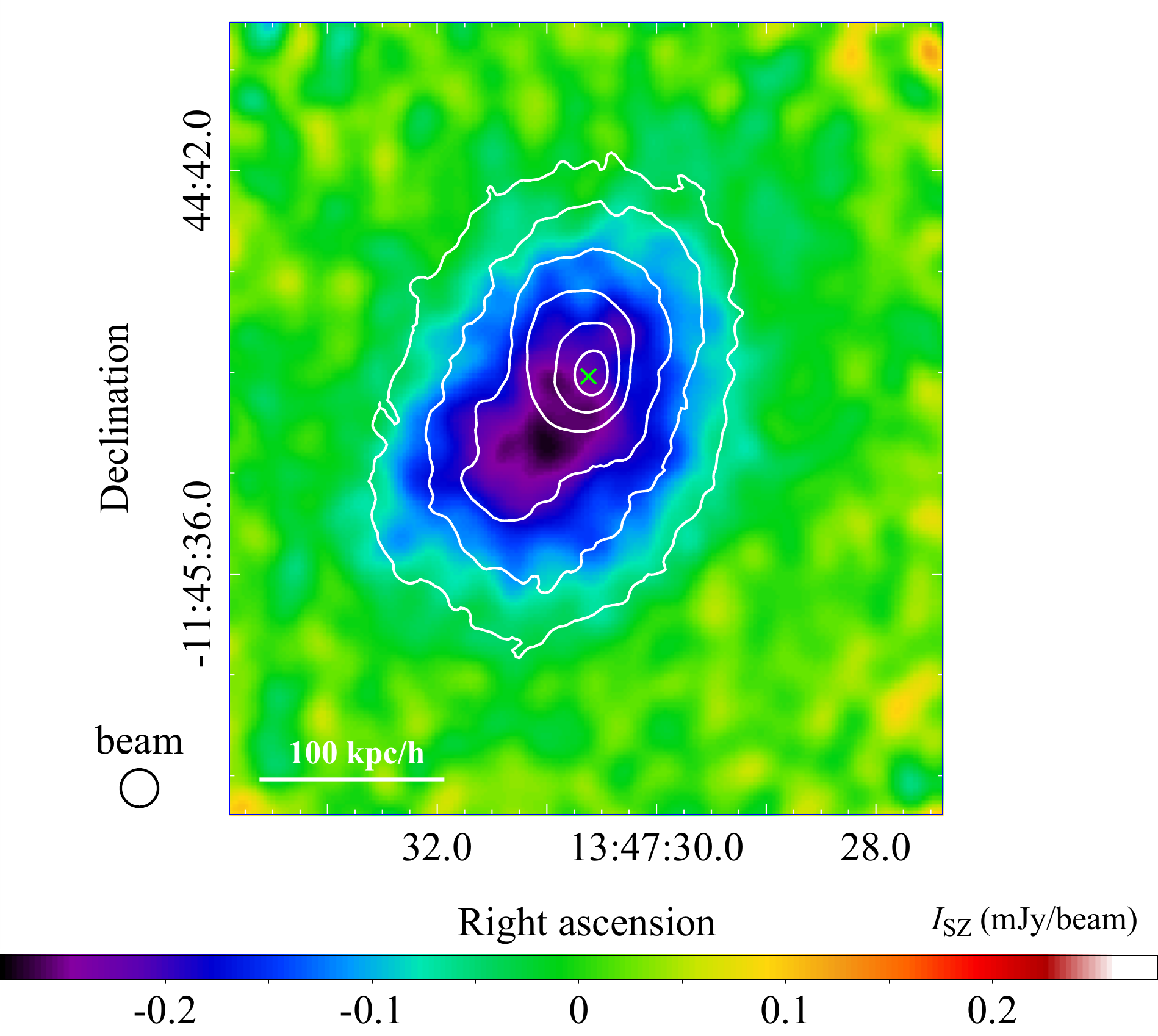}
\includegraphics[width=0.49\textwidth]{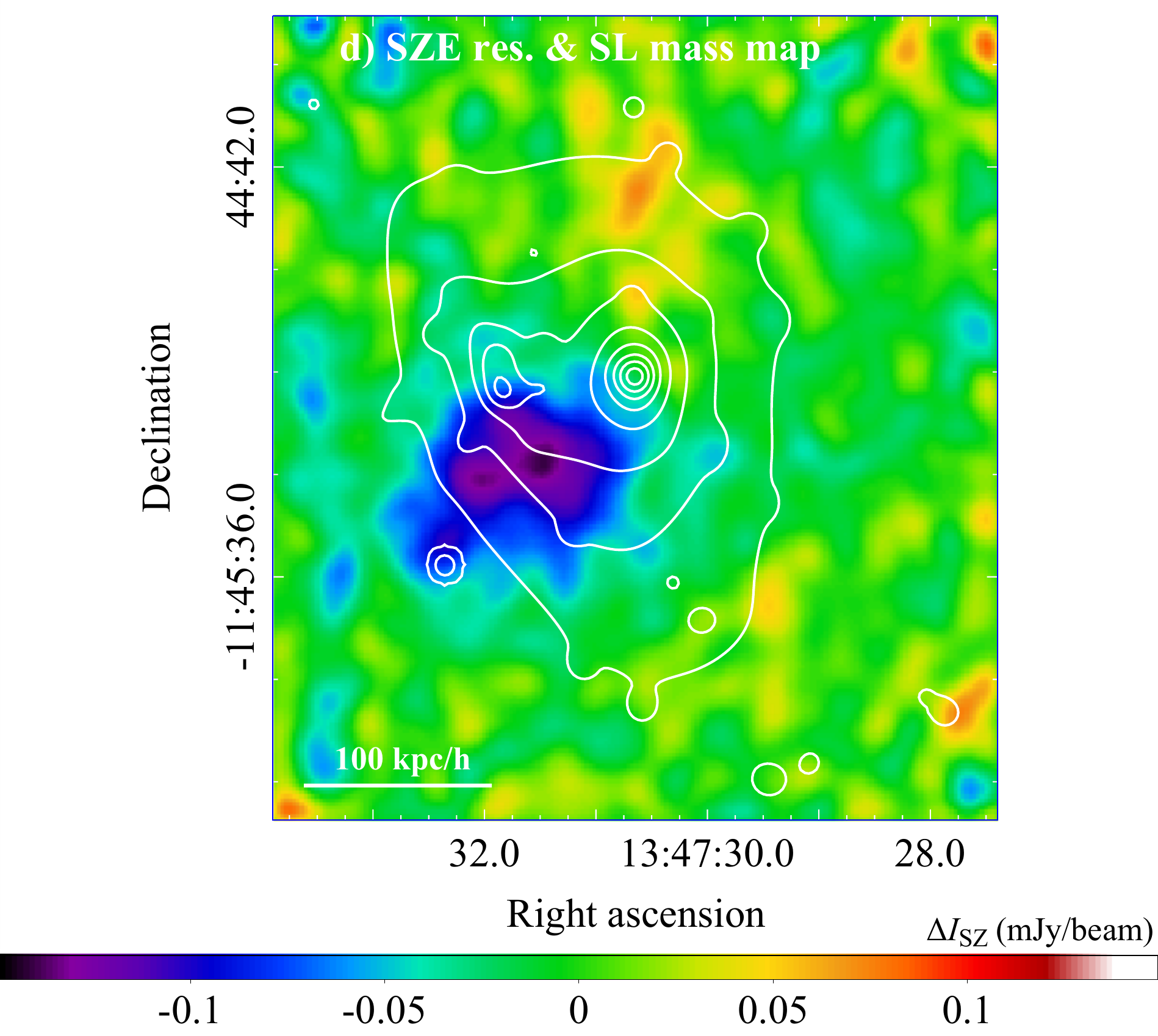}
\end{center}
\caption{ALMA+ACA maps of RX~J1347.5-1145.
\textit{Left panel:} ALMA+ACA image with {\it Chandra} X-ray contours overlaid, showing an offset between the X-ray and SZ surface brightness distributions.
\textit{Right panel:} Residuals in ALMA+ACA image after subtraction of a mean SZ profile computed from the CLEANed map by excluding the region corresponding to the southeastern substructure, with lensing contours overlaid, revealing the location of the dark matter component.
Figures from \cite{Ueda2018}.  See \cite{Ueda2018} for details of the analysis.
}
\label{fig:rxj1347}
\end{figure*}
%%%%%%%%%%%%%%%

\subsection{Probing pressure substructures with tSZ effect}
\label{sec:substructure}

Thanks to the advances of high angular resolution imaging experiments, it is possible to detect and characterise pressure substructures in the ICM through SZ studies. Substructures detected through the tSZ effect are related to gas compression driven by merger events associated with infalling substructures in the ICM. In combination with X-ray observations,  tSZ substructure characterisation has proven useful in furthering our understanding of the underlying ICM physics.
For instance, SZ observations from the DIABOLO \citep{Pointecouteau2001} and NOBA \citep{Komatsu2001,Kitayama04} instruments, respectively installed on the IRAM-30m and Nobeyama-45m telescopes, detected an offset between the dense X-ray core and the pressure peak in the cluster RX~J1347.5-1145 ($z=0.45$), which at that time had been known to be a typical dynamically relaxed cool-core cluster. These observations indicate that RX~J1347.5-1145 is an ongoing merger, and this cluster was later studied by many targeted SZ experiments, such as MUSTANG-1 \citep{Mason2010}, CARMA \citep{Plagge2013}, NIKA \citep{Adam2014}, BOLOCAM \citep{Sayers2016} and ALMA \citep{Kitayama2016}. For illustration, the ALMA image, at a resolution of 5\arcsec, is presented in Figure \ref{fig:rxj1347} and compared to X-ray and strong lensing data. These data were used to characterise the cluster core and the overpressure caused by the merging of a sub-cluster in RX~J1347.5-1145 \citep{Ueda2018}. In combination with X-ray measurements, these SZ data provided the first direct observational evidence for sub-sonic nature of sloshing motion of the cool core. Additionally, a comparison of the offsets between the peaks in the SZ and strong lensing signals was used to infer the self-interaction cross-section of dark matter to be less than $3.7 \ h^{-1}$ cm$^2$ g$^{-1}$.

In addition to the emblematic case of RX~J1347.5-1145, many other clusters were observed to study substructures using the tSZ effect. MUSTANG observations of the massive Frontier Fields cluster MACSJ0717.5+3745 ($z=0.55$), which contains at least 4 merger components \citep{Ma2009,vanWeeren2017}, revealed a high pressure region, presumably resulting from the merger of the 2 most massive components and likely driving and reshaping the southerly radio jet (see the lower-right panel of Figure \ref{fig:macs0744-shock}).  The shocked overpressure region is confirmed in X-ray observations, which indicate the presence of hot ($>$ 18 keV) gas.  An additional SZ feature is seen to the west, in a region of high X-ray surface brightness and lower temperature, indicating the location of an intact cool core.

Furthermore, the {\it Planck} satellite has resolved the tSZ emission from massive nearby systems with a high signal-to-noise ratio, allowing investigations of clusters such as Coma \citep[showing significant steepening in two regions, consistent with the presence of shocks,][]{Planck2013_X}, the Virgo cluster \citep[the largest SZ source in the sky,][]{Planck2016_XL}, and the interacting cluster pair A399 - A401 (presenting a hot gas filament connecting the two objects), and the Shapley super-cluster \citep[][where other nearby clusters are also presented]{Planck2014_XXIX,Planck2016_XXII}.

Ground-based observations using on MUSTANG, NIKA, and NIKA2 have also been instrumental in probing ICM substructures at higher redshifts, with studies ranging from identification of substructures, the impact of substructure on cluster pressure profiles, mass estimates and global morphology, and the study of shocks \citep[see for instance,][as well as the discussion in \S~\ref{sec:profiles} and \S~\ref{sec:shocks}]{Korngut2011,Mroczkowski2012,Adam2015,Young2015,Adam2016,Ruppin17a,Ruppin2018}. 
Notably, the maturity reached in last decade by high spatial resolution SZ imaging has enabled advanced image filtering techniques typically reserved for X-ray to be applied to SZ observations \citep{Adam2018a}. As SZ imaging instrumentation improves and the higher signal to noise images are obtained, we can expect that many of the imaging analysis techniques developed for X-ray studies will be applied to SZ imaging as well.

%%%%%%%%%%%%%%%
\begin{figure*}
\begin{center}
\includegraphics[width=0.48\textwidth]{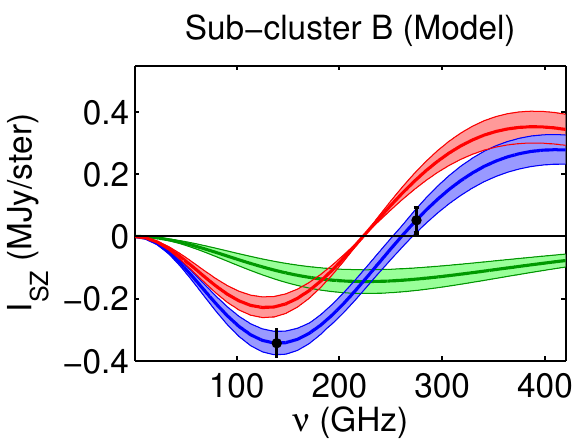}
\includegraphics[width=0.48\textwidth]{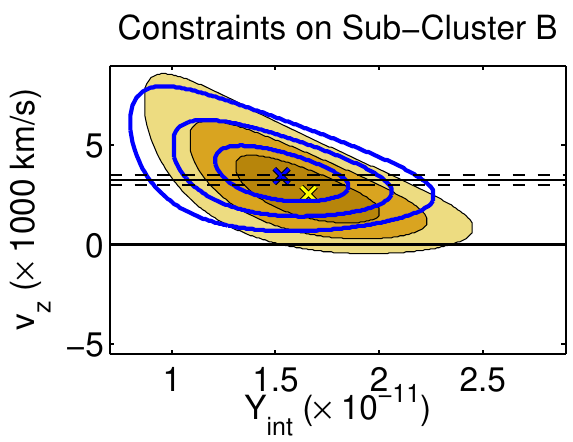}
\includegraphics[trim=0cm 0.7cm 0cm 0cm, clip=true, height=3.0cm]{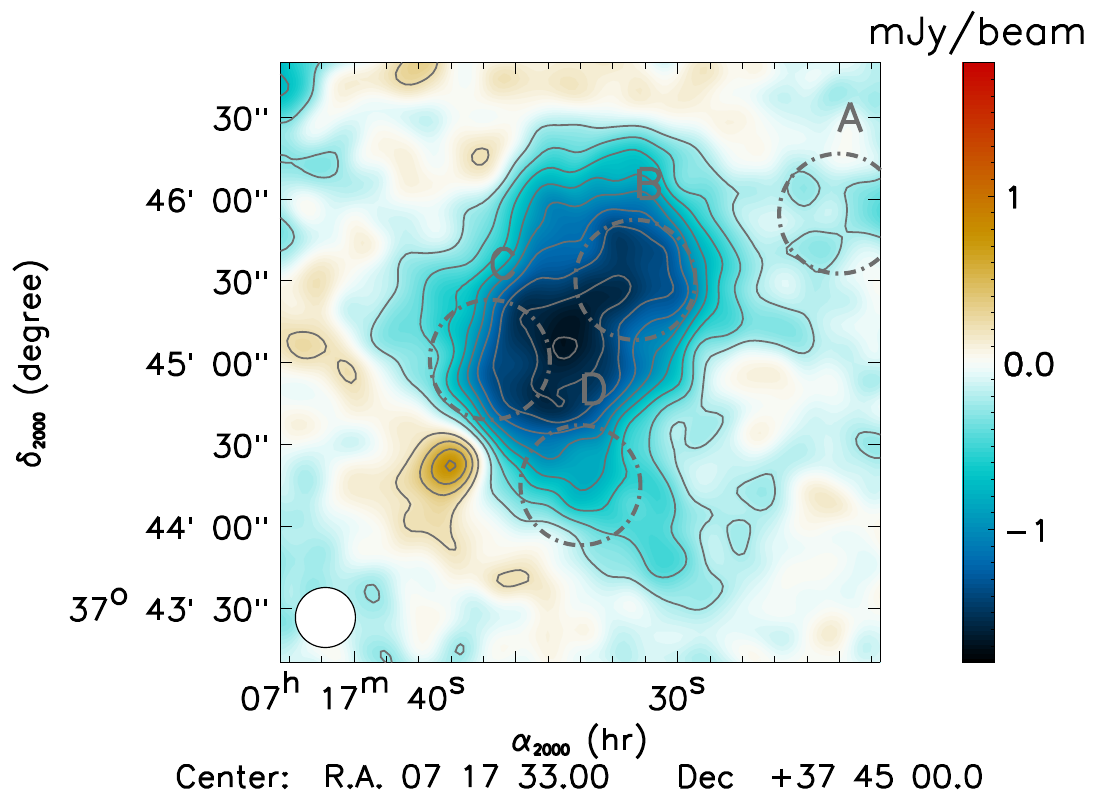}
\includegraphics[trim=3.5cm 0.7cm 0cm 0cm, clip=true, height=3.0cm]{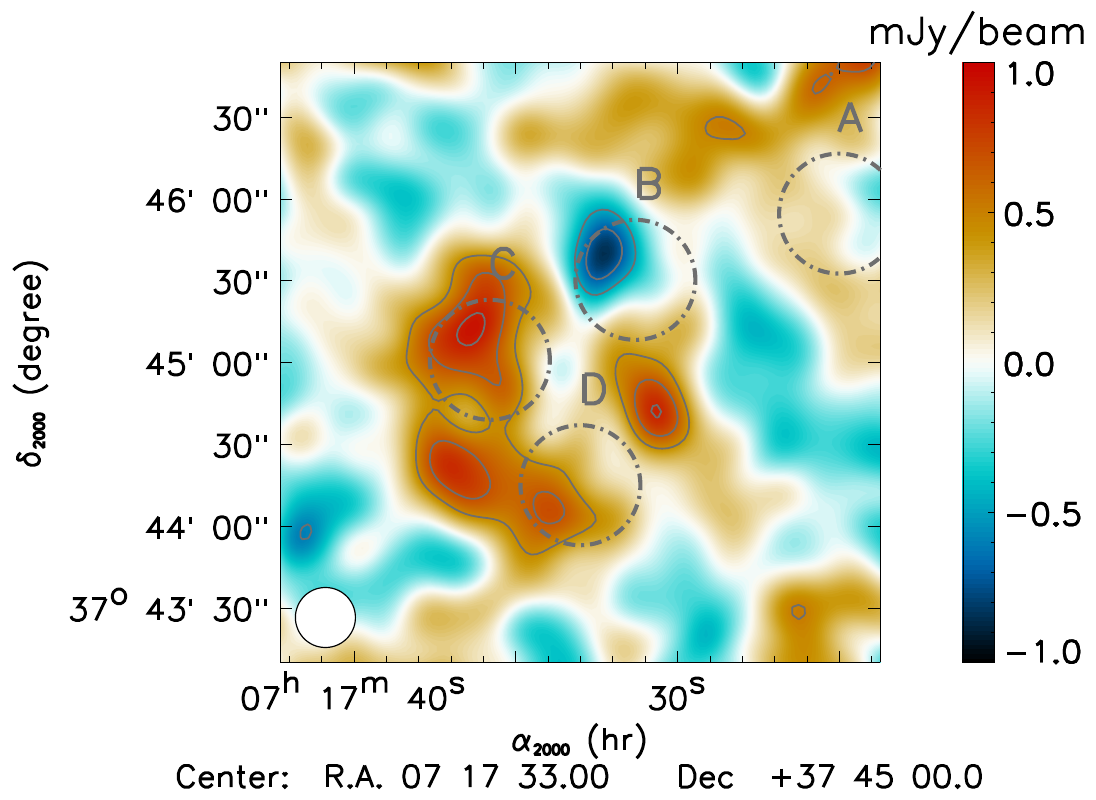}
\includegraphics[trim=3.5cm 0.7cm 0cm 0cm, clip=true, height=3.0cm]{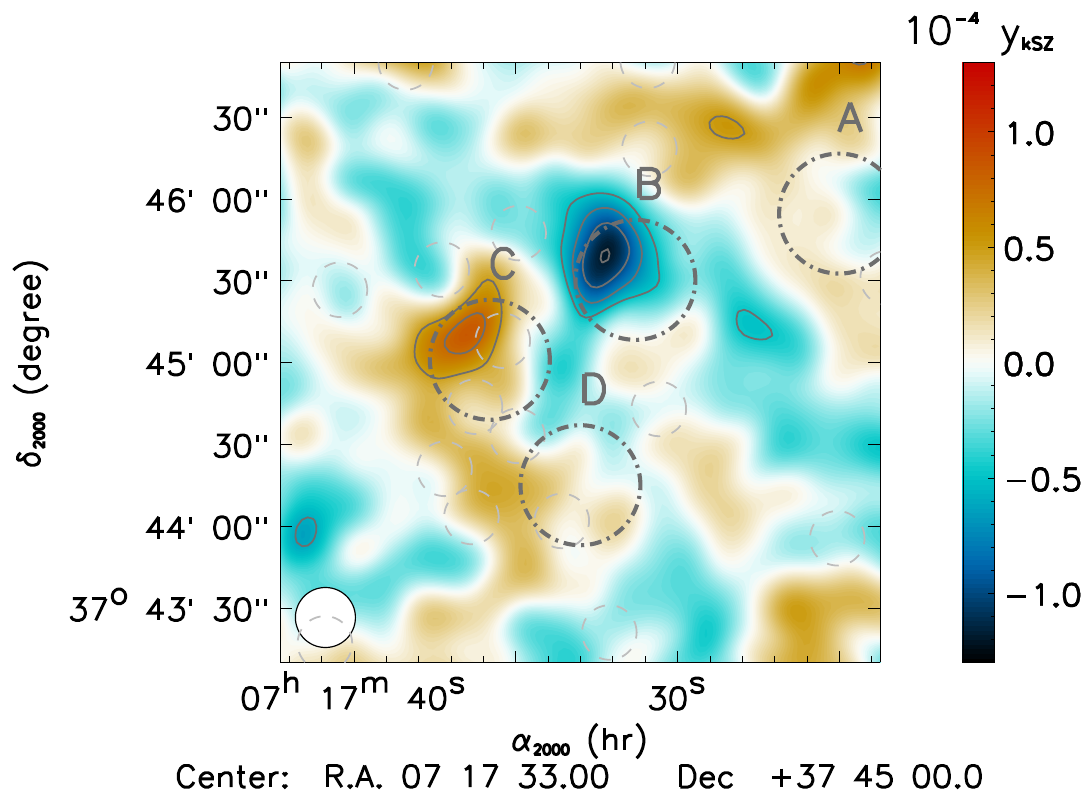}
\end{center}
\caption{Bolocam and NIKA constraints on the kSZ signal towards MACS~J0717.5+3745.  
\textit{Upper-left:} Bolocam constraints on the tSZ (red) and kSZ (green) contribution to the total SZ signal (blue) towards sub-cluster B (the main X-ray dense core, see bottom figures for the localisation).
\textit{Upper-right:} Bolocam constraints on the line-of-sight velocity of sub-cluster B.
\textit{Lower-left:} NIKA 150 GHz raw map.
\textit{Lower-middle:} NIKA 260 GHz raw map.
\textit{Lower-right:} reconstructed kSZ signal ($y_{\rm kSZ}$) by combining NIKA 150 and 260 GHz map, after cleaning for contaminants.
Figures in the upper panels are from \citet{Sayers2013}, and those in the lower panels are from \citet{Adam2017ksz}.
}
\label{fig:macs0717}
\end{figure*}
%%%%%%%%%%%%%%%

\subsection{Probing ICM substructures with resolved kSZ}
\label{sec:resolved_kSZ}

High-spatial resolution maps of the kSZ signal can in principle be used to measure the internal gas motions in the ICM, including bulk and random gas motions \citep{Sunyaev2003,Nagai2003} and rotational motions \citep{Chluba2001, Cooray2002,Chluba2002,Baldi2017,Baldi2018}. Such maps will provide a complementary measure of gas motions to those based on the X-ray surface brightness or pressure fluctuation analyses discussed in \S~\ref{sec:sz_powerspec}, as well as those which will eventually be produced by microcalorimeters on future X-ray missions like the X-ray Imaging and Spectroscopy Mission (XRISM; \cite{Tashiro2018}), {\it Athena} \citep{Nandra2013}, or Lynx \citep{Gaskin2016}, which will be capable of measuring the shift and width of velocity-broadened spectral lines \citep{Inogamov2003}. The different density dependence of the two methods could potentially disentangle the effect of density variations along the line of sight from the measurement of the bulk velocity (see Figure \ref{fig:sz_vs_xray}).

While the first detection of the kSZ signal was performed in a statistical manner, by measuring the mean pairwise momentum of optically identified clusters \citep{Hand2012}, the first kSZ detections in a single cluster were performed by resolving the signal towards MACS~J0717.5+3745 \citep{Mroczkowski2012,Sayers2013,Adam2017ksz}.
At a redshift of $z=0.55$, MACS~J0717.5+3745 is a striking example of a massive merging system and thus is an excellent target within which to search for the kSZ signal. It hosts at least four sub-groups known to present exceptionally large relative line-of-sight velocities from optical measurements, as first noted by \cite{Ma2009}.
Given the large merger velocity ($\beta \approx 0.01$) and the compact nature of the source (a few arcminutes), it is not surprising this has served as the best known source where non-zero velocities can be inferred from the kSZ signal.
Indeed, despite several attempts at detecting the kSZ signature in other sources \citep[e.g.,][]{Holzapfel1997,Benson2003,Sayers2016,Sayers2018}, the kSZ effect signal remains difficult to extract in individual clusters since it follows the same spectral dependence as the primordial CMB.  It can only be identified using the two signals' differing spatial distributions, requiring both high sensitivity and high angular resolution observations. 
Additionally, as the kSZ signal is fainter than the tSZ signal (see Figure~\ref{fig:szspectrum}), it is more prone to contamination from dust and radio emission (see \S \ref{sec:contam} and the discussion in e.g.\ \citealt{Sayers2018}). 

Figure \ref{fig:macs0717} shows the Bolocam \citep{Sayers2013} and NIKA \citep{Adam2017ksz} measurements of MACS~J0717.5+3745. 
In both cases, these analyses took advantage of the dual-band capabilities of the instruments to separate the tSZ from the kSZ contributions, along with ancillary data from e.g.\ {\it Herschel}. Thanks to the higher angular resolution of NIKA with respect to Bolocam (12\arcsec\ and 18\arcsec\ versus 31\arcsec\ and 58\arcsec), they were able to obtain a resolved map of the kSZ signal from the dual-band measurement. 
As noted above, {\it Herschel} observations were essential to account for the contribution from unresolved dusty galaxies, while {\it Chandra} and {\it XMM-Newton} X-ray data were crucial to disentangle the density, temperature, and pressure, in order to extract the gas line-of-sight peculiar velocity, under various modelling assumptions. The main bullet-like core of MACS~J0717.5+3745 was found to move at velocity of $\sim 3000$ km s$^{-1}$ with respect to the CMB reference frame, while no overall net velocity was detected for the whole cluster. Such a high velocity remains consistent with standard \LCDM\ cosmology.

Although still very futuristic, by also adding polarisation information through the pSZ (see \S~\ref{sec:pSZ}) one could in principle reconstruct the full 3D bulk and rotational velocity vectors, thus directly probing the large-scale motions of ICM structures.  Such measurements will require improvements in sensitivity by several orders of magnitude, as well as better spectral coverage for separating and removing contaminating signals.

\subsection{ICM turbulence from pressure fluctuations}
\label{sec:sz_powerspec}

Azimuthally-averaged radial profiles of thermodynamic quantities - the gas pressure, density, temperature or entropy are important proxies for global properties of a cluster, such as its mass. On the other hand, a great deal of information on the ICM dynamics is encoded in the velocity field and in small-scale fluctuations of thermodynamic quantities. Extensive measurements of the former will become available with future X-ray missions like XRISM, {\it Athena}, and Lynx and/or future kSZ imaging, while the latter can  already be measured with the currently operating X-ray observatories and SZ facilities. Cosmological simulations predict that even in relaxed clusters the ICM is not at rest, and the amplitudes of the velocity/density/pressure variations are sensitive to many currently unknown characteristics of the ICM. 

The first attempt to measure pressure fluctuations in the Coma cluster was done in \cite{Schuecker2004}, where  X-ray images and  temperature maps obtained with  {\it XMM-Newton} were combined to get a map of pressure fluctuations. A notable observational complication of this approach is that the accurate temperature measurements require a large number of X-ray photon counts, implying that only low angular resolution maps can be constructed. This reduces the spatial dynamic range that can be probed. It was later realised that it is much easier to deal  with the X-ray images themselves, rather than with the pressure maps, since the X-ray emissivity below $\sim 3$~keV is a direct proxy for the density fluctuations. These density fluctuations could be related to the pressure variations in the turbulent eddies, variations of the gravitational potential, sound waves, bubbles of relativistic plasma, or entropy variations due to the gas displaced from its equilibrium position \citep[][]{Churazov2012}.  Numerical experiments with driven solenoidal turbulence and with the cosmological simulations have shown that for the large-scale subsonic motions, the latter mechanism (entropy perturbations) is likely the dominant source of density variations, which can be related to the scale-dependent amplitude of gas velocities \citep[e.g.,][]{Gaspari2013,Gaspari2014,Zhuravleva2014a}. It was also found that for the cores of relaxed clusters, the isobaric nature of the dominant perturbations can be confirmed observationally \citep{Arevalo2016,Churazov2016,Zhuravleva2016}. This (and similar) approaches have been used to estimate the gas density and velocity power spectra in a number of clusters, with the freedom in choosing the underlying ``unperturbed'' density distribution models being the one of the main sources of uncertainty \citep[e.g.,][]{Zhuravleva2014b,Walker2015,Eckert2017,Zhuravleva2018}.

%%%%%%%%%%%%%%%
\begin{figure*}
\begin{center}
\includegraphics[width=0.49\textwidth]{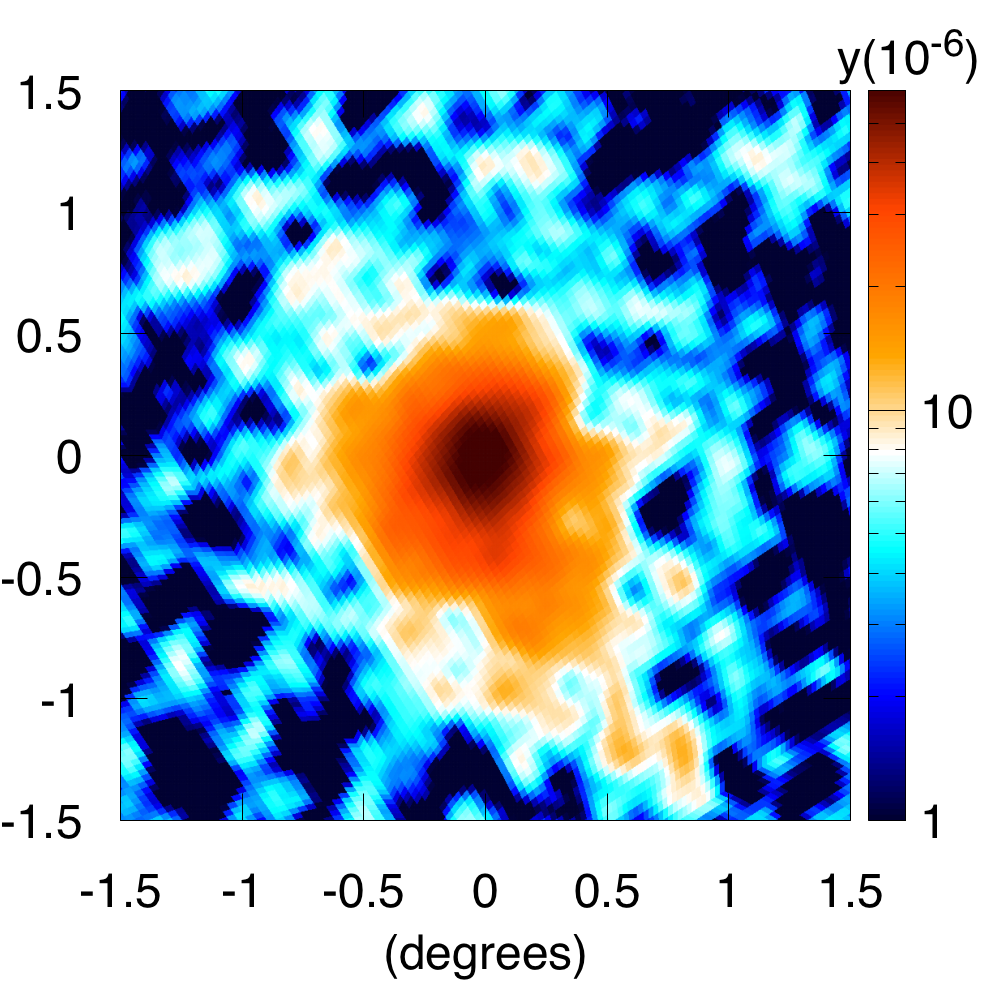}
\includegraphics[width=0.49\textwidth]{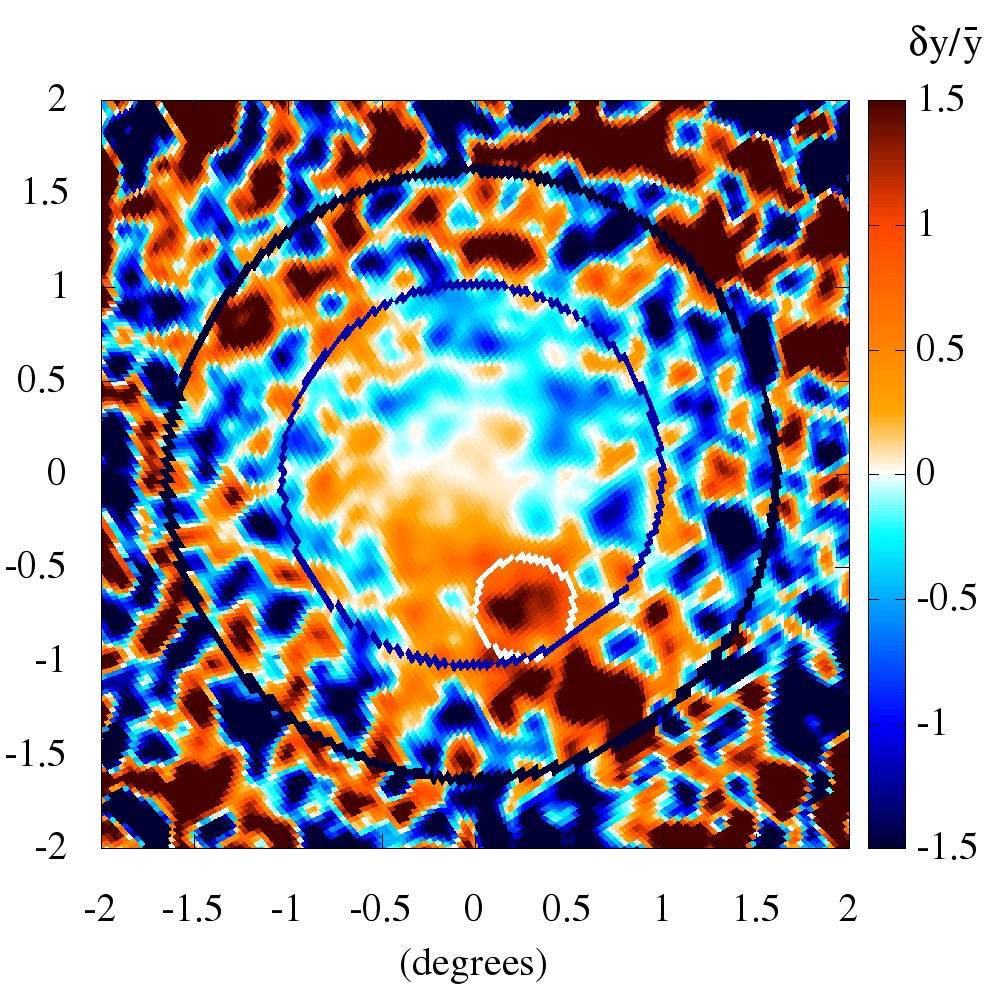}\\
\includegraphics[width=0.75\textwidth]{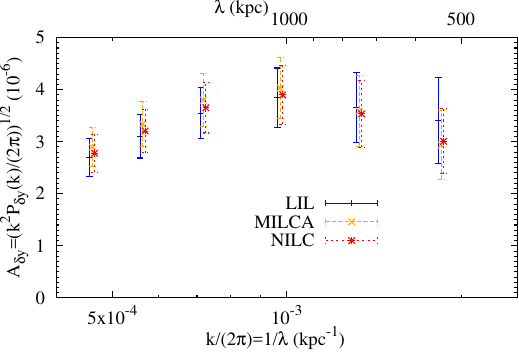}
\end{center}
\caption{
\textit{Upper Left:} The {\it Planck} inferred Compton-$y$ map of the Coma cluster.
\textit{Upper Right:} Residuals in the {\it Planck} Coma map after subtraction of and division by the best fitting model.
\textit{Lower:} Power spectrum of Compton-$y$ fluctuations for three different methods for combining the multi-frequency {\it Planck} data to infer the Compton-$y$ map used in the analysis.
Figures from \citet{Khatri2016}.  
}
\label{fig:coma}
\end{figure*}
%%%%%%%%%%%%%%%

Unlike the X-ray data, tSZ observations are sensitive only to the electron pressure variations.
Therefore, SZ data provide a highly complementary (to X-rays) view of the ICM perturbations, ``cleaned'' from the pure isobaric (entropy) perturbations. Given that in cool cores the isobaric perturbations seem to dominate, we can expect such cores to appear much smoother in deep SZ imaging than in X-rays, although the bubbles of relativistic plasma (see \S \ref{sec:agn_feedback_physics}) and sound waves generated by the AGN \citep[e.g.,][]{Fabian2006} could still contribute. Based on cosmological simulations \citep[e.g.,][]{Lau2009,Zhuravleva2013,Nelson2014}, one can expect on average $\approx$\,10\% pressure fluctuations close to but outside the cores of relaxed systems ($\approx$\,20\% in disturbed systems), progressively increasing toward the outskirts region. Therefore, the role of non-isobaric modes, involving pressure variations, might become more important than the entropy perturbations at large distances from a cluster core. This is especially true for merging systems, where shocks cause large variations in the thermal pressure.

The first exploratory study of SZ fluctuations \citep[][see Figure~\ref{fig:coma}]{Khatri2016} was based on the Coma cluster using {\it Planck} data. The Coma cluster has an angular size of a few degrees, close to the peak anisotropy in the CMB power spectrum.  Thus the analysis necessitates {\it Planck} multi-frequency coverage, in order to separate SZ fluctuations from the CMB primary anisotropies and foregrounds. At the same time, the effective {\it Planck} angular resolution is  $\sim 5-10'$, which limits the available dynamic range. The characteristic amplitude of the pressure variations obtained from the {\it Planck} data is $\sim 30$\% on scales $\sim 15'$ ($\sim 400$~kpc). \cite{Khatri2016} assumed that these pressure variations and the density fluctuations on smaller scales derived from X-ray images \cite[][]{Churazov2012} are part of the same spectrum, and concluded that pressure variations dominate in Coma. It should be noted, that similar to the X-ray analysis, the choice of the ``unperturbed'' global model can affect the results. 
This is because one must divide by the fitted bulk model to estimate the relative fluctuations,  \citep[see e.g.][]{Bonafede2018,DiMascolo2018}. In addition, residuals from primary CMB and foreground fluctuations provide an additional source of uncertainty, which -- as was done in \cite{Khatri2016} --  must be carefully accounted for even when multi-wavelength data are available.

A handy way to combine SZ and X-ray data would be by cross-correlating the deviations seen in SZ and X-ray images relative to the same underlying model used to describe the global distributions of the X-ray and SZ signals \citep[see \S 4.5 in][]{Churazov2016}. This can be used to differentiate between fluctuations having a different nature, e.g., adiabatic versus isobaric fluctuations. Such an approach would work best if the SZ and X-ray images have comparable angular resolutions and comparable S/N per resolution element. For nearby clusters like Coma, the ability to recover large ($\sim 1$ degree) scales is also required, which is challenging for ground-based SZ instruments with few arcminute FoVs (see \S~\ref{sec:interferometric} and \S~\ref{sec:photometric}) and limited frequency coverage, which impacts component separation and the removal of astrophysical contamination (see \ref{sec:contam}). 
On the X-ray side, such widefield data could be provided by forthcoming SRG/eRosita, exploiting its large FoV \citep{Predehl2016,Predehl2017}. 
Both X-rays and SZ data suffer from projection effects, which limit the dynamic range of the spatial scales that can be probed. Indeed, when the size of the perturbation is much smaller than the characteristic size of the cluster, the amplitude of the perturbation is attenuated by averaging along the line of sight. The pressure distribution (SZ) is obviously more extended than the distribution of the density squared (X-rays), implying that SZ data suffer from projection even more than X-rays. As we elaborate in our conclusions (\S \ref{sec:conclusion}), future facilities will have the opportunity to address many of these limitations.
Despite the current limitations, it is clear that studying pressure fluctuations using SZ data alone or in combination with X-ray data is an extremely powerful way of studying the dynamical state of the ICM and its physical properties. 

\subsection{The physics of AGN feedback}
\label{sec:agn_feedback_physics}

%%%%%%%%%%%%%%
\begin{figure}
\begin{center}
\includegraphics[trim=2.4cm 0.3cm 2.4cm 1.6cm, clip=true, width=\textwidth]{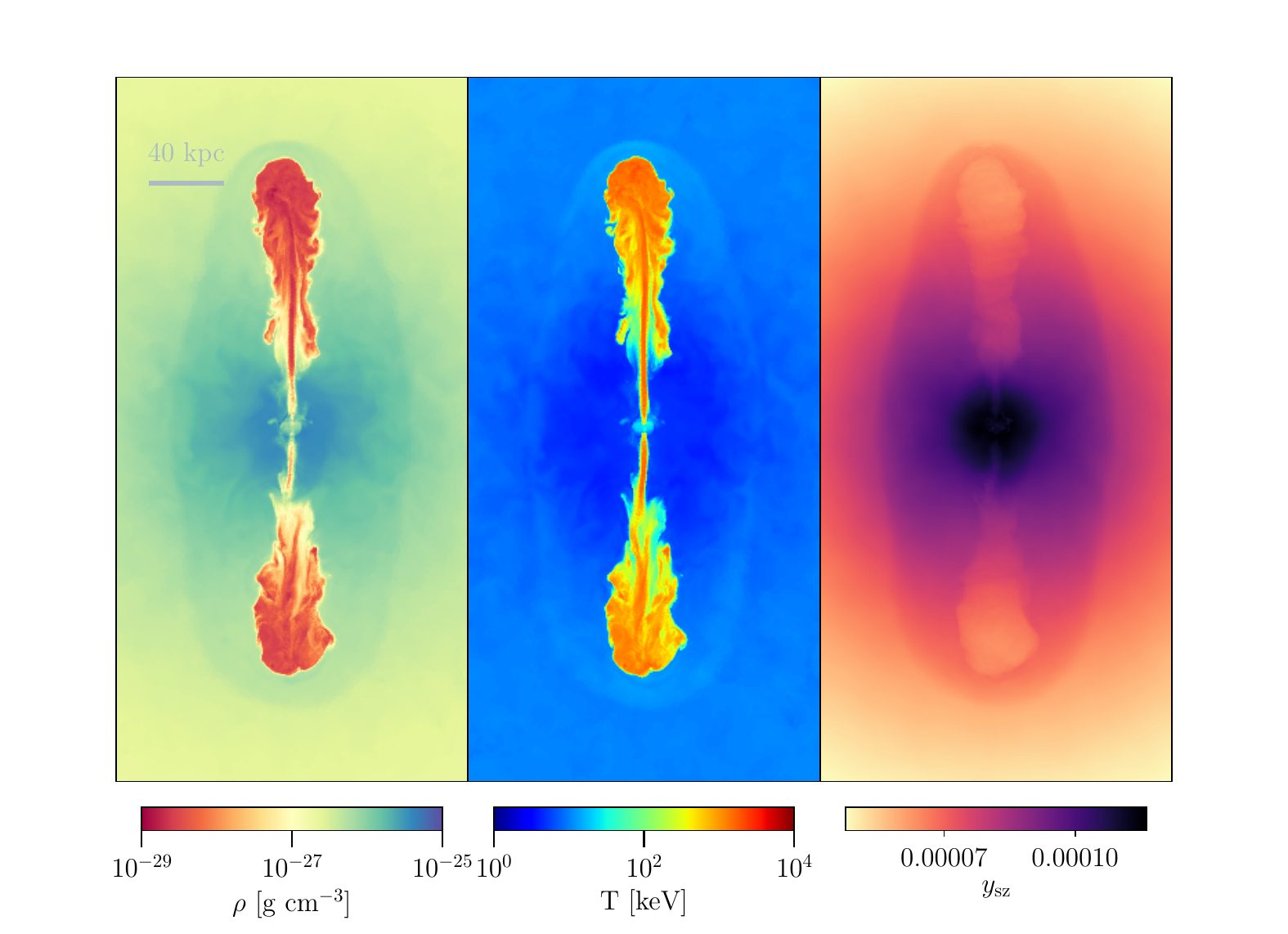}
\end{center}
\caption{Magneto-hydrodynamical simulations of outbursts from an AGN driving a pair of cavities into the ICM.  As discussed in \cite{Ehlert2018}, the simulations include cosmic ray transport.  
The tSZ decrement is suppressed at the bubbles by the ntSZ signal, and enhanced where the outbursts are driving shocks.
Density (left), temperature (center), and Compton $y$ signature from the gas. Figure from \cite{Ehlert2018b}.}
\label{fig:sz_agn_sim}
\end{figure}
%%%%%%%%%%%%%%%

AGN outbursts are the dominant feedback mechanism preventing runaway cooling in galaxy clusters \citep{Gaspari2011,Gaspari2012,Li2015,Yang2019}, and should produce different tSZ and ntSZ effect signatures, both due to the shocks driven by the outbursts and the different possible compositions of the plasma filling the bubbles.
Within cool-core galaxy clusters, AGN outbursts with powers up to $10^{46}$ erg\,s$^{-1}$ and enthalpies of $10^{58}-10^{62}$ erg are observed to produce giant cavities \citep[see e.g.][for reviews]{Blanton2010,McNamara2012,fabian2012}. These outburst driven cavities are seen as low-surface brightness patches in X-ray observations, which are filled with radio emission at low frequencies ($\lesssim$\,10 GHz). 

\begin{figure*}
\begin{center}
\includegraphics[width=0.465\textwidth]{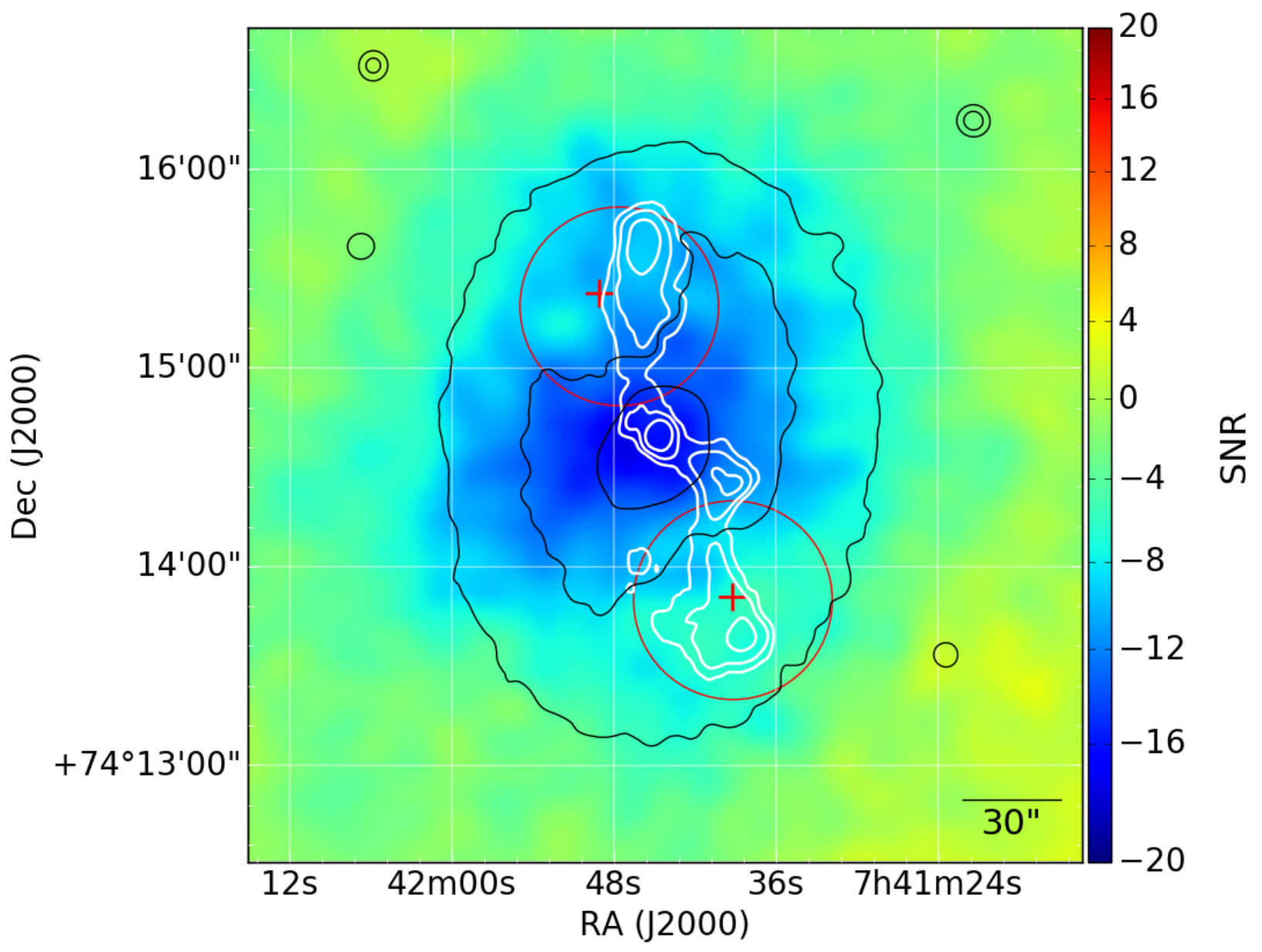}%
\includegraphics[width=0.52\textwidth]{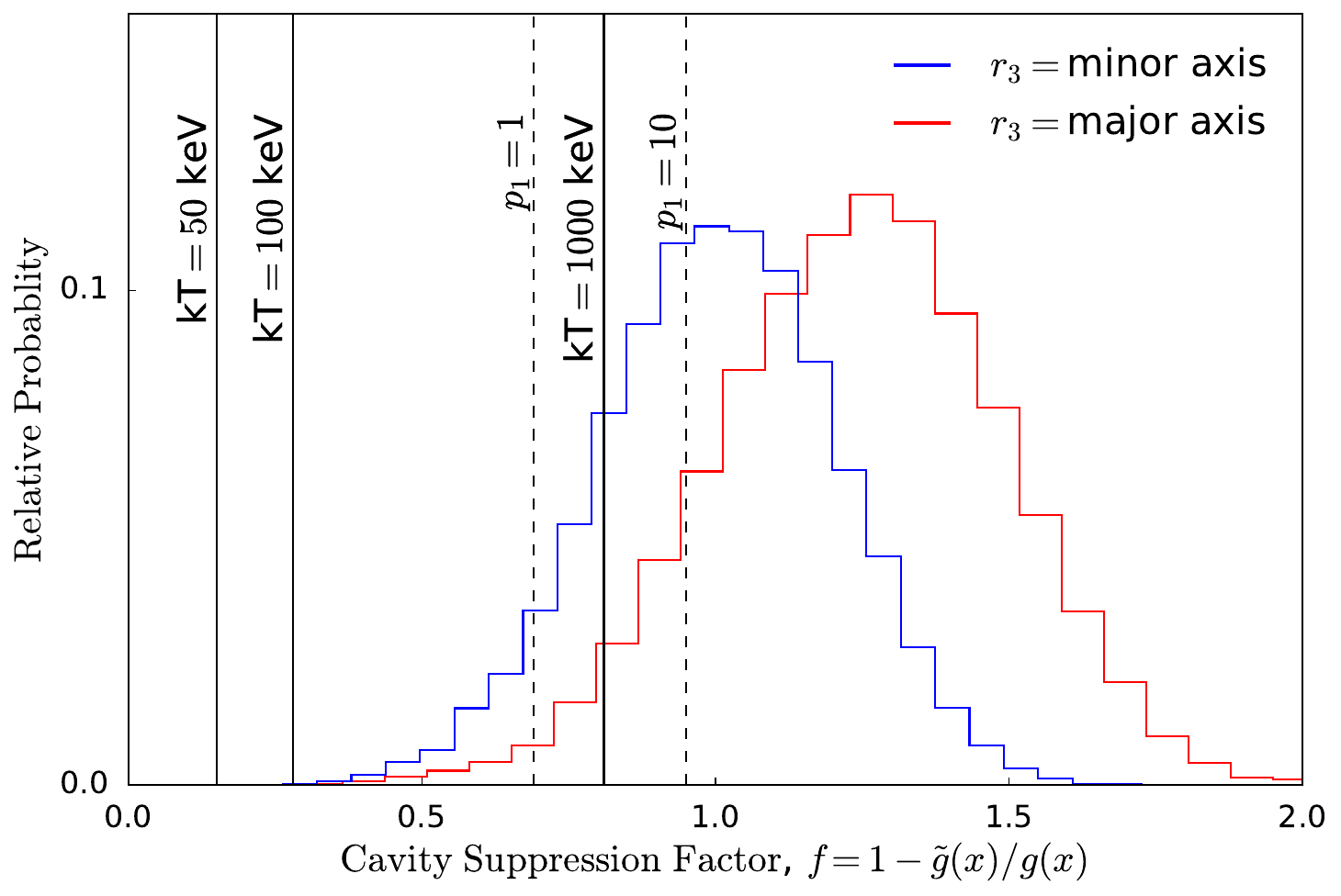}
\end{center}
\caption{\textit{Left:} CARMA 30 GHz map of the bubble cluster MS0735.6+7421 ($z=0.21$). The colour image shows the CARMA tSZ map in signal-to-noise units, white contours are VLA 327 MHz data from synchrotron emission, and the black contours show {\it Chandra} X-ray image ($0.5-7$ keV).
\textit{Right:} Posterior probability distributions of the cavity suppression factor (see text) from the analysis of CARMA visibility data. The two coloured histograms denote two different orientations for the elliptical cluster model, and the temperature and minimum electron momentum values denote the limits on cavity suppression from thermal and non-thermal SZ effects, respectively. Figures from \citet{Abdulla2018}.}
\label{fig:ms0735}
\end{figure*}

In Figure \ref{fig:sz_agn_sim}, we show the results of magneto-hydrodynamical simulations, reported in \citet{Ehlert2018b}, of AGN outbursts including cosmic ray support.
These cavities drive shocks, and are thought to be supported by magnetic fields, ultra-hot thermal gas, and cosmic rays, although the relative proportions of these sources of pressure are still highly unknown.  
By exploiting the unique spectral signatures of various SZ effects discussed in \S~\ref{sec:sz_theory}, frequency-resolved SZ observations have the unique potential to constrain the nature of the pressure support in cavities \citep[see e.g.,][]{Pfrommer2005,Prokhorov2012}.
For example, as discussed in \cite{Pfrommer2005}:
\begin{enumerate}
\item If magnetic fields provide significant support to the cavity, it should be close to equipartition with a relativistic pressure component, and the SZ decrement will have $< 1/8$ the strength of standard nonrelativistic thermal gas. 
\item If very hot ($\gtrsim 50$~keV) gas provides most of the pressure, the strength of the tSZ effect decrement will be suppressed by more than 25\% due to relativistic corrections at fixed $y$-parameter.
\item  If non-thermal relativistic cosmic ray protons or electrons dominate the pressure support, the tSZ effect signal from a bubble will be negligible (see Figure \ref{fig:ms0735}).
\end{enumerate}
In most such systems, the simplest assumption of equipartition between relativistic and magnetic support is unlikely, as the internal pressure falls far short of the work required to carve out the cavities by more than an order of magnitude.

%%%%%%%%%%%%%%%
%%%%%%%%%%%%%%%

Recently, \citet{Abdulla2018} reported on a 30~GHz observation of the bubbles in MS0735.6+7421, the highest energy pair of bubbles currently known \citep{mcnamara2007, Vantyghem2014}, using the 23-element CARMA interferometer (see Figure \ref{fig:ms0735}).  Using a model based on the X-ray results in \cite{Vantyghem2014}, they find the joint tSZ signature from the bubble pair is consistent with zero at $\sim5\sigma$. 
The right panel in Figure~\ref{fig:ms0735} shows this result quantitatively. The lack of tSZ signal in the cavities is modelled in terms of a ``cavity suppression factor'', which at a value of unity will imply there is minimal tSZ contribution and the cavities are pressure supported by either non-thermal relativistic electrons, thermal electrons from a very hot gas ($\Te \gtrsim 1000$~keV), or magnetic fields. This cavity suppression factor is obtained directly from an MCMC analysis of the CARMA visibility data, assuming a specific cluster geometry, resulting in the two histograms shown in the figure for two different line-of-sight orientations. The level of cavity suppression that can be achieved by a thermal plasma in pressure equilibrium with the surroundings and having a specific temperature, or by a non-thermal electron distribution with certain minimum momentum (here $p=\beta\gamma$, where $\beta=\varv/c$ and $\gamma=1/\sqrt{1-\beta^2}=E_{\rm e}/\me c^2$), are marked by the different solid and dashed lines, respectively.

Tentatively, the \citet{Abdulla2018} result can be considered as an argument for non-thermal pressure support within AGN cavities.  This is because it is difficult to make the case for a $\gtrsim 1000$~keV thermal plasma filling the cavities since it is well beyond the pair-production energy of electrons. In other words, the CARMA observations have provided indirect evidence for cosmic ray transport by AGN bubbles. 
The streaming of cosmic ray protons from a central AGN into the dense ICM cores has been theorised to be a gentle source of heating that can solve the cooling flow problem (e.g., \citealt{Guo-Oh2008}, \citealt{Fujita2012}, \citealt{Pfrommer2013}, \citealt{Wiener2013}). The transport of cosmic rays by AGN bubbles can aid this streaming process as the bubbles rise through the ambient medium at the sound speed, overcoming the limitation posed by Alfv\`en waves (\citealt{Ruszkowski2017}, \citealt{Ehlert2018}).

Clearly, data from single-frequency observations cannot be used to fully model the non-thermal SZ spectrum (see \S \ref{sec:ntSZ_detail}), as more frequencies are necessary to separate the tSZ, kSZ, and rSZ effect contributions. Future observations combining multi-frequency data with few arcsecond angular resolution can be expected to provide this much sought-after proof of cosmic ray heating in the ICM.

\subsection{Large-scale impact of AGN feedback with stacked SZ measurements}
\label{sec:agn_feedback_stacked}

The SZ effect has the potential to constrain the large-scale impact of feedback from AGN. As introduced in \S \ref{sec:agn_feedback_physics}, feedback processes are thought to play a key role in suppressing the growth of new star-forming disks around isolated massive galaxies \citep[e.g.,][]{scannapieco2004,granato2004,croton2006,thacker2006,dimatteo2008} and preventing the formation of large amounts of $\leq 1$ keV gas in cool-core galaxy clusters \citep[e.g.,][]{McNamara2012,fabian2012}. 

%%%%%%%%%%%%%%%
\begin{figure}
\begin{center}
\includegraphics[trim=0.6in 0.6in 0.6in 0.2in, width=0.8\textwidth]{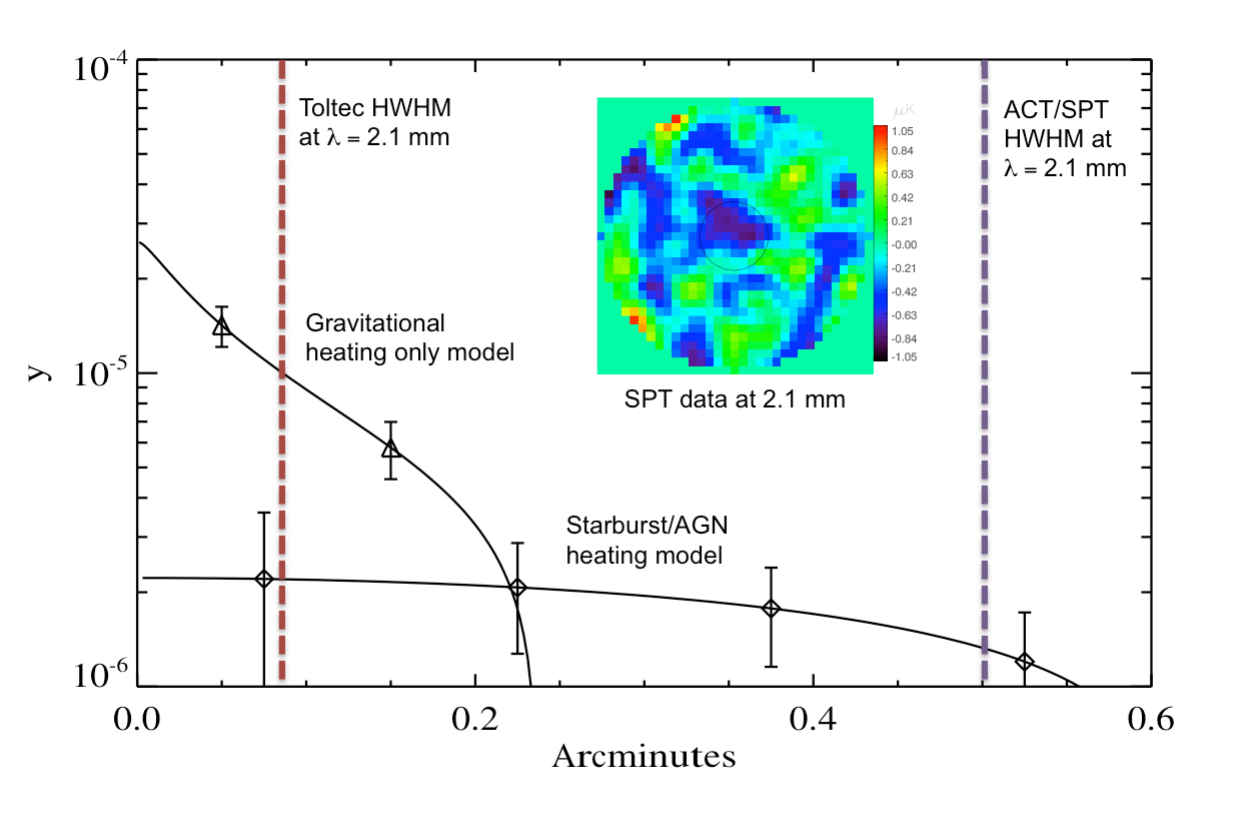}
\end{center}
\caption{
{\it Main:} Simulated Compton $y$ vs. angular radius for a $10^{12} \Msol$ elliptical galaxy at $z=1$. The peaked inner curve corresponds to a model where the halo gas is gravitationally-heated ,and has an NFW profile consistent with gas at the virial temperature.  The gas is in hydrostatic balance in a host halo with a total mass $5\times10^{13}\Msol$ \citep[see][]{Marconi2003}.
 The outer component corresponds to a model in which the halo gas is heated by energy injection at the centre, and has a Sedov profile \citep[i.e. a Sedov-Taylor blast wave; see e.g.][]{Sedov1946,Sedov1959}. 
 The Sedov profile has an energy input of $3\times10^{61}$~ergs,  which is consistent with typical theoretical feedback estimates for such galaxies \citep[e.g.][]{scannapieco2004}. The model is observed at a time of 1~Gyr after the initial energy input.
 The errors are the forecasted noise when combining 2000 galaxies in the TolTEC (see \S \ref{sec:toltec}) wide field survey. The dashed vertical lines correspond to the HWHM beam widths of TolTEC and SPT at 2.1~mm. 
{\it Inset:} Co-added 2.1~mm image of 3394 massive, quiescent $0.5<z<1$ galaxies from SPT at 150~GHz. 
Although the angular resolution of SPT is $\sim$1\arcmin\ and the depth at  2.1~mm is approximately 17 $\mu$K-arcmin, co-adding nevertheless allows for a $2\sigma$ measurement of the SZ effect from the CGM around galaxies outside of clusters.
The image is $8\arcmin\times8\arcmin$, and the black circle represents a 1\arcmin\ radial aperture. Main figure from Scannapieco \& Mauskopf in prep. Inset image from \citealt{spacek2016}.
}
\label{fig:sz_agn_feedback}
\end{figure}
%%%%%%%%%%%%%%%

In many theoretical models, energetic AGN outflows due to broad absorption-line winds or radio jets heat the circumgalactic medium and suppress the cooling needed to form further generations of stars and AGN. This suppression is redshift dependent, as the higher-redshift circumgalactic medium is more dense and rapidly radiating.  Therefore, at high redshift, a more energetic outflow driven by a larger AGN is required for quenching to be effective.  This provides an explanation for the large observed drop in the characteristic mass of star-forming galaxies since z=2 \citep[e.g.,][]{arnouts2005,treu2005,schaye2010}.
However, the details of such feedback remain extremely uncertain, and direct measurements of AGN kinetic energy input around isolated galaxies are extremely difficult \citep[e.g.,][]{dekool2001,chartas2007,borguet2013}. 

Measurements of the tSZ effect offer a method to overcome these difficulties. From Equation \eqref{eq:Comptony}, and as noted previously, we see that the tSZ signal integrated over the area of sky around any object of interest is a calorimetric measure of electron thermal energy. This means that one can potentially use the tSZ effect to measure the total thermal energy output from AGN into the ICM.
Searching for this signal was first suggested by  \cite{natarajan1999}, who roughly estimated the SZ effect from AGN and argued that it would be detectable even outside of galaxy clusters.  More detailed calculations were carried out in e.g. \cite{yamada1999},\cite{yamada2001}, and \cite{Chatterjee2007}, including estimates of the redshift dependence of the signal and its global impact on the CMB. Such global calculations were refined in \cite{scannapieco2008} and \cite{Chatterjee2008}, who used large-scale cosmological simulations to compute the impact of AGN on the CMB power spectrum, and showed that SZ measurements  would be able to constrain AGN feedback by appropriately stacking measurements around quasars and early-type galaxies.  
In Figure \ref{fig:sz_agn_feedback}, we show the results of a simple estimate of the difference in the radial SZ profile around elliptical galaxies.
The predictions show that the signal is expected to be within reach of current and near future instruments (specifically, the figure compares the predictions for TolTEC, discussed in \S \ref{sec:toltec}).

Recent work by \citet{Soergel2017} finds only low-significance evidence of the tSZ signal associated with AGN feedback, based on an analysis of {\it Planck} and AKARI all-sky data and SDSS QSO catalogue. These authors illustrate the importance of modelling the FIR emission from AGN simultaneously, and show how neglecting it may have biased previous studies.

Constraints on the widespread impact of AGN through galaxy stacking have recently been obtained by several groups.
\cite{Chatterjee2010} found a tentative detection of quasar feedback using the Sloan Digital Sky Survey (SDSS) to select galaxies for use in stacking Wilkinson Microwave Anisotropy Probe (WMAP) data, though the significance of AGN feedback in their measurements is disputed \citep{Ruan2015}. \cite{Hand2011} stacked ACT measurements around 2300 SDSS-selected luminous red galaxies and found a $\approx 3\sigma$ detection. 
\cite{Planck2013_Y} investigated the relationship between tSZ signal and stellar mass with significant results above stellar masses $\approx 10^{11} M_\odot$ \citep[see also][]{Greco2015}.
\cite{Gralla2014} stacked ACT data to make a $5\sigma$ detection of the thermal SZ signal from radio galaxies.  
\cite{Ruan2015} also stacked {\it Planck} tSZ  maps centred on the locations of 26,686 SDSS quasars and estimated the mean thermal energies in gas surrounding them to be $\geq 10^{62}$ ergs \cite[for a counterexample, see][]{Cen2015}.  
Similarly, \cite{Crichton2016} stacked  $>$ 17,000 radio-quiet quasars from the SDSS in ACT data and found $3\sigma$ evidence for the presence of associated thermalised gas.

The inset panel of Figure \ref{fig:sz_agn_feedback} illustrates the results of stacking 150 GHz data around 3394 large passive galaxies with an average stellar mass of $1.5 \times 10^{11} M_\odot$ located outside of large galaxy clusters in the data publicly released by the SPT collaboration \citep{spacek2016}. Note that at 150 GHz the SZ signal appears as a decrement, and this distinguishes it from all emission processes that can contaminate the signal.  The signal was detected with a noise level of 0.7$\mu$K at 150 GHz. These measurements, along with similar stacked measurements from ACT \citep{spacek2017}, are sensitive enough to place constraints on models of non-gravitational heating \citep{spacek2018}. 

%%%%%%%%%%%%%%%%%%%%%%%%%%%%%%%%%%%%%%%%%%%%%%%%%%%%%%%%%%%%%%%%%%

\subsection{Probing the missing baryons with stacked measurements of the SZ effect}
\label{sec:missingbaryons2}

%%%%%%%%%%%%%%%
\begin{figure*}
\begin{center}
\includegraphics[width=\textwidth]{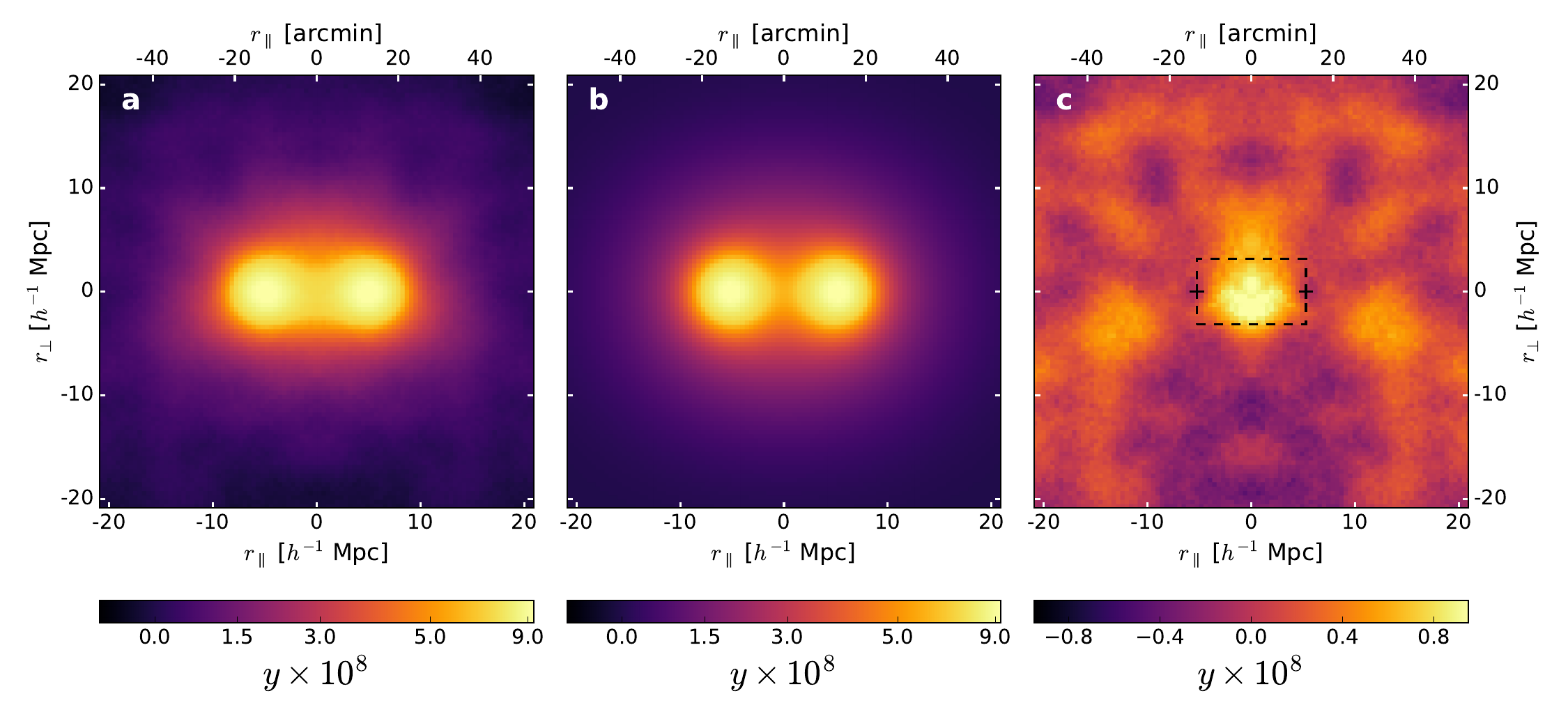}
\end{center}
\caption{
{\it Left:} Stacked Compton-$y$ signal from $\sim 10^6$ galaxy pairs, symmetrised about the x-axis.
{\it Middle:} Galaxy Compton-$y$ model fit to the off-axis regions of the stacked signal.
{\it Right:} Residual image after subtraction of the model, showing a $\sim 5-\sigma$ excess where cosmic web filamentary structure is expected.  
The asymmetry in the y direction in the residual image is low significance and likely due to reprojection of the {\it Planck} HEALPIX maps into Cartesian coordinates.
Analysis details are presented in \citet{deGraaff2017}.  Figure from \citet{deGraaff2017}. 
}
\label{fig:whim}
\end{figure*}
%%%%%%%%%%%%%%%

A cosmic web of filamentary structures connecting clusters, groups, and galaxies is an essential prediction of nearly all large-volume cosmological simulations, and is expected to be found at temperatures of $10^5-10^7$~K \citep[][]{Cen1999,shull12}. The emission is thus expected to peak at energies generally too soft to be observed in the X-ray, and the associated densities indicate these regions should have a very low X-ray surface brightness \citep{Bregman2009}.  At the same time, this gas is too warm to be easily detected through absorption along the line of sight to AGN \citep{Nicastro2008}, although most searches so far have focused on this approach. 

Another method is to cross-correlate measurements of the tSZ and kSZ effect with other tracers of large-scale structure such as lensing maps \citep{VanWaerbeke2014,Battaglia2015b,Atrio2017}, galaxy surveys
\citep{Hernandez2015}, or the dispersion measures of fast radio bursts \citep{Munoz2018}. Stacked measurements not only improve the signal to noise by increasing the effective integration time, but can also reduce certain forms of astrophysical contamination, as discussed in \S \ref{sec:contam}. 

Recent stacking analyses of \textit{Planck} data have provided the first putative direct detection of this material using the the tSZ effect \citep{deGraaff2017,Tanimura2019}. In both works, large catalogues ($\sim 10^6$) of galaxies were used to identify pairs of galaxies separated by $\sim 10$~Mpc on the sky. These pairs were then scaled and stacked symmetrically, and models for the galactic SZ signatures were fit to the signal perpendicular to the filament direction and subtracted as shown in Figure \ref{fig:whim}. 
The results of both analyses showed a $\sim 5\sigma$ residual between the pairs.  If a temperature $\Te \sim 10^6$ K is assumed, consistent with simulations, a significant fraction of the `missing' baryons are recovered when accounting for the filling fraction of this cosmic web. We caution that the temperatures in simulations range by 2 orders of magnitude, however, and that the result for the baryon fraction recovered is linearly sensitive to the assumed temperature. A similar stacking analysis was also used in \cite{Tanimura2018} to reveal filamentary structures in supercluster environments.

The {\it Planck} all-sky Compton $y$ map has a 10\arcmin\ angular resolution, which dilutes the signal over a larger area than it likely subtends.  Future higher resolution SZ/CMB observations will suffer less from beam dilution, improving these constraints. Furthermore, this probe is expected to yield more significant results from stacking and cross-correlation studies, as well as through the analysis of the tSZ power spectrum at low multipoles ($\ell < 100$). By stacking in different mass-bins, one may furthermore be able to calibrate the temperature-mass relation using the rSZ \citep[e.g.,][]{Erler2018}, thereby providing another handle on the hydrostatic mass bias.

%%%%%%%%%%%%%%%%%%%%%%%%%%%%%%%%%%%%%%%%%%%%%%%%%%%%%%%%%%%%%%%%%%

\section{Observational considerations}
\label{sec:obscons}

The SZ effect is an attractive probe of the hot ICM for three main reasons: it has redshift independent surface brightness, observations can be performed from the ground, and the volumetrically-integrated tSZ signature is proportional to thermal energy, making it a useful probe for ICM calorimetry or detection of a cluster with an observable that tracks mass well.  
With the sensitive new instruments now online or planned for the near future (see \S \ref{sec:instrumentation}), the integration times often rival those of X-ray observations, particularly for high-$z$ systems and in the case where X-ray spectroscopy is necessary to perform calorimetric determinations of the cluster gas.
However, there are several practical matters to take into account when studying cluster astrophysics using SZ observations.  
Here we briefly discuss interferometric and single-dish photometric observations, and contamination and confusion from mm/submm astrophysical sources.

\subsection{Interferometric measurements}
\label{sec:interferometric}

As noted in the introduction, interferometric arrays have a long history of use in detecting and imaging the tSZ effect. In contrast to single dish telescopes, interferometers have the appeal that they provide high resolution measurements without the requirement of building large aperture elements (see \S \ref{sec:photometric}), and that the uncorrelated atmospheric signal is naturally filtered from the observation.   This comes with the expense that the interferometric array behaves as a frequency-dependent spatial filter, and rarely attains the mapping speed of a single dish.

The point source sensitivity of an interferometer with $M$ identical array elements scales as \citep[see e.g.][]{tms,CondonRansom2016}:
\begin{equation}
\label{eq:interferometric_sensitivity}
\sigma_{\rm rms} = \frac{2 \kB T_{\rm sys}}{A_{\rm e} \sqrt{N_{\rm bl} t \Delta \nu}},
\end{equation}
where $N_{\rm bl} = M (M-1)/2$ is the number of baselines in the observation, $A_{\rm e}$ is the effective area of the array elements (i.e.,\ optical efficiency $\eta$ times collecting area $A_{\rm c}$; $A_{\rm e} = \eta A_{\rm c}$), $T_{\rm sys}$ is the total system noise temperature including atmospheric and instrumental contributions, $\Delta \nu$ is the bandwidth, and $t$ is the integration time.  Thus large gains for continuum signals like the SZ effect can be made by increasing the number of interferometric array elements, reducing the instrumental noise, and increasing the instrument bandwidth.

The most fundamental property of an interferometric observation of the SZ effect (or any extended source) is that it samples the Fourier transform of a portion of the sky multiplied by the response pattern of the antennas \citep[see e.g.][]{tms,CondonRansom2016}.  
For an array with homogeneous elements, this response pattern is simply a Gaussian with size equal to the diffraction limit of each antenna. Thus, for an interferometer, the primary beam determines the usable FoV of an observation, rather than the angular resolution of the measurement.  This, in turn, is determined by the sampling in `$uv$-space' (so-called for the Fourier space coordinates $u$ and $v$).
The average response pattern is called the synthesised beam (often approximated as an elliptical Gaussian), which depends both on the $uv$-coverage and the weighting scheme used for creating an image from the visibilities. It is therefore cleanest to model the data in $uv$-space, as discussed for the interferometric data presented in, e.g., \S~\ref{sec:profiles} and \S~\ref{sec:shocks},

\begin{figure}
\begin{center}
\includegraphics[width=1.00\textwidth]{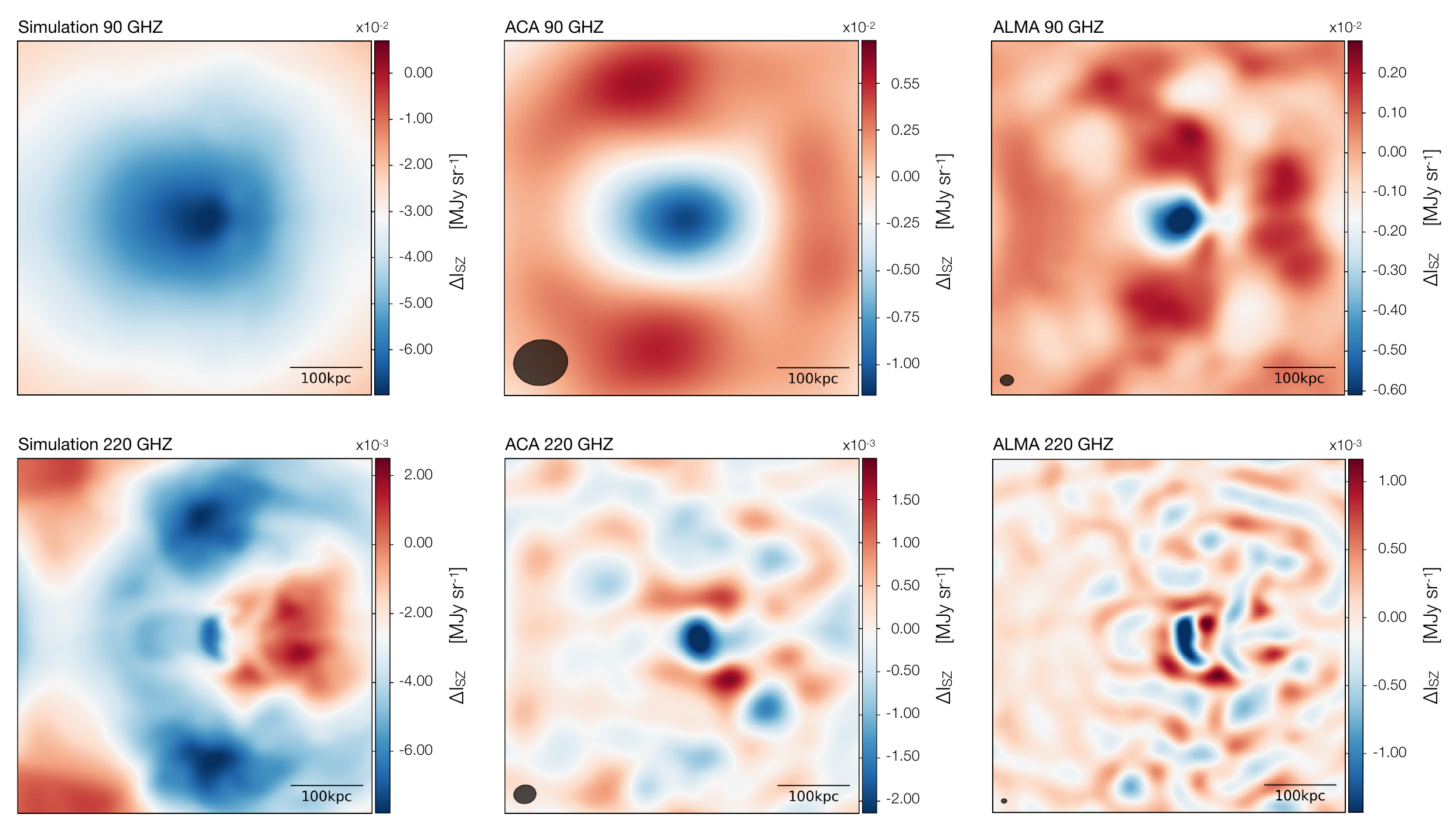}
\end{center}
\caption{Mock ACA and ALMA observations at 90 and 220 GHz of the simulated cluster of Figure~\ref{fig:sz_vs_xray}. The maps were generated applying a natural weighting scheme to the raw visibilities and without correcting for the incomplete sampling of the $uv$-space through deconvolution. The resulting synthesised beams for each of the observations are also reported.
The top row shows the 90 GHz input map, a mock ACA observation, and a mock ALMA observation from left to right.  The bottom row is the same, but for 220 GHz. 
The simulated observations are noise-free in order to highlight solely the filtering effects of the interferometer.}
\label{fig:simalma}
\end{figure}

In Figure \ref{fig:simalma}, we show idealised, noiseless mock ACA and ALMA Band 3 and 6 observations (centred at 90 and 220 GHz respectively).
These show the ability of ALMA/ACA to probe SZ structure at high resolution, but at the expense of having no sensitivity to scales larger than the full width half maximum width of the primary beam.  For ALMA, this is $56\arcsec/(\nu/90~\rm GHz)$, much smaller than the typical extents of a massive galaxy cluster at any redshift.  We discuss the ALMA/ACA interferometers in detail in \S \ref{sec:alma_B3}.

\subsection{Photometric measurements using direct-detection instruments}
\label{sec:photometric}

While many pioneering SZ studies were performed using interferometers, there has been a nearly universal shift towards the use of photometric imaging arrays for SZ observations, particularly in the case of wide-field SZ surveys where large mapping speeds are required. Currently, state of the art instruments for targeted observations contain a few thousand individual detectors covering 1--3 separate broadband ($\Delta \nu/\nu \sim 20$\%) spectral channels placed within the atmospheric windows centred at roughly 90, 150, and 260~GHz. Coupled to large aperture telescopes (see e.g.\ \S~\ref{sec:mustang2}, \S~\ref{sec:nika2}, and \S~\ref{sec:toltec}), current and near-future instruments obtain $\sim 10\arcsec$ angular resolution over a $\sim 5\arcmin$ FoV.  As angular resolution $\theta$ at wavelength $\lambda$ scales as the inverse of the primary mirror diameter $D$ (see \S \ref{app:beam} for details), achieving much higher resolution than this requires either an interferometer (\S \ref{sec:interferometric}) or a much larger ($>100$~m) single-dish large-aperture telescope.

To reasonable approximation, most modern photometric arrays are limited by photon statistics (i.e., the intrinsic detector and instrument noise is negligible), and the point source sensitivity scales as (e.g.\ \citealt{Richards1994,Bryan2018}):
\begin{equation}
    \sigma_{\rm rms} = \frac{\sqrt{4 \eta \, \kB T_{\rm load} \, h \nu_0 + 4 (\eta \, \kB T_{\rm load})^2}}{\eta A_{\rm c} \! \sqrt{q N_{\rm det} t \Delta \nu}},
\label{eq:photometric_sensitivity}
\end{equation}
where $\eta$ is the optical efficiency, $T_{\rm load}$ is the effective temperature of the total photon loading on the detector, $\nu_0$ is the observing band-centre, $q$ is the number of polarisations to which the detector is sensitive, $A_{\rm c}$ is the collecting area of the single-dish telescope, and $N_{\rm det}$ is the number of detectors.
The first term in the numerator represents photon shot noise (random arrival of photons), while the second represents photon Bose noise (photon bunching), and the two terms tend to be approximately equal for ground-based receivers at relevant SZ frequencies.  Comparing Equations \eqref{eq:interferometric_sensitivity} and \eqref{eq:photometric_sensitivity}, it is clear the noise improves with detector count as $\sqrt{N_{\rm det}}$ for photometric arrays, while the noise improves with baseline number as $\sqrt{N_{\rm bl}}$ for interferometers.

At the observing frequencies where photometric arrays operate, spatial and temporal variations of the column depth of water vapour in the atmosphere produce a signal with a peak-to-peak amplitude of $\lesssim$1~K over long integrations, many orders of magnitude brighter than the SZ effect. To reasonable approximation, these variations occur within a single layer in the atmosphere with a fixed spatial pattern that moves across the telescope's FoV at an angular rate of $\sim 30\arcmin$/sec (e.g., \citealt{Lay2000, Sayers2010}). The power spectrum of these variations rises sharply at large angular scales, and tends to dominate the detector data time-streams at frequencies below $\sim 1$~Hz for telescope scan rates slower than or comparable to the atmospheric angular drift speed. Therefore, faster telescope scanning, at least up to the point when it is much faster than the atmospheric drift, mitigates the atmospheric contamination by modulating the astronomical signal to higher frequencies in the data time-streams above the dominant atmospheric signal at $\lesssim 1$~Hz. To achieve this fast scanning while keeping the cluster within the instrument's FoV, the telescope is often moved in a Lissajous `daisy' pattern \citep[e.g.][]{Korngut2011} centred on the cluster, although linear raster scans are still used in some cases \citep[e.g.][]{Adam2018}.

For a single-dish large-aperture telescope, all of the detectors in the focal plane look through a nearly identical column of atmosphere, and so the atmospheric signal at any given instant in time is approximately common in all of the detectors. As a result, this unwanted signal is generally subtracted via a common-mode template(s), usually in combination with a high-pass filter applied to the data time-streams. The largest astronomical scales recoverable after this common-mode removal are thus comparable to the instrument's FoV.  An example of this is given in the discussion of MUSTANG-2 (\S \ref{sec:mustang2}). 

Looking forward, advancements in detector technologies, particularly kinetic inductance detectors (KIDs), hold the promise of far larger detector counts when coupled to high-throughput telescopes. Such advancements would allow for broader spectral coverage (e.g., 6 bands between 90--400~GHz) and significantly larger FoV ($\gtrsim 1^\circ$). While there are prospects for wide FoV telescopes with improved mapping speeds and ability to recovered larger angular scales  (see \S \ref{sec:atlast}), there is no immediate prospect for the larger telescope apertures ($>100$ meters) required for substantial gains in angular resolution. For that, an interferometer is required (\S \ref{sec:interferometric}).

\subsection{Astrophysical contamination}
\label{sec:contam}

As the sensitivity of SZ effect measurements has improved, it has become increasingly important to mitigate contamination from unwanted astrophysical sources coincident on the sky with the cluster. 

\vspace{2mm} \noindent {\it Radio sources:} Early SZ effect observations, mainly at $\le 30$~GHz, often included non-negligible synchrotron emission from field and cluster-member galaxies, particularly the BCG (e.g., \citealt{Cooray1998, Coble2007}). While synchrotron emission is much less problematic at the higher frequencies typical of modern SZ observations ($\ge 90$~GHz), it can still pose a challenge to higher resolution measurements (e.g., \citealt{Mason2010,Romero2017}). However, since the luminosity function of synchrotron sources is relatively flat (e.g., \citealt{deZotti2010}), the additional contamination in future, deeper SZ effect observations is expected to be modest. Furthermore, the typical synchrotron spectrum, which decreases as a power-law with frequency, is readily separable from both the tSZ and kSZ effect spectra and their relativistic corrections.  Ultimately, the level of contamination must be assessed as new deeper observations are acquired.

\vspace{2mm} \noindent {\it Primary CMB Anisotropies:} Another astrophysical contaminant comes from the primary anisotropies in the CMB, which become significant ($\gtrsim 1\mu$K) at angular scales larger than $\gtrsim 5\arcmin$ \citep{Planck2016_XI}. This is well inside the virial diameter of most clusters, and these CMB anisotropies can therefore impact measurements of large-scale cluster features in SZ images (e.g., \citealt{Plagge2010}). Fortunately, the CMB spectrum can easily be distinguished from the tSZ effect spectrum, and multi-band observations can effectively separate the two signals (e.g., \citealt{Planck2016_XXII}). The kSZ effect, however, has an identical spectrum to the CMB, and so the only way to separate the two signals is via their spatial templates. At small angular scales where the CMB is relatively smooth, for example when measuring the kinematic SZ signal towards sub-structures in the ICM, this spatial separation can work quite well (e.g., \citealt{Sayers2013,Romero2017,Adam2017ksz,Ruppin2018}). However, separating the full-cluster-scale kSZ effect signal from the CMB is in general significantly more difficult (e.g., \citealt{Planck2014_XIII, Planck2018_LIII}).

\vspace{2mm} \noindent {\it Dusty Galaxies:} Looking towards future SZ effect measurements, thermal dust emission from background galaxies may be the most problematic astrophysical contaminant. Unlike the luminosity function for synchrotron emission, the luminosity function for the thermal dust emission is quite steep (e.g., \citealt{Glenn2010}), and so the relative amount of contamination grows quickly with deeper SZ effect observations. The average dust temperature in these galaxies is $\sim 40$~K (e.g., \citealt{Chapman2010}), and their redshift distribution peaks near $z \sim 1$--3 (e.g., \citealt{Smith2017}), resulting in a typical observed spectrum with an effective temperature of $\sim 10$--20~K. While such a spectrum is highly degenerate with the shape of both the tSZ and kSZ signal at lower frequencies ($\nu \lesssim 100$~GHz), it can be distinguished with coverage extending to higher frequencies (e.g., \citealt{Sayers2013, Erler2018}). Furthermore, because of the typical sub-arcsecond angular size of these galaxies, they will be unresolved in nearly all SZ effect observations, potentially facilitating spatial separation from any resolved SZ effect features. 

While in principal the spatial and spectral properties of these dusty galaxies make them highly distinguishable from the SZ signal, there are many practical challenges to their subtraction from SZ images. Their steep luminosity function results in an effective noise floor due to source confusion that is only a modest function of angular resolution for $\textrm{FWHMs} \gtrsim 5\arcsec$ (e.g., \citealt{Bethermin2011}). Therefore, even relatively high resolution SZ images are likely to contain a non-negligible signal from multiple dusty galaxies per resolution element. Even in the simplest scenario of a single dust temperature per galaxy, this will result in a complex multi-temperature dust spectrum within each resolution element (e.g., \citealt{Roseboom2010}), and it is unclear how well such a spectrum can be modelled even with broad spectral coverage. Lensing of this galaxy population by the cluster can add complicated spatial variations that present an additional challenge (e.g., \citealt{Zemcov2013,Sayers2018}). Further study on this topic is required, but SZ sensitivities significantly better than the confusion limit from dusty galaxies may be extremely difficult to achieve. One potential solution to this problem is ancillary observations from a very high resolution facility such as ALMA, although such observations would be prohibitively expensive for large cluster samples.

\vspace{2mm} \noindent {\it Galactic Diffuse Emission:} Mainly for clusters located at low Galactic latitudes, the diffuse signal from within our galaxy due to synchrotron, free-free, spinning dust, and thermal dust emission can present another possible contaminant \citep{Planck2016_XXV, Planck2016_XLVIII}. However, even when these signals are non-negligible, they tend to be relatively smooth on the angular scales relevant for clusters, and can be effectively removed by tools such as matched filtering. As a result, contamination from Galactic emission is unlikely to be a major impediment to upcoming SZ effect studies.

\vspace{2mm} \noindent {\it SZ confusion:} Since the SZ effect signals contain the contributions from all free electrons along the line of sight, with effectively no redshift dependence, the SZ effects produced by the intervening structures (i.e., ionised gas in and around galaxies, groups, clusters, and cosmic web of filaments) in the background and foreground of the cluster will contaminate the SZ signal of the cluster. This SZ confusion is small for massive clusters, but it could become significant for low-mass clusters and groups (\citealt{White2002}; see also \citealt{voit2001} for the confusion in the X-ray band). Furthermore, the SZ contribution from filaments in the cosmic web is expected at the level of 8\%-10\% \citep{Hallman2007}. The SZ confusion can be a significant issue for stacking and cross-correlation analyses as well. 

%%%%%%%%%%%%%%%%%%%%%%%%%%%%%%%%%%%%%%%%%%%%%%%%%%%%%%%%%%%%%%%%%%

\section{Current and future instrumentation}
\label{sec:instrumentation}

The section provides an overview of current-generation instruments (e.g.,\ MUSTANG-2, NIKA2, ALMA+ACA), from which we provided a science highlights already, and presents a few key aspects of planned and proposed future instruments.  
Here we caution the reader when comparing measured mapping speeds of current instruments with forecasts for background-limited performance with future instruments.  The background-limited mapping speeds have historically been a factor of several times too optimistic in some cases. 

\subsection{Current instrumentation}

In this section we summarise the current generation of  instrumentation that have measured the tSZ decrement at subarcminute resolution.  Table \ref{tab:real_mapping_speeds} lists the salient aspects of each.

\subsubsection{ALMA/ACA}
\label{sec:alma_B3}
The Atacama Large Millimeter/Submillimeter Array (ALMA) is an observatory consisting of 3 arrays: fifty 12-meter antennas (usually called ``ALMA''), twelve 7-meter antennas known as the Morita / Atacama Compact Array (ACA), and four 12-meter antennas known as the `Total Power Array' (TPA), which share the same design as the 12-meter elements of ALMA.  

The best imaging fidelity comes from the 12-meter ALMA main array due to the large collecting area, which achieves low noise, and the superb sampling of the sky in $uv$-space made possible by having up to 1225 baselines.  However, the 12-meter array has the smallest primary beam (and hence FoV, see \S \ref{sec:interferometric}), and only recovers scales $\sim 0.5^\prime$ at its lowest current frequency (see right hand panels of Figure \ref{fig:simalma}).  

The ACA observes in a more compact configuration and, due to its smaller 7-meter elements, has a FoV 1.7 times larger than that of the main array at the same given frequency.  As a trade off, the ACA has less collecting area, lower resolution ($\sim13\arcsec$ in Band 3), and sparser sampling of the sky in $uv$-space (up to 66 baselines), making imaging more challenging (see middle panels of Figure \ref{fig:simalma}).

And finally, the TPA, which nominally holds the promise of single dish total power measurements, is unfortunately not suitable for SZ observations as it fails to recover continuum signals like the tSZ and kSZ, and -- unlike most single dish telescopes in operation -- only samples the sky with a single beam per telescope.  This is due in large part to the small FoV of the 12-meter elements; the antenna design was not optimised for focal plane arrays.  Additionally, the overlap in Fourier modes sampled by a 12-meter is generally insufficient to provide good joint imaging, and several studies have shown a single dish should be at least 3 times the diameter of the interferometric elements with which the data will be combined \citep[see e.g.][]{Frayer2017}.  A new, large aperture ($>36$ meter), wide FoV telescope could finally provide this capability, improving dramatically the spatial dynamic range of ALMA (e.g.\ AtLAST, \S \ref{sec:atlast}).

Upgrades to ALMA in the next few years will deliver bands covering 35-51 GHz (`ALMA Band 1') and 67-90 GHz (`ALMA Band 2'), with the possibility the latter will be extended to cover 67-116 GHz.  
Both bands are firmly at frequencies best suited for probing the tSZ (see Figure \ref{fig:szspectrum}).
ALMA Band 1 will improve ALMA's ability to probe larger angular scales by a factor of 2.5.  While smaller than the instantaneous FoV bolometer arrays like MUSTANG-2 (\S \ref{sec:mustang2}), NIKA2 (\S \ref{sec:nika2}), and TolTEC (\S \ref{sec:toltec}), the larger FoV and the ability of ALMA to jointly image a target in multiple array configurations at the same frequency will further improve imaging fidelity over a broad spatial dynamic range. 
Additionally, ALMA Band 2 will deliver wider bandwidths ($\geq 16$~GHz) that anticipate upgrades to the ALMA correlator, and covers frequencies where the tSZ is a factor of $\sim4\times$ stronger than it is in Band 1.  Such capabilities will dramatically improve sensitivity to pressure substructures, shocks, and fluctuations in the tSZ.  

\subsubsection{MUSTANG-2}
\label{sec:mustang2}
The 2nd generation Multiplexed SQUID/TEC Array at Ninety Gigahertz (MUSTANG-2) is a continuum bolometer camera on the 100-meter Green Bank Telescope (GBT) in Green Bank, West Virginia (USA).  It consists of 215 feedhorn coupled polarisation insensitive detectors in the atmospheric window centred about 90 GHz (3.3 mm), useful primarily for probing the tSZ decrement. The 100-meter aperture of the GBT yields a 9\arcsec\ resolution, positioning it as one of the highest resolution bolometer cameras to probe the thermal SZ effect.
Several examples of MUSTANG-1 results are shown in, e.g, \S~\ref{sec:ICMthermodynamics} and \S~\ref{sec:ICMstructures}. The upgrade from MUSTANG-1 to MUSTANG-2, while bringing newer, more sensitive detectors, was also motivated in large part by the need to recover larger scales.  

MUSTANG-2 is now being demonstrated to recover scales comparable to its 4.2\arcmin\ FoV, while MUSTANG-1 was limited to scales similar to its 42\arcsec\ FoV (see discussion of photometric arrays in \S \ref{sec:photometric}).  

\subsubsection{NIKA2}
\label{sec:nika2}
NIKA2 \citep{Adam2018,Catalano2018} is a millimetre camera, consisting of dichroic arrays of KIDs operating at 150~mK.  It was installed at the focus of the IRAM 30-meter telescope in September 2015. NIKA2 observes the sky at 150 and 260~GHz with a wide FoV, 6.5\arcmin, a high angular resolution (17.7\arcsec\ and 11.2\arcsec\ at 150 and 260 GHz, respectively), and state-of-the-art sensitivity (8 and 30~$\rm{mJy}.s^{1/2}$, respectively). 
It features 616 polarisation-insensitive detectors at 150~GHz, and two arrays of 1140 detectors each at 260~GHz to provide polarisation measurements in this band.
The camera's performance at the IRAM 30-meter telescope is described in \citet{Adam2018,Catalano2014}.
The NIKA2 camera is well-suited for high-resolution SZ observations 
for several reasons. First, it is a dual-band camera operating at frequencies for which the thermal SZ signal is negative and slightly positive respectively. The 260-GHz map may be used for the detection for point sources or as a template of the atmospheric noise. Then, a high angular resolution and a large FoV enable NIKA2 to well match intermediate and high redshift clusters.

\subsubsection{A comparison of current instrumentation}

\begin{figure}
\begin{center}
\includegraphics[width=0.9\textwidth]{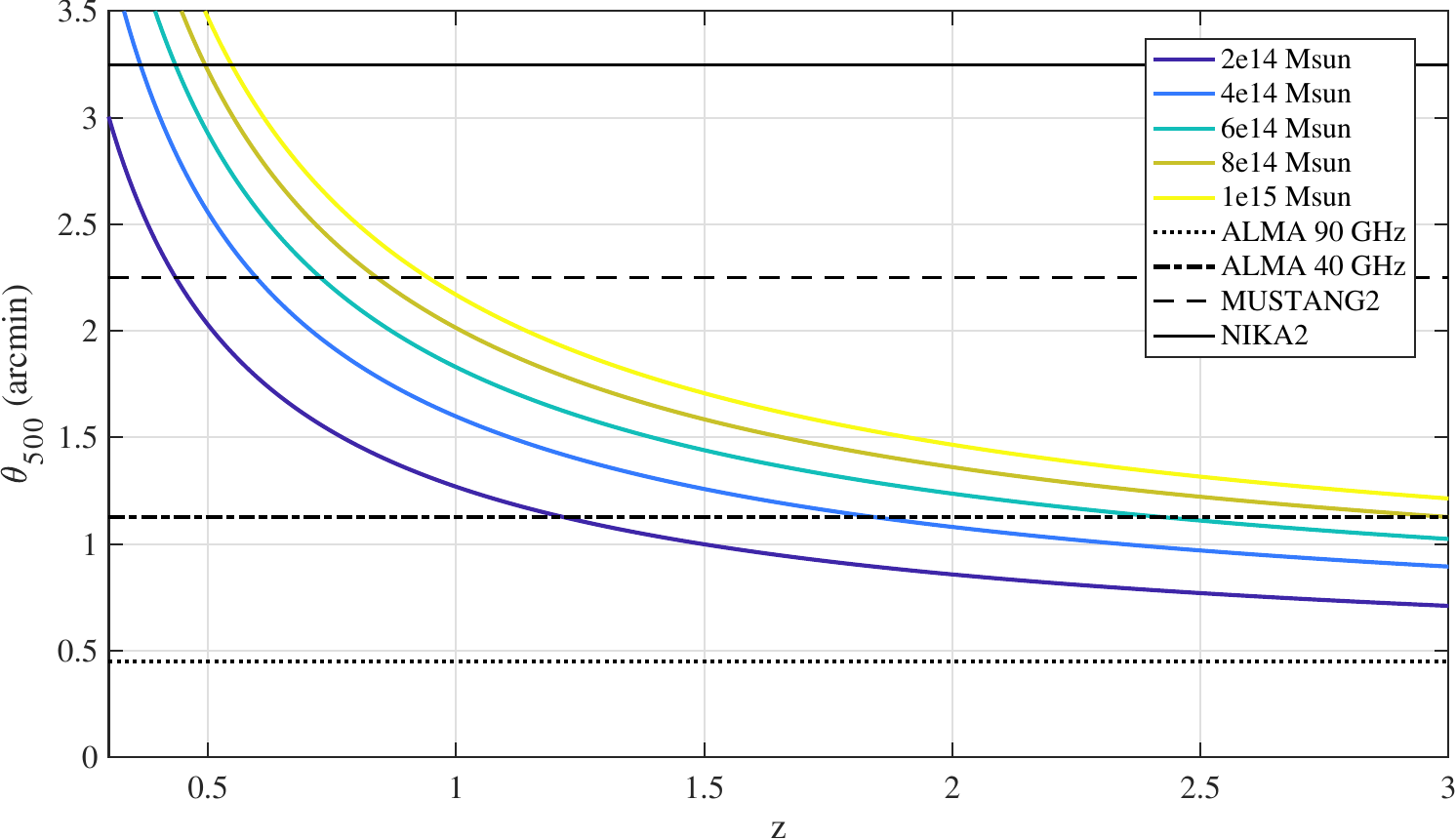}
\end{center}
\caption{The radial angular extent corresponding to $r_{500}$, $\theta_{500}$, of a galaxy cluster as a function of redshift, plotted for a range of masses.  
The radial scales accessible to a few instruments, namely the ALMA 12-meter array in Bands 1 and 3 (see \S \ref{sec:alma_B3}), MUSTANG2 on the 100-meter GBT (see \S \ref{sec:mustang2}), and NIKA2 on the IRAM 30-meter (\S \ref{sec:nika2}), are plotted for comparison. 
For nearly all massive clusters at $z \gtrsim 1$, instruments with $\approx 4.5\arcmin$ FoV are able to recover scales comparable to $\theta_{500}$, but provide only a filtered view of scales larger than this radius.
Note that TolTEC on the 50-meter LMT (\S \ref{sec:toltec}) has a $\sim 4\arcmin$ FoV, comparable to that of MUSTANG-2.
Note also that we have assumed for ALMA that we recover scales as large as the primary beam, while typical scales recovered in the compact configuration are $0.6\times$ that.
And finally, note that the scales accessible to the ACA 7-meter array are a factor of $1.7\times$ larger than those accessible to the ALMA 12-meter array at the same frequency.}
\label{fig:cluster_sizes}
\end{figure}

The typical scale used in cosmological applications is $r_{500}$, a convention adopted due to the limited ability of state of the art X-ray instruments to probe the full extent of a cluster in reasonable exposure times. 
 One of the new frontiers for cluster astrophysics is to extend the measurements out to and beyond the virial radius ($\gtrsim 3 \times r_{500}$; see e.g.\ \citet{Walker2019}) and avoid loss of signal due to spatial filtering, which implies it will be important to build more sensitive, multi-band instruments and telescopes that can access tens of arcminute scales even for $z>0.5$. The cluster angular scale corresponding to $r_{500}$ is plotted as a function of redshift to highlight the limitations of current instruments in Figure \ref{fig:cluster_sizes}.  Over-plotted are the radial scales accessible for a few of the instruments discussed in \S \ref{sec:instrumentation}.
 
 % Table of demonstrated mapping speeds
\begin{table}
\caption{
Measured mapping/imaging speeds for current generation subarcminute resolution instruments.  The photometric mapping and interferometric imaging speeds are reported for the average central depth for the map size reported.  For example, in order for the ALMA 12-meter array in Band 3 to recover a depth of 84 $\mu$K over a region as large as a typical NIKA2 or MUSTANG-2 map, one would require $\sim$55-60 hours of integration time.
Note that the map depth generally scales as $t^{-1/2}$ only when not near the confusion limit.
Also, note that the ALMA sensitivity calculator yields sensitivities in Rayleigh-Jeans brightness temperature, rather than $\Delta\Tcmb$ (see \S \ref{app:Tb}).
%$T_{\rm b}=I_\nu \lambda^2 / 2 \kB$, which at frequencies $> 40$~GHz can diverge significantly from $\Delta \Tcmb$.  Following \cite{Finkbeiner1999}, the conversion factor is $\Delta \Tcmb/T_{\rm b} = (\expf{x}-1)^2/(x^2 \expf{x})$, where $x = h\nu/\kB \Tcmb \approx \nu /56.8$ GHz.
}
\label{tab:real_mapping_speeds}       % Give a unique label
\begin{tabular}{lcccccc}
\hline\noalign{\smallskip}
Instrument       & $\nu$ & Resolution & FoV        & Speed in $\Delta \rm I_\nu$ & Speed in $\Delta$T  &  Map Size \\
(name)           & (GHz) & (\arcsec)  & (\arcmin)  & $(\mu \rm Jy~hr^{0.5})$ & ($\mu \rm K~hr^{0.5})$  & (arcmin$^2$) \\
\noalign{\smallskip}\hline\noalign{\smallskip}
ALMA 12-m Band 3 &84-116& $<5\arcsec$  & $<$0.9\arcmin &   11 &   84 & 0.6\\
ACA Band 3       &84-116& $13\arcsec$  & $<$1.4\arcmin &  148 &  162 & 1.5\\
MUSTANG-2        & 90     & $9.5\arcsec$ & 4.5\arcmin    &   56 &   86 & 33\\
NIKA2 (band1)    & 150    & 18\arcsec    & 6.5\arcmin    &  444 &  192 & 36\\
NIKA2 (band2)    & 260    & 11.2\arcsec    & 6.5\arcmin    & 2350 & 2221 & 36\\
\noalign{\smallskip}\hline
\end{tabular}
\end{table}

%%%%%%%%%%%%%%%%%%%%%%%%%%%%%%%%%%%%%%%%%%%%%%%%%%%%%%%%%%%%%%%%%%

\subsection{Future instrumentation}

\subsubsection{TolTEC}
\label{sec:toltec}

TolTEC is a new multi-chroic photometric camera being built for the 50-meter Large Millimeter-wave Telescope (LMT) at Sierra Negra, Mexico for operation in 2019.   The optical design simultaneously couples the $4\arcmin$~FoV onto 6,300 polarisation-sensitive detectors, divided into three focal planes.  Filter passbands  will be centred at 150, 220, and 280 GHz, spanning the null in the tSZ effect and peak in the kSZ.  The large telescope on which this camera will be mounted will result in significant angular resolution gains from previous generation instruments, leading to a 150 GHz FWHM beam size of 9.5\arcsec.  This is sufficient to measure the full profile of the hot gas on the scale of individual large galaxies. The camera is also expected to acheive mapping speeds of 10 deg$^2$/mJy$^2$/hour or better at 150 GHz, depending on the coherence scale of atmospheric fluctuations at the site \citep{Bryan2018}.

%%%%%%%%%%%%%%%%%%%%%%%%%%%%%%%%%%%%%%%%%%%%%%%%%%%%%%%%%%%%%%%%%%
\subsubsection{KISS}
\label{sec:kiss}

The KID Interferometer Spectrometer Surveyor (KISS) is a KID-based MPI spectroscopy camera to be installed at the QUIJOTE 1 telescope at the Teide observatory (at 2390 meters a.s.l. in Tenerife) in November 2018. KISS, which shares a heritage with CONCERTO (\S \ref{sec:concerto}), will operate 600 spatial pixels in the frequency range spanning 70--260~GHz with spectroscopic resolution of $\sim$3~GHz. With a resolution of $\sim3\arcmin$, KISS will target observations of low redshift ($z <0.2$) clusters to map their thermodynamic properties via the tSZ effect, and their velocity and mass via the kSZ effect.

\subsubsection{CONCERTO}
\label{sec:concerto}

CONCERTO\footnote{\url{https://people.lam.fr/lagache.guilaine/CONCERTO.html}}, the CarbON CII line in post-rEionisation and Reionisation epoch \citep{Lagache2018} is a 5000-element KID-based camera, styled off of NIKA2, coupled to an Martin-Puplet interferometer to provide 1.5 GHz resolution spectroscopy in the atmospheric windows between 125-360 GHz. The instrument will be installed on the APEX telescope, at 5100 meters a.s.l., approximately 100 meters higher than the plateau populated by ALMA.
While designed primarily to target [CII] 158 micron emission from high redshift ($4<z<8$) galaxies, it will be capable of targeted tSZ and kSZ observations.  

%%%%%%%%%%%%%%%%%%%%%%%%%%%%%%%%%%%%%%%%%%%%%%%%%%%%%%%%%%%%%%%%%%

\subsubsection{TIME}
\label{sec:time}

The TIME instrument \citep{Crites2014} is designed to measure highly-redshifted 158 micron [CII] (1.9 THz) line emission.  However, due to its wide frequency coverage and 150 GHz continuum photometry channel, the instrument will be capable of measuring the tSZ and kSZ in targeted clusters of galaxies.  The instrument is a 183-336 GHz spectrometer with a photometer channel centred at 150 GHz. The spectrometer will have a resolution of approximately 100 with 60 spectral elements spanning the 183-336 GHz wavelength range. The instrument will have 16 spatial elements each with a 0.4\arcmin\ beam arranged in a single row.  One challenge that normally limits kSZ measurements is effective removal of atmospheric fluctuations. The TIME spectrometer architecture is ideal for removing this source of contamination. Development of the instrument is well underway and is slated to be installed at the Arizona Radio Telescope (ARO) 12-meter prototype ALMA antenna in 2019.

%%%%%%%%%%%%%%%%%%%%%%%%%%%%%%%%%%%%%%%%%%%%%%%%%%%%%%%%%%%%%%%%%%

\subsubsection{Prime-Cam / CCAT-prime}
\label{sec:ccatp}

CCAT-prime will be a 6-m aperture submillimetre telescope located at 5600 meter altitude in Chile and beginning its expected operations in 2021 \citep{Stacey2018}. Prime-Cam is proposed as the first-light bolometer instrument for this telescope, that will cover the frequency range between 190 and 450 GHz with multi-chroic TES detectors, plus host a broadband KIDs array at 860 GHz \citep{Vavagiakis2018}. The high-altitude location means excellent atmospheric transmission at submillimetre wavelengths, which will place CCAT-prime in a unique position to explore the increment of the tSZ spectrum. The expected sensitivity in the $220-400$ GHz bands will be an order-of-magnitude or more better than the {\it Planck} all-sky data (Table 3, \citealt{Erler2018}), thanks to the roughly 10,000 detectors in each band hosted inside a wide focal plane (roughly $1.5^{\circ}$ FoV). Combined with mm-wavelength (tSZ decrement) data either from separate optical tubes in the Prime-Cam or from other CMB surveys like advACT and Simons Observatory, CCAT-prime will provide broad coverage of the tSZ spectrum similar to {\it Planck} within a $\sim 10,000$ deg$^2$ sky area, which is expected to facilitate the separation of the tSZ, kSZ, and rSZ contributions in hundreds of high-mass clusters (\citealt{Mittal2018}, \citealt{Erler2018}). In the 860 GHz band, the angular resolution will be 15 times better than {\it Planck}, providing significant improvements in modelling the CIB and subtracting the contribution from cluster-centric dust emission (e.g., \citealt{Melin2018dust}).

%%%%%%%%%%%%%%%%%%%%%%%%%%%%%%%%%%%%%%%%%%%%%%%%%%%%%%%%%%%%%%%%%%

\subsubsection{Upgrades to current CMB survey instruments}
\label{sec:CMBupgrades}

Advanced ACTpol, which is the current generation instrument on ACT, and SPT-3G, the current generation camera on SPT, will also see some upgrades that will allow them to detect on the order of thousands of SZ selected clusters. 
Both are in the field operating.  
SPT-3G for example is predicted to find $\sim$5000 clusters at a signal-to-noise $\geq$ 4.5 \citep{Benson2014}.
 The ACT 6-meter will join the Simons Observatory, and both SPT and ACT will likely become part of CMB-S4 (\S \ref{sec:futureCMB}).
 
%%%%%%%%%%%%%%%%%%%%%%%%%%%%%%%%%%%%%%%%%%%%%%%%%%%%%%%%%%%%%%%%%%

\subsubsection{Future CMB survey instruments: Simons Observatory and CMB-S4}
\label{sec:futureCMB}
 
The next decade of CMB survey instruments, while not showing any dramatic improvements in spatial resolution, will continue to progress to large detector counts and additional bands from $\sim$30-300 GHz. 
First and foremost, the Simons Observatory \citep[S.O.;][]{simons} will combine several existing CMB experiments in the Atacama desert, and add a new 6-meter telescope with a similar optical design to CCAT-prime (\S \ref{sec:ccatp}).  The S.O.\ large aperture telescope receiver (LATR) will feature $\sim$60,000 detectors spanning 6 broad receiver bands, centred at 27, 39, 93, 145, 225 and 280 GHz.
Combined with CCAT-prime, this will deliver strong constraints on the kSZ effect and on relativistic corrections to the tSZ effect.  The broad multi-band capabilities should significantly advance the ability to separate SZ components from each other and contamination by radio synchrotron and dusty sources. It has an anticipated first light in 2021.

Looking further ahead, instrument developments for the S.O. anticipate many of the CMB survey improvements expected for CMB-S4 \citep{cmbs4_2016}.  The details of the CMB-S4 project are still being determined, but CMB-S4 will likely add up to three 6-meter antennas of similar design as the S.O.\ and CCAT-prime 6-m, and several more lower resolution 1-meter class antennas. CMB-S4 has an anticipated start date circa 2028.  Its dramatic leap forward relies on detector counts 1-2 orders of magnitude larger than S.O.\ and CCAT-prime (\S \ref{sec:ccatp}), finally reaching megapixel counts.

%%%%%%%%%%%%%%%%%%%%%%%%%%%%%%%%%%%%%%%%%%%%%%%%%%%%%%%%%%%%%%%%%%

\subsubsection{AtLAST}\label{sec:atlast}

The Atacama Large Aperture Submillimeter/millimeter Telescope (AtLAST)\footnote{\url{http://atlast-telescope.org/}} is a broad international community-driven effort to build a 50-meter class single dish observatory with over 1 square degree instantaneous FoV in the Atacama desert in northern Chile.
It will likely be located at $\approx 5100$ meters a.s.l., close to the ALMA site, which has demonstrated performance over all full ALMA bands (35-950 GHz) and would facilitate accessibility and operational logistics \citep{de_breuck_carlos_2018_1158848}.

Conservative estimates assuming a 1 megapixel photometric array for AtLAST indicate its mapping speed will be well over $10^5 \times$ faster than the full ALMA observatory (c.f.\ Equations \ref{eq:interferometric_sensitivity} and \ref{eq:photometric_sensitivity}).
Currently no submm facilities larger than 12 meters are operational in the southern hemisphere, and AtLAST will serve as a powerful complement to the many lower-resolution primary CMB and SZ survey telescopes located in or planned for the southern hemisphere (see \S \ref{sec:futureCMB}).
At the same wavelengths as the LMT 50-meter (\S \ref{sec:toltec}), AtLAST will have the same spatial resolution.  However the large ($>1~\rm degree^2$) FoV, planned higher surface accuracy, and better atmospheric transmission from $\gtrsim 5100~\rm meters$ a.s.l.\ will allow for simultaneous measurements in the submillimetre ($\sim 350-950$~GHz), to better probe contamination from dusty sources and the increment in the SZ effects.  AtLAST will thus build upon the results with CCAT-prime (\S \ref{sec:ccatp}) and deliver twice the resolution of the original CCAT 25-meter.

%%%%%%%%%%%%%%%%%%%%%%%%%%%%%%%%%%%%%%%%%%%%%%%%%%%%%%%%%%%%%%%%%%

\subsubsection{CSST}\label{sec:csst}

The Chajnantor Sub/millimeter Survey Telescope (CSST) is a Caltech-led effort to build an inexpensive 30-meter class single-dish survey telescope operating between 90--420~GHz. The concept is based upon a novel telescope design that would allow for a significant reduction in cost in exchange for a somewhat limited range of motion ($\sim$1~radian, \citealt{Padin2014}). CSST will seek a site on the Chajnantor plateau. The nominal first-light instrument would be a photometric camera with 6 observing bands filling a $1\deg$ FoV. As a survey telescope, CSST would have access to $\sim$20,000~$\deg^2$ (half the sky) and would be able to map $\sim$1,000~$\deg^2$/year to the confusion limit due to dusty sources at 350~GHz. The actual survey design would likely consist of a ``wedding cake'' to varying depths. While pointed cluster observations would also be possible, the survey data themselves would provide extremely deep SZ maps of a very large sample of clusters, thus enabling a range of novel SZ science by significantly expanding the spectral coverage available from other large-aperture telescopes like GBT, IRAM, and LMT while providing significantly better angular resolution than any of the planned CMB survey telesopes. 

%%%%%%%%%%%%%%%%%%%%%%%%%%%%%%%%%%%%%%%%%%%%%%%%%%%%%%%%%%%%%%%%%%

\subsubsection{Future space missions}
\label{sec:space}
There has also been significant discussion of and planning for SZ science from space- and balloon-borne platforms. {\it Planck} has clearly demonstrated that wide frequency coverage, control of systematics, and coverage of a large sky fraction are some of the key ingredients for obtaining and doing science with large samples of SZ clusters \citep[e.g.,][]{Planck2011_VII, Planck2016_XXVII}. 

Subarcminute angular resolution at mm wavelengths is challenging for space (and near-space), as the size of the primary mirror strongly drives the cost. 
However, future CMB mission concepts such as {CORE} \citep{Delabrouille2018, Remazeilles2018} and {PICO} could take a dramatic leap forward in sensitivity and spectral coverage versus that of {\it Planck}, enabling an increase in the number of detected SZ clusters to a ${\rm few}\times 10^5$ \citep{Melin2018JCAP}. This will open the way for detailed SZ studies with unprecedented precision, allow a vast number of cross-correlation studies, and provide additional constraints on cosmological parameters and neutrino mass and number \citep[e.g., see][]{Hill2013, Bolliet2018, Makiya2018, Salvati2018, simons}.

Higher angular resolution and wide spectral coverage across the mm, submm, and FIR from space could be achieved with the Millimetron Space Observatory (MSO) \citep{Smirnov2012,Smirnov2018}. This concept uses a 10-meter folded mirror approach to overcome some of the payload limitations of the launch vehicle.  The mirror will be cooled to $\sim4.5~\rm K$ to reduce the noise contribution due to thermal emission  (which is already low, due to the low intrinsic emissivity of the mirror). MSO will target individual clusters, resolving them at subarcminute resolution with the goal of extracting the rSZ signals and reconstructing cluster profiles.

Finally, even low-resolution all-sky CMB measurements using spectrometer concepts such as PIXIE \citep{Kogut2011PIXIE, Kogut2016SPIE}, PRISTINE, and the spectrometer on board of {\it CMB-Bharat} could provide a new census of the hot baryons in the Universe by constraining the average all-sky $y$-parameter and $y$-weighted temperature \citep{Hill2015}. These concepts use a Fourier-Transform-Spectrometer (FTS) approach to sample the absolute CMB sky intensity in hundreds of frequency channels covering a few$\times (10\,{\rm GHz} - 1\,{\rm THz})$. A similar FTS concept was used by COBE/FIRAS to measure the CMB energy spectrum in the 1990's, and by {\it Olimpo} \citep{Masi2003, Schillaci2014}, a balloon-borne SZ experiment targeting clusters that had its first flight in mid-2018.

%%%%%%%%%%%%%%%%%%%%%%%%%%%%%%%%%%%%%%%%%%%%%%%%%%%%%%%%%%%%%%%%%%
\section{Conclusions}
\label{sec:conclusion}

SZ surveys and a few pioneering instruments have made tremendous progress in the past two decades, and detections and imaging of the SZ effect from massive systems has now become routine.
With upcoming advances brought by new instruments, observing across the centimetre/millimetre/submillimetre spectrum ($\sim 30~\rm GHz - 1~\rm THz$) and delivering higher spatial resolution over larger FoV, this progress will continue.
It will soon be possible to characterise the intra-group, circumgalactic, and warm/hot intergalactic medium through SZ observations.
In order to be able to probe small scales and achieve better spatial fidelity than currently possible, any future SZ facility must not only have high angular resolution, but also a large FoV and broad frequency coverage over most of the mm/submm bands.  Fortunately, these are generic requirements to future mm/submm facilities, not endemic to galaxy clusters and SZ measurements.
New facilities should aim for imaging of scales ranging from sub-kpc to several Mpc, and perhaps larger for the warm-hot component of the cosmic web.

In the near term, our SZ window on the warm and hot Universe could potentially benefit by performing a few key, deep ($\gg100\sigma$) SZ observations, analogous to deep X-ray and HST observations, to address specific topics in astrophysics.
For comparison, the most transformative results with {\it Chandra} have often come out of  very deep ($\geq 400$~ksec) and `visionary' ($\sim 2$~Msec)\footnote{Recall that 100 hours is 360~ksec, and most SZ observations are only a few to tens of hours.} observations of shocks, cold fronts, AGN outbursts, and nearby clusters \citep[see e.g.][]{mcnamara2007,Markevitch2007,Blanton2011,Zhuravleva2014b,Walker2018b}.
Comparably deep SZ observations should exploit the key strength common to the SZ effects -- redshift-independent surface brightness -- by e.g. probing the evolution of clusters and proto-clusters in the first half of cosmic history.
SZ observations have only begun the scratch the surface, and one can imagine using the tSZ to image the ICM farther into the cluster outskirts, to the accretion shocks and perhaps into the surrounding filaments.

Two decades ago, measurements of mK level temperature fluctuations were state of the art; now, large surveys and deep, targeted observations are achieving RMS values in the few $\mu$K range.
Our ability to probe the tSZ effect will further advance, and what is now possible through stacking and radial binning will eventually lead to direct imaging of the intra-group, circum-galactic, and warm-hot intergalactic media.
Apart from relativistic effects, which provide a handle on ICM temperature, the tSZ has no formal band dependence and can probe gas that is nearly invisible in the soft and medium X-ray bands.

Our ability to probe clusters and other structures through the kSZ will also improve through current and near future instrumentation advances, making kSZ detections and studies of gas motions and cluster rotation more routine. Nonetheless, given the faintness of the kSZ signal in most clusters, such measurements will remain  challenging. Astrophysical contaminants such as Galactic cirrus, dust, and radio and submm galaxies (\S \ref{sec:contam}) remain difficult to separate and remove fully, due to limited spectral leverage and poor signal to noise of the individual components. 
Additionally, systems in which the kSZ signal is strong are often merging objects in which temperatures locally boosted by shocks may require non-negligible relativistic corrections.  It is clear that to significantly advance the field, instruments going far beyond the state of the art -- 2-3 colour instruments plus sparse ancillary radio/submm data -- will be necessary to constrain all of the SZ contributions as well as the contaminants.\footnote{We note that much of the kSZ and rSZ work reported here relied on {\it Herschel} observations for submm source subtraction, and the number of clusters observed by {\it Herschel} during its lifetime is quite limited.  On the submm observational side, CCAT-prime (\S~\ref{sec:ccatp}) may be the clearest near-term successor to {\it Herschel}, with nearly twice the resolution and a longer expected project lifetime, while e.g.\ ALMA (\S~\ref{sec:alma_B3}) or AtLAST (\S~\ref{sec:atlast}) could constrain the flux densities of compact sources directly in each band of interest.}

Disentangling the velocity of the gas from its optical depth will likely also remain a challenge for kSZ analyses, as it is difficult to model the electron distribution in complex systems.  A promising avenue will open when rSZ can be used to make spatially-resolved measurements of ICM temperature.
At this point, SZ characterisation of the motions and bulk properties of the ICM and warm/hot baryonic components in the Universe may be less directly dependent upon X-ray data.  However, we note that direct measurements of the mass-weighted temperature $\Tmw$ may remain elusive, as rSZ yields a Compton $y$-weighted temperature unless higher order moments of the temperature-field become accessible (see discussion in \S~\ref{sec:XSZ}).

Near future instrumentation will also soon allow us to measure the non-thermal SZ effect (\S~\ref{sec:ntSZ_detail}), which has the unique capability of measuring the composition of AGN driven bubbles/cavities (\S~\ref{sec:agn_feedback_physics}) and, potentially, could even allow one to constrain the nature of dark matter. Both have been subjects of longstanding interest across many branches of astrophysics and cosmology, including large scale simulations and the fields of galaxy evolution and black hole growth.
Looking forward to the more distant future, perhaps we can expect stacking and radial averaging to bring more exotic aspects of the SZ, such as the polarised and multiple scattering effects (\S~\ref{sec:pSZ} and \S~\ref{sec:msSZ}), within reach. 

\bigskip

\begin{acknowledgements}
We thank ISSI for the opportunity to provide this invited review.
T.M.\ is supported for scientific activities by ESO's Directorate for Science.
D.N.\ acknowledges Yale University for granting a triennial leave and the Max-Planck-Institut f\"ur Astrophysik for hospitality when this work was carried out. 
J.C.\ is supported by the Royal Society as a Royal Society University Research Fellow at the University of Manchester, U.K.
R.A.\ acknowledges support from Spanish Ministerio de Econom\'ia and Competitividad (MINECO) through grant number AYA2015-66211-C2-2. 
K.B.\ acknowledges partial funding from the Transregio programme TRR33 of the Deutsche Forschungsgemeinschaft (DFG). 
A.T.C.\ is supported by the National Science Foundation Astronomy and Astrophysics Postdoctoral Fellowship under Grant No.\ 1602677.
F.M., L.P., J.F.M.P., and F.R.\ acknowledge funding from the French ANR under the contract ANR-15-CE31-0017 and from the ENIGMASS LabEx. 
\end{acknowledgements}

\begin{appendix}

\section{Appendix}
\label{appendix}

This appendix covers some of the practical aspects common to observations of the SZ effect.  What follows are observational considerations like beam size, beam solid angle, and how to compare sensitivity in common units like Jy/beam, $\mu$K, and  $\mu \rm \Kcmb-\rm arcmin$.

\subsection{Beam Size}
\label{app:beam}

After frequency and sensitivity, the primary observational considerations are resolution and scales recovered.  The scales recovered differ for observations with an interferometric array and for photometric measurements with a bolometric array on a single dish, and are considered in \S \ref{sec:interferometric} and \S \ref{sec:photometric}, respectively.

The full width at half maximum $\theta_{\mbox{\tiny FWHM}}$ for the main beam of a diffraction limited telescope can be approximated as a Gaussian,
\begin{equation}
\label{eq:fwhm}
\theta_{\mbox{\tiny FWHM}} \approx 1.22 \frac{\lambda}{D},
\end{equation}
for wavelength $\lambda$ and dish diameter $D$. 
Any power outside of this main beam, such as that due to `side lobes' (see e.g.\ \cite{tms}),  is often referred to as the `error beam,' and can often be characterised with a wider Gaussian of lower amplitude.
In practice, $\theta_{\mbox{\tiny FWHM}}$ is often slightly larger due to under-illumination of the primary and imperfect focusing.
The solid angle $\Omega_{\mbox{\tiny bm}}$ subtended by a beam with $\theta_{\mbox{\tiny FWHM}}$ is
\begin{equation}
\label{eq:solidangle}
\Omega_{\mbox{\tiny bm}} = \frac{\pi (\theta_{\mbox{\tiny FWHM}}/2)^2} {\ln(2)}.
\end{equation}
For instance, the 100-m Green Bank Telescope operating at 90~GHz (3.3 mm) would have a diffraction limited FWHM of 8.38\arcsec and a corresponding beam volume $\Omega_{\mbox{\tiny bm}} = 80$ square arcseconds; however, in practice the under-illumination yields a main beam which, if fitted by single Gaussian would have $\theta_{\rm FWHM} \approx 9.5$\arcsec. Surface imperfections create an error beam, such that the total beam volume is better estimated as $\Omega_{\mbox{\tiny bm}} = 120$ square arcseconds.

\subsection{Surface brightness}
\label{app:sb}

The surface brightness $S_\nu$ in units of flux density per beam (in $Jy/bm$) is related to intensity $\Delta I_\nu$ (Equations \ref{eq:dI_tsz}, \ref{eq:dI_ksz}, \& \ref{eq:dI_general_beta})
by integrating over the beam solid angle:
\begin{equation}
\label{eq:SB}
\Delta S_\nu =\int \! \Delta I_\nu \, \id\Omega 
= \langle \Delta I_\nu \rangle \, \Omega_{\mbox{\tiny bm}}.
\end{equation}
Generically, one can convert between intensity ${\Delta I}$ and a change in the CMB temperature ${\Delta\Tcmb}$ using the derivative of the blackbody function.  The ratio is 
\begin{equation}
\label{eq:ItoTcmb}
\frac{\Delta I}{\Delta\Tcmb} = \frac{I_0}{\Tcmb} 
\frac{x^4 \expf{x} }{(\expf{x}-1)^2},
\end{equation}
where the primary CMB intensity normalisation $I_0$ was defined in Eq.~\eqref{eq:I0_cmb}.

\subsection{CMB survey noise}
\label{app:uKarcmin}

The RMS noise in maps made using arcminute-resolution CMB instruments, such as those from ACT and SPT, are often compared in units of $\mu\Kcmb-\rm arcmin$, defined as the RMS of the CMB temperature fluctuations $\Delta\Tcmb$ within a map created with pixels that each subtend a solid angle of 1 arcmin$^2$.
To convert this figure to the RMS of CMB temperature fluctuations within a given instrument's beam,
%(e.g.,\ in $\mu\Kcmb$ for use in Eqs.\ \ref{eq:dT_tsz} and \ref{eq:dT_ksz}),
%To convert this figure to units of the CMB temperature fluctuation, 
one would divide by the square root of the beam solid angle in square arcminutes (Eq.~\ref{eq:solidangle}).
For example, a map made from an instrument with $\Omega_{\mbox{\tiny bm}} = 120$ square arcseconds (0.033 arcmin$^2$) with an RMS noise of $10~\mu\Kcmb$ per beam  would correspond to $1.8~\mu\Kcmb-\rm arcmin$.  
This conversion assumes the noise properties in the maps are Gaussian and uncorrelated on the scales being binned, which is a simplification that is not generically applicable.

\subsection{Rayleigh-Jeans brightness temperature}
\label{app:Tb}

Many instruments report sensitivities in Rayleigh-Jeans brightness temperature
\begin{equation}
\label{eq:Trj}
\Delta T_{\rm b} \equiv \Delta I_\nu \lambda^2 / 2 \kB, 
\end{equation}
which at $\nu > 40$~GHz can diverge significantly from the temperature decrement in the CMB, $\Delta T$, in units of \Kcmb.  
This relation can also be expressed in surface brightness units (e.g. using Eq.~\ref{eq:SB}), as
\begin{equation}
\label{eq:SB_Trj}
\Delta S_\nu  = 2 \kB \Delta T_{\rm b} \Omega_{\mbox{\tiny bm}} / \lambda^2. 
\end{equation}

Brightness temperature can be trivially converted to units more directly applicable to CMB and SZ measurements.
Following \cite{Finkbeiner1999}, this conversion, called the `Planck correction factor', is 
\begin{equation}
\label{eq:planck_corr}
\frac{\Delta\Tcmb}{\Delta T_{\rm b}} = \frac{(\expf{x}-1)^2}{x^2 \expf{x}},
\end{equation}
where $x$ is defined as it was for Equation \ref{eq:dI_tsz}.
In Table \ref{tab:PlackCorr}, we provide computations of the ratio of $\Delta\Tcmb/\Delta T_{\rm b}$ for a few representative frequencies.

\begin{table}[h]
\caption{Correction factors for converting surface brightness temperature $T_{\rm b}$ to $\Delta\Tcmb$ and the classical (non-relativistic) tSZ functions $g(x)$ and $f(x)$ in Equations \ref{eq:dI_tsz} and \ref{eq:dT_tsz}.}
\label{tab:PlackCorr}       % Give a unique label
\begin{tabular}{ccccccccccc}
\hline\noalign{\smallskip}
{$\nu$ (GHz)}     & 30   & 40   & 90    & 150  & 220  & 280  & 350  & 400  & 640 & 870 \\
{$\Delta \Tcmb / \Delta T_{\rm b}$} & 1.02 & 1.04 & 1.23 & 1.74 & 3.07 & 5.61 & 12.4 & 23.0 & 616 & 19126 \\
$g(x)$ & -0.53 & -0.91 & -3.27 & -3.83  & 0.18  & 4.33 & 6.68  & 6.58 &  1.50 &  0.14 \\
$f(x)$  & -1.95 &  -1.92  & -1.60 &  -0.95 & 0.04 & 1.00 &   2.19  &  3.05  &  7.27  & 11.3 \\
\noalign{\smallskip}\hline
\end{tabular}
\end{table}

\subsection{Lists of tables and figures}

\listoftables
\listoffigures

\end{appendix}

\bigskip

\bibliographystyle{apj}      % basic ApJ style, author-year citations
%\bibliographystyle{aps-nameyear}      % American Physical Society (APS) style, author-year citations


\bibliography{sz_bibliography}   % name your BibTeX data base

\end{document}